\documentclass{aa}
\usepackage{graphicx}
\usepackage{txfonts}
\usepackage{enumerate}
\usepackage{aalongtable}

\newcommand{\msun}{$M_{\odot}$}

\begin{document}
   \title{A SCUBA imaging survey of ultracompact HII regions}

   \subtitle{The environments of massive star formation\thanks{A full version of Fig.~\ref{fig:scuboth} is only available in
   electronic form at \texttt{http://www.edpsciences.org}}}

   \author{M.A.~Thompson\inst{1,\,2}, J.~Hatchell\inst{3}, A.J.~Walsh\inst{4}, G.H.~Macdonald\inst{2}
    \and T.J.~Millar\inst{5,\,6}}

   \authorrunning{M.A.~Thompson et al.}
   \titlerunning{A SCUBA imaging survey of UC H{\sc ii} regions}
   
   \offprints{M.A.~Thompson\\ \email{mat@star.herts.ac.uk}}

   \institute{Centre for Astrophysics Research, Science \& Technology Research Institute,
              University of Hertfordshire, College Lane, Hatfield, AL10 9AB, UK
              \and Centre for Astrophysics \& Planetary Science,
              School of Physical Sciences, University of Kent, Canterbury, CT2 7NR, UK
              \and School of Physics, University of Exeter, Stocker Road, Exeter EX4 4QL, UK
              \and School of Physics, University of New South Wales, NSW, 2052, Australia.
              \and Jodrell Bank Centre for Astrophysics, School of Physics and Astronomy, 
	      University of Manchester, P.~O.~Box 88, Manchester, M60 1QD, UK
	      \and School of Mathematics and Physics, Queen's University Belfast, Belfast BT7 1NN, UK
              }

   \date{}

   \abstract{We present a SCUBA submillimetre (450 \& 850 $\mu$m) survey of the environment of 105 IRAS point
   sources, selected from the Wood \& Churchwell (\cite{wc89a}) and Kurtz, Churchwell \& Wood
   (\cite{kcw94}) radio
   ultracompact (UC) H{\sc ii} region surveys. We detected a total of 155 sub-mm clumps associated with the IRAS point
   sources and identified three distinct types of object: ultracompact cm-wave sources that are not
associated with any sub-mm emission (sub-mm quiet objects), sub-mm clumps that are associated with
ultracompact cm-wave sources (radio-loud clumps); and sub-mm clumps that are not associated with any known
ultracompact cm-wave sources (radio-quiet clumps).  90\% of the sample of IRAS point sources were found to be
   associated with strong sub-mm emission. We consider the sub-mm colours, morphologies and
   distance-scaled fluxes of the sample of sub-mm clumps and show that the sub-mm quiet objects are unlikely to
   represent embedded UC H{\sc ii} regions unless they are located at large heliocentric distances. Many of the
   2\farcm5 SCUBA fields contain more than one sub-mm clump, with an average number of
companions (the companion clump fraction) of 0.90. The
   clumps are more strongly clustered than other candidate HMPOs and the mean clump surface density exhibits a broken
   power-law distribution with a break at 3 pc. We demonstrate that the sub-mm and cm-wave fluxes of the majority
   of radio-loud clumps are in excellent agreement with the standard model of ultracompact H{\sc ii} regions. We
   speculate on the nature of the radio-quiet sub-mm clumps and, whilst we do not yet have
   sufficient data to conclude that they are in a pre-UC H{\sc ii} region phase, we argue that their characteristics
   are suggestive of such a stage.}

   \maketitle
%

\section{Introduction}

Ultracompact (UC) H{\sc ii} regions are perhaps one of the most reliable tracers of recent
massive star formation.  UC H{\sc ii} regions are defined as dense, compact bubbles of
photoionised gas less than 0.1 pc in diameter, surrounding and excited by massive young
stellar objects (YSOs).  The estimated age of UC H{\sc ii} regions is between 10$^{4}$--10$^{5}$
years, inferred from their spatial diameters and the typical expansion rates of H{\sc ii}
regions.  Ultracompact H{\sc ii} regions should thus trace the earliest embedded phases of
massive star formation, wherein the YSOs have just begun to ionise their surrounding natal
molecular clouds. For recent reviews of UC H{\sc ii} regions see Churchwell (\cite{c2002}),
Kurtz et al.~(\cite{kcchw00}) or Garay \& Lizano (\cite{gl99}).

Most UC H{\sc ii} regions have to date been identified by their centimetre-wavelength free-free
emission, due to the ability of radio-wavelength radiation to penetrate the dense molecular
gas and dust cores within which the UC H{\sc ii} regions are embedded. There are two types of radio
surveys that have been used to find UC H{\sc ii} regions, each relying on snapshot radio
interferometry either targeted toward colour-selected IRAS point sources  (Wood \& Churchwell
\cite{wc89a}; Kurtz, Churchwell \& Wood \cite{kcw94}) or over unbiased regions of the Galactic
Plane  (Becker et al.~\cite{b90}, \cite{b94}; Zoonematkermani et al.~\cite{z90}; Giveon et
al.~\cite{gbh2005}). The targeted
surveys rely upon the Wood \& Churchwell (\cite{wc89b}) IRAS colour selection criteria 
for UC H{\sc ii} regions, which propose that UC H{\sc ii} regions possess distinctive far-infared colours
and can be identified by a discrete region in a 25/12 $\mu$m versus 60/12 $\mu$m colour-colour
diagram. 

Few UC H{\sc ii} regions have been mapped at a sufficient angular scale or resolution to investigate
their dense molecular surroundings and this means that we are relatively ignorant of the
location of the UC H{\sc ii}s within their embedding cores, the global physical properties of the
cores and the wider star-forming neighbourhood. Improving our understanding of the enviroments
in which massive stars form is crucial to our comprehension of massive star formation and the
impact that massive stars have upon their surroundings. In particular it is becoming apparent
that a number of UC H{\sc ii} regions are in fact bright compact components of more extended
emission (Kurtz et al.~\cite{kwho99}; Kim \& Koo \cite{kk01}). One possible explanation for
these extended regions is that the hierarchichal structure of molecular clouds can give rise
to H{\sc ii} regions with both ultracompact and extended components  if the O star exciting the H{\sc ii}
region is displaced from the centre of its embedding core (Kim \& Koo \cite{kk01},
\cite{kk03}). 

Recently developed millimetre and sub-millimetre wavelength bolometers such as SCUBA (Holland et
al.~\cite{scuba}), SHARC (Hunter, Benford \& Serabyn \cite{hbs96}), MAMBO (Kreysa et al.~\cite{mambo}) and SIMBA
(Nyman et al.~\cite{simba}) have made it possible to rapidly map dust emission from extended regions over 
arcminute scales at resolutions of 10--20\arcsec. A number of sub-millimetre and millimetre continuum mapping
studies of massive star-forming regions have been carried out (e.g.~Hill et al.~\cite{hill2005}; Walsh et
al.~\cite{walsh03}; Mueller et al.~\cite{msej02}; Hatchell et al.~\cite{hfmtm00}; Hunter et al.~\cite{hcwcb00})
but the largest surveys have so far been aimed at H$_{2}$O (Mueller et al.~\cite{msej02}) or CH$_{3}$OH masers
(Walsh et al.~\cite{walsh03}; Hill et al.~\cite{hill2005}) rather than UC H{\sc ii} regions. Although the
individual populations of H$_{2}$O masers, CH$_{3}$OH masers and UC H{\sc ii} regions are known to overlap
somewhat (e.g.~Walsh et  al.~\cite{wbh98}, Palagi et al.~\cite{pcc93}),  without a dedicated large-scale UC H{\sc
ii} region survey it is difficult to understand the potentially different physical characteristics of masing
versus non-masing regions. Nevertheless, surveys of this type are beginning to reveal the density distributions
and physical properties of massive star-forming cores. Additional dust clumps located near to those containing the
active star formation traced by masers or UC H{\sc ii}s are also being discovered (Hill et al.~\cite{hill2005};
Walsh et al.~\cite{walsh03}; Hatchell et al.~\cite{hfmtm00}) and the implication is that these clumps may
represent earlier evolutionary stages than the maser cores or UC H{\sc ii} regions.

In order to more fully understand the environments of UC H{\sc ii} regions, constrain the physical properties of
their embedding dusty clumps and search for nearby associated clumps, we have carried out a sub-mm SCUBA imaging
survey of 105 UC H{\sc ii} regions from the Wood \& Churchwell (\cite{wc89a}) and Kurtz, Churchwell \& Wood
(\cite{kcw94}) radio catalogues. In this paper, the first of a series reporting our study, we present a
description of the survey  procedure,  the SCUBA images of the UC H{\sc ii} regions and their sub-mm fluxes, plus
a statistical analysis of their sub-mm colours, morphologies, clustering properties and likely physical natures.
A detailed treatment of the physical properties (e.g.~temperature, mass, density profile) of each of the sub-mm
sources detected in the survey and their correlation with the presence of  hot molecular cores will be presented
in subsequent publications.

In Sect.~\ref{sect:obs} we outline the sample selection, observational and data reduction
procedure. The 450 \& 850 $\mu$m images and the fluxes of the sources detected in the survey are
presented in Sect.~\ref{sect:results}. We investigate the sub-mm detection statistics of the UC H{\sc ii}
regions and their positional association with sub-mm cores, the morphology of the sub-mm clumps and their
clustering properties in Sect.~\ref{sect:analysis}. Finally,
in Sect.~\ref{sect:concl} we summarise the initial results of the survey and its implications for
the future study of massive star-forming regions.


\section{Observational procedure and data reduction}
\label{sect:obs}

\subsection{Sample selection}
\label{subsect:sample}

The sample of UC H{\sc ii} regions to be observed was drawn from the radio catalogues of Wood \&
Churchwell (\cite{wc89a}, hereafter WC89a), and Kurtz, Churchwell \& Wood (\cite{kcw94},
hereafter KCW94). Each of these catalogues consists of a list of compact radio sources
observed with the VLA  towards IRAS FIR point sources.  Although each radio survey 
consisted of single VLA pointings toward individual IRAS sources their initial survey
selection criteria differ somewhat. The WC89a survey selected their target objects from
known compact H{\sc ii} regions (Wink, Altenhoff \& Mezger \cite{wam82};  Wood, Churchwell \&
Salter \cite{wcs88}; Wood et al.~\cite{whf88}) that were likely to contain ultracompact
components as suggested by their long-wavelength spectra. The KCW94 survey on the other
hand selected their targets from a sample of bright IRAS point sources (S$_{100 \mu \rm m}
>$ 1000 Jy) that solely fulfilled the Wood \& Churchwell (\cite{wc89b}) FIR colour criteria
for UC H{\sc ii} regions. KCW94 made no attempt to screen their sample for well-known objects of
other types (e.g. planetary nebulae, compact or extended H{\sc ii} regions) and as will be seen
later certain of the KCW94 objects are not true UC H{\sc ii} regions.

The combined WC89a and KCW94 surveys consist of observations of 143 individual IRAS point sources, of
which radio emission from UC H{\sc ii}s was detected toward 101 IRAS point sources.  Many of the IRAS
sources display multiple radio components: out of 101 IRAS sources with associated radio emission  a
total of 150 discrete ultracompact radio components are identified in the WC89a and KCW94 surveys.
Our initial SCUBA sample was selected from the 143 individual IRAS sources observed by WC89a and
KCW94. As the SCUBA ``instantaneous'' field of view (FOV) is very similar to the primary beam of the
VLA at 2cm and the IRAS 100 $\mu$m beam FWHM we chose to observe each IRAS point source with a single
SCUBA field. 

There are 42 IRAS sources from the WC89a and KCW94 surveys towards which no radio emission was detected
at the time of the WC89a \& KCW94 surveys. It is possible that these objects may either represent H{\sc ii}
regions extended over scales $\ge$ 10\arcsec, which would have been missed by the snapshot
interferometric observations of the radio surveys, UC H{\sc ii} regions that were too faint to be detected in
the original surveys, or potential massive protostellar objects in an evolutionary phase prior to that
of UC H{\sc ii} regions. We searched the SIMBAD database of astronomical catalogues for more recent radio
observations of these fields and discovered that ultracompact radio components had been found toward 21
of these IRAS point sources (predominantly from the 5 GHz galactic plane survey of Becker et
al.~\cite{b94} and Giveon et al.~\cite{gbh2005}). In compiling this list of ultracompact radio
components we matched each survey against the others to determine a unified source list. Where two
(or more) surveys report a radio source of similar brightness within 5\arcsec\ of one another we
consolidated  these sources as a single detection. Positions and fluxes are used from the survey with
the highest signal-to-noise ratio (generally the WC89a or KCW94 surveys). Thus our initial sample
comprises 122 IRAS point sources with associated compact centimetre-wave emission ("UC H{\sc ii} regions") 
and 21 IRAS point sources that are not associated with any detected compact centimetre-wave
components.   The latter 21 IRAS sources were included in our sample so that we could compare their
sub-mm fluxes and morphologies with those dust cores known to contain UC H{\sc ii} regions.

Whilst the initial sample to be observed consisted of 143 IRAS point sources, due to the manner in
which our observations were carried out (flexibly scheduled mode, which is described further in
Sect.~\ref{subsect:op}) it was not possible to complete observations of the entire sample. We were
able to map 105 out of the total number of 143 IRAS point sources, of which 94/122 UC H{\sc ii}-associated
IRAS point sources and 9/21 non-UC H{\sc ii}-associated point sources were mapped. The mapped fields and
their coordinates are given in Table \ref{tbl:samplelist}. In terms of the fields containing known
ultracompact radio components our survey is thus $\sim$ 80\% complete. As the selection of the mapped
fields from within the larger sample was essentially random (depending upon hour angle, weather
conditions, instrument availability etc.) we are confident that there is little selection bias within
our final list of UC H{\sc ii} regions.

\begin{table*}
\begin{minipage}{\linewidth}
\caption{UC H{\sc ii} fields observed during the survey, their central coordinates and estimated distance to the UC H{\sc ii}s in each field.
Fields indicated with a dagger ($^{\dag}$) show no evidence for UC H{\sc ii} regions in the
WC89a and KCW94 surveys, or in the literature search described in Sect.~\ref{subsect:sample}. Fields
indicated by an asterisk ($^{*}$) were subject to the 450 $\mu$m calibration problems described in
Sect.~\ref{subsect:dr}. In cases where
their is an ambiguity between the near and far kinematic distance both values are listed. 
Where the distance to the UC H{\sc ii} in the field is not known or there is no known UC H{\sc ii} in the field an ellipsis is given.} 
\label{tbl:samplelist}
\begin{tabular}{lccrr|lccrr}\hline\hline
UC H{\sc ii}& \multicolumn{2}{c}{Field centre (J2000)} & Dist. & Dist. & 
UC H{\sc ii} & \multicolumn{2}{c}{Field centre (J2000)} & Dist. & Dist. \\ 
 field name  & RA  & Dec. & (kpc) &Ref. & field name &RA  & Dec. & (kpc) &  Ref. \\\hline
G1.13$-$0.11   & 17:48:41.0   &  $-$28:01:42.4    &  8.5  & 1  &   G31.40$-$0.26  &   18:49:34.7 &   $-$01:29:08.4   & 7.3   & 16  \\
G4.41+0.13     & 17:55:17.6   &  $-$25:05:01.3    & \ldots   &   &    G32.80+0.19  &   18:50:30.9 &   $-$00:01:59.3   & 12.9   & 17  \\
G5.48$-$0.24   & 17:59:02.7   & $-$24:20:54.1     &  14.3  &  2 &   G32.96+0.28$^{\dag}$  &   18:50:30.3 &   +00:09:04.7   &  \ldots  &   \\
G5.89$-$0.39   & 18:00:30.4   &  $-$24:04:00.2    &  2.0  &  3 &    G33.13$-$0.09  &   18:52:09.9 &   +00:08:39.7   &  7.1  &  7 \\
G5.97$-$1.17   & 18:03:40.5   & $-$24:22:44.3     &  2.7  &  4 &   G33.92+0.11  &   18:52:50.2 &   +00:55:29.4   &  7.1  & 5  \\
G6.55$-$0.10   & 18:00:50.0   & $-$23:20:32.5     & 16.7  &  2 &    G34.26+0.15  &  18:53:18.6 &   +01:14:57.7   &  4.0 &   4   \\  
G8.14+0.23     & 18:03:00.7   & $-$21:48:08.4     &  4.2  &  2 &    G35.02+0.35  &  18:54:04.2  &  +02:01:33.9    & 10.0 & 18   \\ 
G8.67$-$0.36   & 18:06:19.0   & $-$21:37:32.3     &  4.8  &  5 &    G35.05$-$0.52$^{*}$  &  18:57:09.0  &  +01:39:03.4    & 12.7 & 2 \\ 
G9.62+0.20     & 18:06:13.5   & $-$20:31:47.3     &  5.7  &  6 &    G35.57+0.07$^{*}$  &  18:56:00.7  &  +02:22:52.2    & 10.2 & 18   \\ 
G9.88$-$0.75   & 18:10:17.1   & $-$20:45:42.5     &  3.9  &  7 &    G35.58$-$0.03$^{*}$  &  18:56:23.5  &  +02:20:37.8    & 10.1 & 18 \\ 
G10.10+0.74    &  18:05:13.1  &   $-$19:50:34.7   &  4.4  &  5 &    G37.36$-$0.23$^{*}$  &  19:00:20.7  &  +03:50:15.6    & \ldots &  \\ 
G10.15$-$0.34  &  18:09:21.0  &   $-$20:19:30.9   &  6.0  &  2 &    G37.55$-$0.11$^{*}$  &  19:00:16.0  &  +04:03:15.1    & 9.9 & 4 \\ 
G10.30$-$0.15  &  18:08:56.2  &   $-$20:05:53.4   &  6.0  &  2 &    G37.77$-$0.20$^{*}$  &  19:01:00.4  &  +04:12:51.4    & 8.8 & 18  \\ 
G10.62$-$0.38  &  18:10:28.7  &   $-$19:55:51.7   &  4.8  &  5 &    G37.87$-$0.40$^{*}$  &  19:01:53.0  &  +04:12:50.1    & 9.2 & 19 \\ 
G10.84$-$2.59  &  18:19:11.9  &   $-$20:47:33.6   &  1.9  &  7 &    G39.25$-$0.06$^{*}$  &  19:03:13.7  &  +05:35:36.9    & 11.4 & 18 \\ 
G11.11$-$0.34  &  18:11:33.3  &   $-$19:30:38.9   &  5.2  &  7 &    G41.13$-$0.32$^{\dag}\,^{*}$  &  19:07:36.3  &  +07:08:40.3    & \ldots & \\ 
G11.94$-$0.62  &  18:14:01.0  &   $-$18:53:25.0   &  5.2  &  2 &    G41.52+0.04$^{*}$  &  19:07:03.7  &  +07:39:04.1    & \ldots &    \\ 
G12.43$-$0.05  &  18:12:54.6  &   $-$18:11:08.4   & 16.7  &  2 &    G41.71+0.11$^{*}$  &  19:07:09.8  &  +07:51:36.1    & \ldots &    \\ 
G12.90$-$0.25$^{\dag}$  &  18:14:37.3  &   $-$17:52:02.5   &  4.0  &  8 &    G41.74+0.10$^{*}$  &  19:07:15.5  &  +07:52:43.9    & \ldots &   \\ 
G13.19+0.04     &  18:14:05.6  &   $-$17:28:38.7   & 4.4/12.1 & 9 &    G42.42$-$0.27  &  19:09:49.5  &  +08:18:43.8    &        5.2 & 10 \\ 
G13.87+0.28   &  18:14:35.6  &   $-$16:45:37.5   &  4.5  & 10  &    G42.90+0.57  &  19:07:42.0  &  +09:07:12.4    & \ldots &    \\ 
G15.04$-$0.68  &  18:20:24.8  &   $-$16:11:35.0   &  2.1  &  2 &    G43.18$-$0.52  &  19:12:08.7  &  +08:52:08.8    & 4.6 & 2   \\ 
G18.15$-$0.28  &  18:25:02.3  &   $-$13:15:50.8   &  4.2  &  7 &    G43.24$-$0.05  &  19:10:35.0  &  +09:08:31.9    & 11.7 & 18 \\ 
G18.30$-$0.39  &  18:25:42.3  &   $-$13:10:19.9   &  2.9  &  7 &    G43.26$-$0.18$^{\dag}$  &  19:11:05.9  &  +09:05:58.0    & \ldots & \\ 
G19.07$-$0.27  &  18:26:47.0  &   $-$12:26:26.5   &  5.4  &  2 &    G43.80$-$0.13  &  19:11:53.3  &  +09:35:46.3    & 9.0 & 10  \\ 
G19.49+0.14    &  18:26:03.0  &   $-$11:52:34.4   & 2.0/14.0 & 9,11  &    G43.82+0.39$^{\dag}$  &  19:10:03.7  &  +09:51:23.7    & \ldots &     \\ 
G19.61$-$0.23  &  18:27:38.1  &   $-$11:56:39.5   &  4.5  &  2 &    G44.26+0.10  &  19:11:56.6  &  +10:07:01.2    & \ldots &    \\ 
G20.08$-$0.14  &  18:28:10.5  &   $-$11:28:48.6   & 12.3  &  5 &    G45.07+0.13  &  19:13:22.1  &  +10:50:53.4    & 6.0 & 4 \\ 
G20.99+0.09    &  18:29:04.1  &   $-$10:34:16.2   & 1.8/14.1 & 12  &    G45.12+0.13  &  19:13:27.9  &  +10:53:36.7    & 6.9 & 4 \\ 
G22.76$-$0.49$^{\dag}$  &  18:34:28.4  &   $-$09:16:59.7   & 4.7/10.9 & 12  &    G45.20+0.74$^{\dag}$  &  19:11:25.0  &  +11:14:37.4    & \ldots &      \\ 
G23.46$-$0.20  &  18:34:44.9  &   $-$08:31:07.4   &  9.0  & 2  &    G45.40$-$0.72$^{\dag}$  &  19:17:03.8  &  +10:44:33.8    & \ldots & \\ 
G23.71+0.17    &  18:33:53.5  &   $-$08:07:13.8   &  8.9  & 2  &    G45.47+0.05$^{*}$  &  19:14:25.8  &  +11:09:25.9    & 6.0 & 2 \\ 
G23.87$-$0.12  &  18:35:13.4  &   $-$08:06:52.4   &  4.6/10.9  & 12  &    G69.54$-$0.98  &  20:10:09.1  &  +31:31:34.4    & 3.0 & 16 \\ 
G23.96+0.15    &  18:34:25.6  &   $-$07:54:46.8   &  6.0  &  2 &    G70.29+1.60  &  20:01:45.6  &  +33:32:44.1    & 8.3 & 17 \\ 
G24.39+0.07$^{\dag}$    &  18:35:30.6  &   $-$07:33:44.2   & \ldots   &   &    G76.18+0.13  &  20:23:55.7  & +37:38:10.5    & 4.4 & 7 \\ 
G24.47+0.49    &  18:34:09.2  &   $-$07:18:04.0   &  5.7/9.8  & 12  &    G76.38$-$0.62  &  20:27:26.5  &  +37:22:47.9    & 1.0  & 7 \\ 
G24.68$-$0.16  &  18:36:51.6  &   $-$07:24:52.3   &  6.3/9.1  & 9  &    G77.97$-$0.01  &  20:29:36.4  &  +39:01:17.5    & 4.4 & 7 \\ 
G25.38$-$0.18  &  18:38:14.3  &   $-$06:47:53.3   &  3.7  & 13  &    G78.44+2.66  &  20:19:39.3  &  +40:56:30.4    & 3.3 & 7    \\ 
G25.65+1.05    &  18:34:19.8  &   $-$05:59:44.2   &  3.0  & 14  &    G79.30+0.28  &  20:32:29.3  &  +40:16:05.5    & 1.0 & 7    \\ 
G25.72+0.05    &  18:38:02.8  &   $-$06:23:47.3   & 14.0  &  2 &    G79.32+1.31  &  20:28:12.4  &  +40:52:27.7    & 1.5 & 7     \\ 
G26.10$-$0.07  &  18:39:11.0  &   $-$06:06:22.2   &  \ldots  &   &    G80.87+0.42  &  20:36:52.6  &  +41:36:32.6    & 2.1 & 7      \\  
G26.54+0.42    &  18:38:15.6  &   $-$05:29:37.2   & 5.2/10.0 & 12  &    G106.80+5.31 &  22:19:18.2  &  +63:18:46.3    & 0.9 & 7      \\  
G27.28+0.15    &  18:40:35.2  &   $-$04:57:44.0   & 15.2  & 2  &    G109.87+2.11 &  22:56:19.1  &  +62:01:57.4    & 0.7 & 7     \\  
G27.49+0.19    &  18:40:49.0  &   $-$04:45:15.1   & 2.5/12.6   & 12  &    G110.21+2.63 &  22:57:05.2  &  +62:37:44.3    & 0.7 & 7   \\  
G28.20$-$0.05  &  18:42:58.2  &   $-$04:14:59.8   &  9.1  &  7 &    G111.28$-$0.66 &  23:15:31.5  &  +61:07:09.3    & 2.5 & 7      \\  
G28.29$-$0.36  &  18:44:18.6  &   $-$04:17:53.1   &  3.3  & 7  &    G111.61+0.37 &  23:16:04.8  &  +60:02:59.7    & 5.2 & 7      \\  
G28.60$-$0.36  &  18:44:49.0  &   $-$04:01:24.9   &  5.2/9.7  & 12  &    G133.95+1.06 &  02:27:01.1  &  +61:52:13.7    & 3.0 & 7      \\  
G28.80+0.17$^{*}$    &  18:43:16.7  &   $-$03:35:45.5   &  6.4/8.5  & 12  &    G138.30+1.56 &  03:01:32.3  &  +60:29:11.8    & 3.8 & 7      \\  
G28.83$-$0.23$^{\dag}$  &  18:44:45.4  &   $-$03:45:18.1   &  5.1/9.8  & 9  &    G188.77+1.07$^{*}$ &  06:09:13.8  &  +21:53:12.7    & \ldots &      \\  
G29.96$-$0.02  &  18:46:04.1  &   $-$02:39:21.5   &  7.4  & 5  &    G188.79+1.03$^{*}$ &  06:09:07.8  &  +21:50:38.7    & 4.1 &  11    \\  
G30.54+0.02    &  18:46:59.3  &   $-$02:07:24.5   & 13.8  & 2  &    G188.95+0.92$^{*}$ &  06:08:53.9  &  +21:38:36.7    & 2.2 &  7    \\  
G30.78$-$0.02  &  18:47:35.6  &   $-$01:55:25.9   &  5.5  & 15  &    G213.88$-$11.84$^{*}$&  06:10:51.0  &  $-$06:11:54.1     & 1.0 & 7         \\ 
G31.28+0.06    &  18:48:12.3  &   $-$01:25:48.3   &  7.2  &  7 &                 &     &    &     &       \\ \hline
\end{tabular}

Distance references are: (1) Mehringer et al.~(\cite{mgl98}). 
(2) Wood \& Churchwell (\cite{wc89a}).
(3) Acord, Churchwell \& Wood (\cite{acw98}).
(4) Churchwell, Walmsley \& Cesaroni (\cite{cwc90}).
(5) Fish et al.~(\cite{frw03}).
(6) Hofner et al.~(\cite{hkcwc94}).
(7) Kurtz, Churchwell \& Wood (\cite{kcw94}).
(8) van der Tak et al.~(\cite{vve2000}).
(9) kinematic distance evaluated from CH$_{3}$OH maser velocity given by Szymczak, Hrynek
\& Kus (\cite{shk2000}).
(10) Harju et al.~(\cite{hlb98}).
(11) kinematic distance evaluated  from CH$_{3}$OH maser velocity given by Caswell et
al.~(\cite{caswell95}).
(12) kinematic distance evaluated from Hydrogen radio recombination line velocity given by Kuchar \& Clarke
(\cite{kc97}).
(13) Blum, Conti \& Damineli (\cite{bcd00}). 
(14) Zavagno et al.~(\cite{zdnc02}).
(15) Motte, Schilke \& Lis (\cite{msl03}).
(16) Palagi et al.~(\cite{pcc93}).
(17) Araya et al.~(\cite{ahck03}).
(18) Watson et al.~(\cite{was2003}).
(19) Codella et al.~(\cite{cfn94}).
\end{minipage}
\end{table*}

\subsection{Observational procedure}
\label{subsect:op}

Our observations were carried out using the SCUBA (Holland et al.~\cite{scuba})
common-user bolometer array in operation at the James Clerk Maxwell Telescope
(JCMT\footnote{The
JCMT is operated by the Joint Astronomy Centre on behalf of PPARC for the United
Kingdom, the Netherlands Organisation of Scientific Research, and the National Research
Council of Canada.}). The observations were performed in a flexibly-scheduled
mode whereby observations are not scheduled over a pre-defined period but are carried out
over a 6-month observing semester  according to the appropriate weather (atmospheric
opacity) band, the visibility of the sources from the telescope and the scientific
priority of the observations. The survey data were thus obtained by visiting observers
over several separate periods across the 98B (Aug--Jan) and 99A (Feb--July) observing semesters at the JCMT.

SCUBA is comprised of two bolometer arrays, a short-wave array of 91 pixels optimised for
operation at 450 $\mu$m and a long-wave array of 37 pixels optimised for operation at 850
$\mu$m. Both arrays simultaneously sample  a similar field of view (approx 2\arcmin\
square), although the spacing between individual bolometers on the array means that not
all of the field of view is sampled  instantaneously.  To fill in the gaps in spatial
coverage the telescope secondary mirror is moved in a 64-point pattern (``jiggling''),
whilst also chopping at a frequency of 1 Hz to remove the sky emission. This procedure is
commonly known as a ``jiggle-map''  and provides  maps with full spatial sampling at both
wavelengths.

As the SCUBA FOV is approximately the same size as the VLA primary beam at 2cm and the
IRAS 100 $\mu$m beam area we obtained a single jiggle-map centred at the coordinates of
either the UC H{\sc ii} region radio emission or the position of the IRAS source (where no radio
information was available). For the UC H{\sc ii} regions contained in the KCW94 survey we
mistakenly used the IRAS positions rather than the radio positions, which resulted in the
array being slightly offset from the main sub-mm emission. However, the SCUBA FOV was
found to be larger than the positional offsets introduced and thus there is little effect upon
the data in question. Additional maps with positional offsets were obtained for a number
of sources where the mapped emission was affected by noisy bolometers or by being located
at the edge of the FOV.

Each source was integrated on for three of the 64-point SCUBA jiggle cycles (also known as SCUBA
``integrations'') for a total on-source integration time of 192 seconds. This typically resulted in 1-$\sigma$
r.m.s.~noises of $\sim 1.5$ Jy per 8\arcsec\ FWHM beam at 450 $\mu$m and  $\sim 0.1$ Jy per 14\arcsec\ FWHM beam
at 850 $\mu$m. All sources were observed with a chop throw of 120\arcsec\ in azimuth. As SCUBA is located at the
Nasmyth focus of the JCMT and does not possess a beam rotator, this means that the chop positions are also a
function of elevation. 

Regular pointing checks were made and the average pointing offset was found to be
$\le$5\arcsec. Hourly skydips were carried out at approximately the same azimuth as the
observations to estimate the atmospheric zenith optical depth at at 450 and 850 $\mu$m.
Values derived from the skydips were compared to and found consistent with  those
extrapolated from the fixed-azimuth measurements at 225 GHz made every 10 minutes by the
CSO tipping radiometer. Absolute flux calibration and beam maps of the primary flux
calibrator Uranus were also obtained at least once per observing session. 

\subsection{Data reduction}
\label{subsect:dr}

The data were reduced using a combination of the automated SCUBA reduction pipeline ORACDR
(Economou et al.~\cite{oracdr}),  the SCUBA reduction package SURF (Jenness \& Lightfoot
\cite{surf}) and the Starlink image analysis package KAPPA (Currie \& Bell \cite{kappa}). The
reduction procedure for 450 and 850 $\mu$m data was essentially the same and proceeded along the
following outline. Initially the chopping and nodding positions were subtracted from the on-source data to
form a time-ordered series of sky-subtracted bolometer measurements. Given that the UC H{\sc ii}s are
likely to be embedded in larger giant molecular cloud complexes, it is likely that extended
emission from these regions is present in the chopping positions and is subtracted from the
on-source sub-mm fluxes by this procedure. The fluxes quoted in this paper should thus be
strictly regarded as lower limits to the true sub-mm fluxes. 

The time-ordered bolometer data were then corrected for atmospheric extinction using an
optical depth value interpolated from skydips carried out before and after the jiggle-map.
At this stage  bolometers with a mean noise in excess of 100 nV were blanked and transient
bolometer  noise spikes were removed by applying a  5$\sigma$ clip to the data. As several
of the sources were found to be strongly centrally-peaked, care was taken to avoid the
removal of the central pixel by this  despiking procedure. Residual sky variations between
individual bolometers were removed by specifying emission-free bolometers and using the
SURF task \emph{remsky}. The time-ordered data were then regridded to J2000 sky
coordinates with the SURF task \emph{rebin}. 

Absolute flux calibration was carried out using the calibration maps of Uranus. Predicted fluxes
for Uranus were estimated using the values given by the Starlink package FLUXES (Privett,
Jenness \& Matthews \cite{fluxes}) and on the JCMT calibrator webpage respectively. Flux
correction factors (FCFs) for each wavelength were then determined by dividing the predicted
flux by the measured peak value of the calibrator. By monitoring the variation in FCF over each
observing period it was found that the absolute flux calibration was accurate to 30\% at 450
$\mu$m  and 10\% at 850 $\mu$m. These errors in calibration are predominantly due to variations
of the line-of-sight optical depth on timescales smaller than can be accounted for by skydips.
Each jiggle-map was calibrated in units of Jy/beam by multiplying by the appropriate FCF. The
FWHMs and peak values of the  telescope main and error beams were determined by fitting two
Gaussians to azimuthal averages of the maps of the primary calibrator (Uranus). 

A number of images, mainly taken on a single night, were found to be essentially uncalibratable
at 450 $\mu$m due to either large values of or rapid variations in the 450 $\mu$m  atmospheric
optical depth. Although there is a close linear relation between the  450 \& 850 $\mu$m zenith
optical depth (Archibald et al.~\cite{ath02}), the atmosphere is typically more unstable and
much more optically thick (by a factor of 4 or more) at 450 $\mu$m. The result of this is that
whilst the 450 $\mu$m data was unusable for the affected periods, the 850 $\mu$m data was
relatively unaffected. We do not present the affected 450 $\mu$m data from these periods, which
is a total of 18 images.  
 
 \begin{figure*}[p]
 \begin{minipage}{\linewidth}
 \includegraphics[scale=0.35,angle=-90]{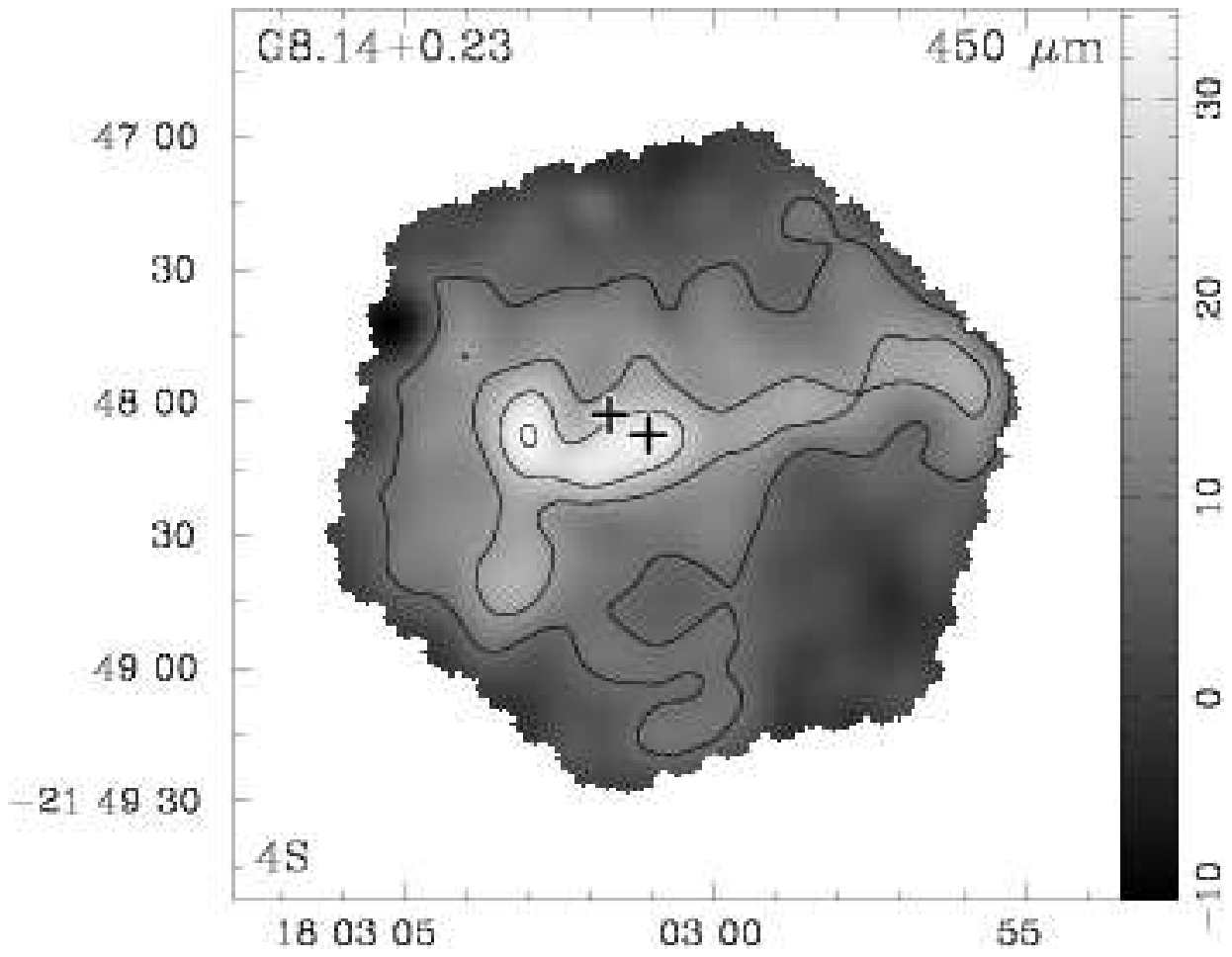}
 \includegraphics[scale=0.35,angle=-90]{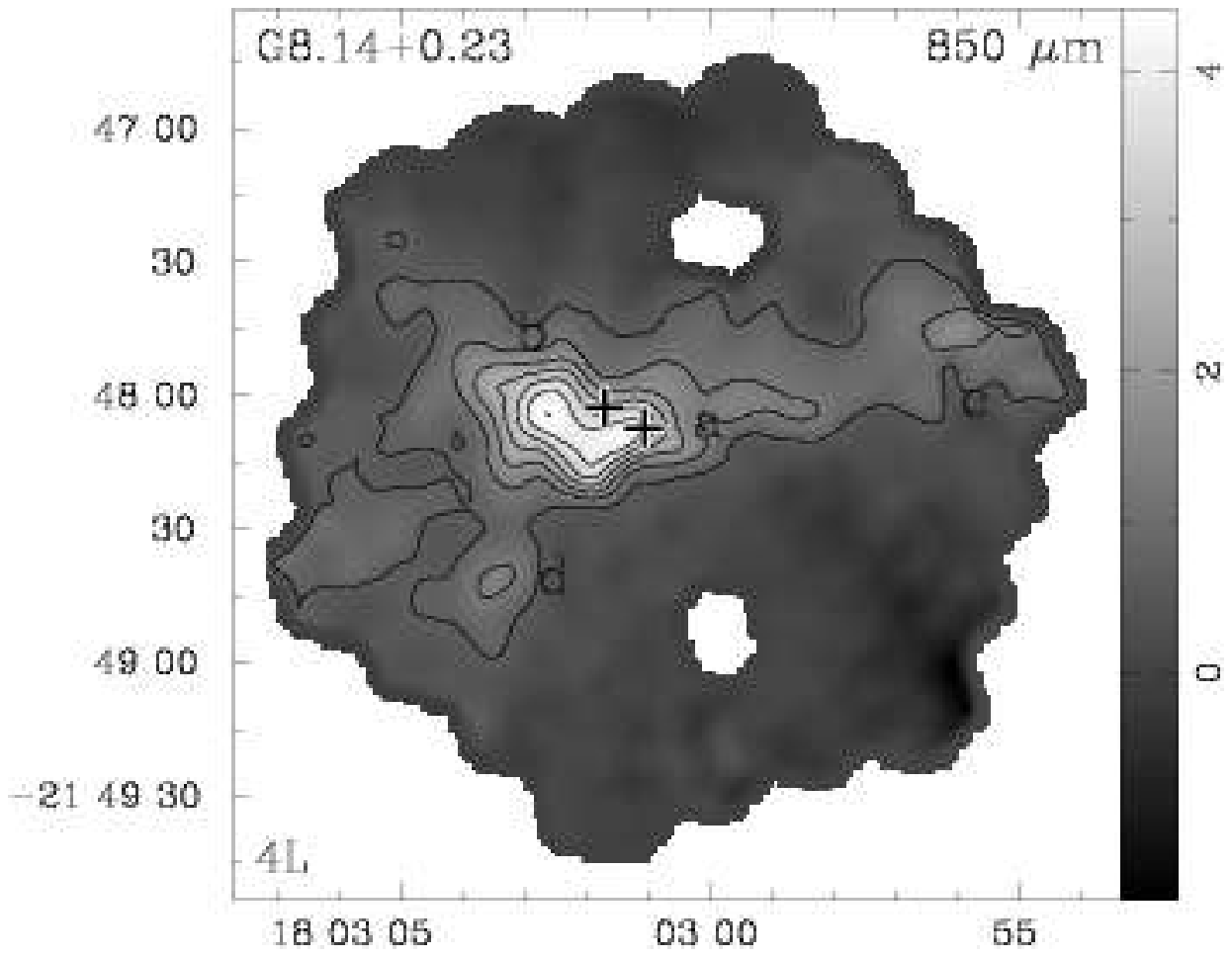}
 \includegraphics[scale=0.35,angle=-90]{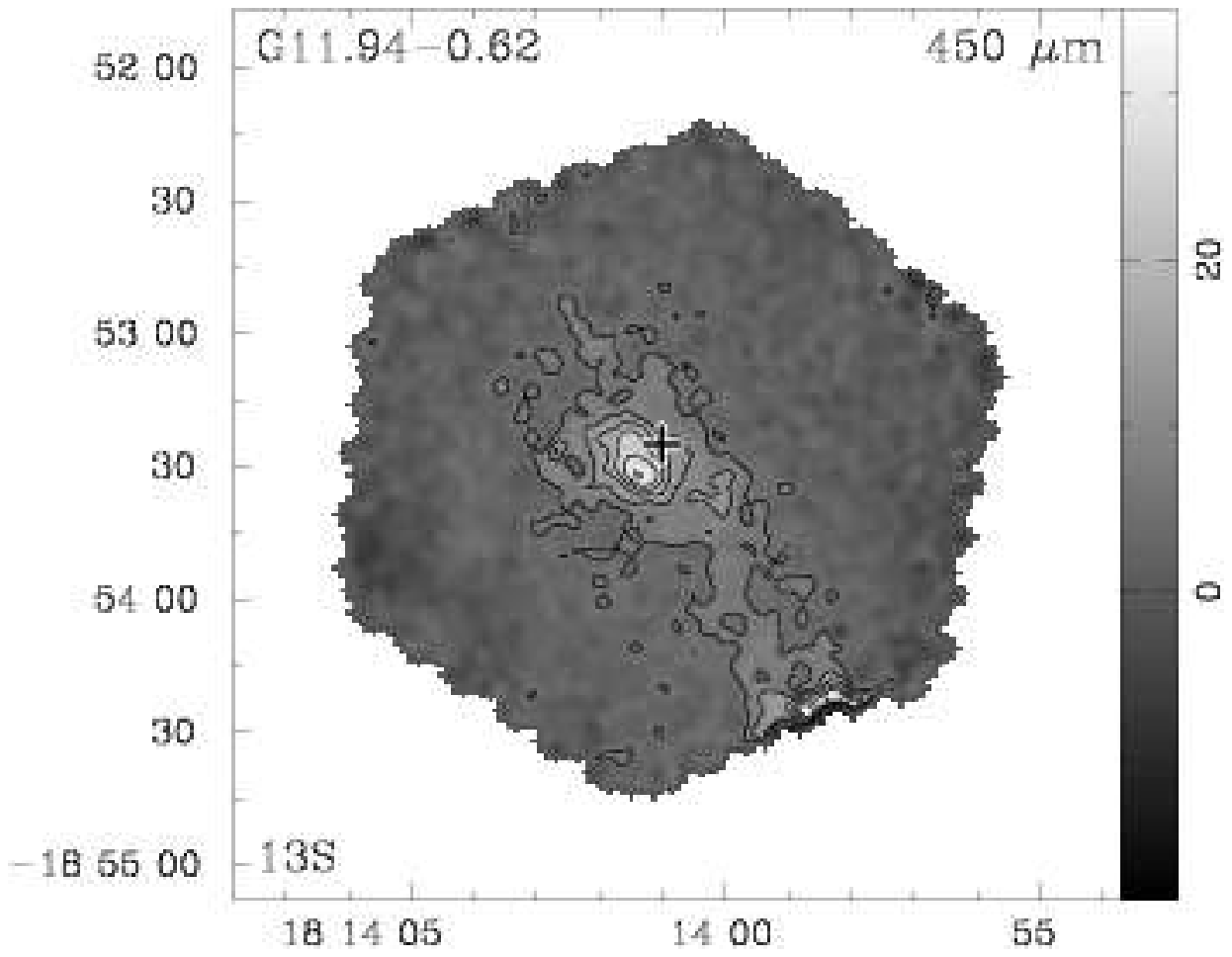}
 \includegraphics[scale=0.35,angle=-90]{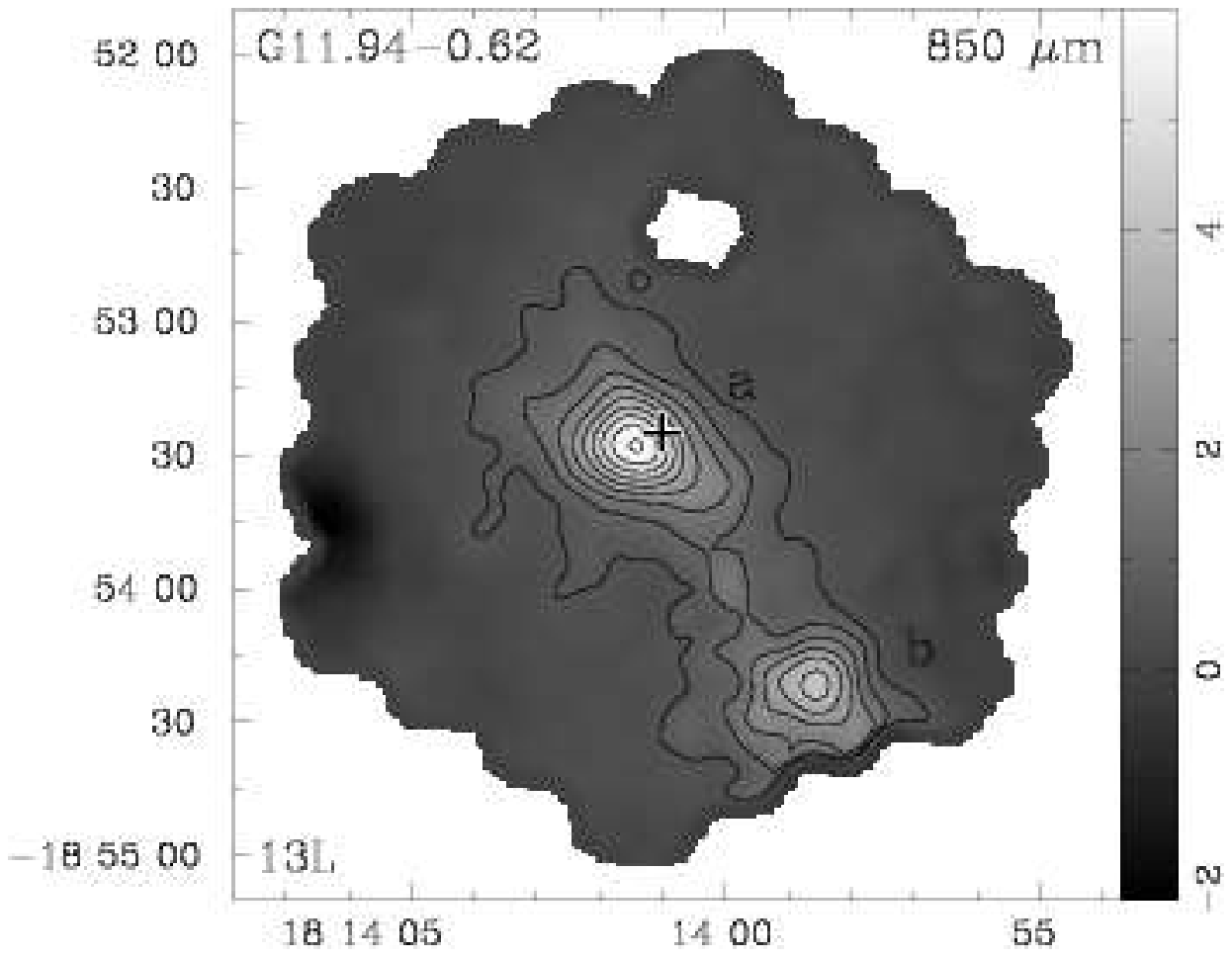} \\
 
 \includegraphics[scale=0.35,angle=-90]{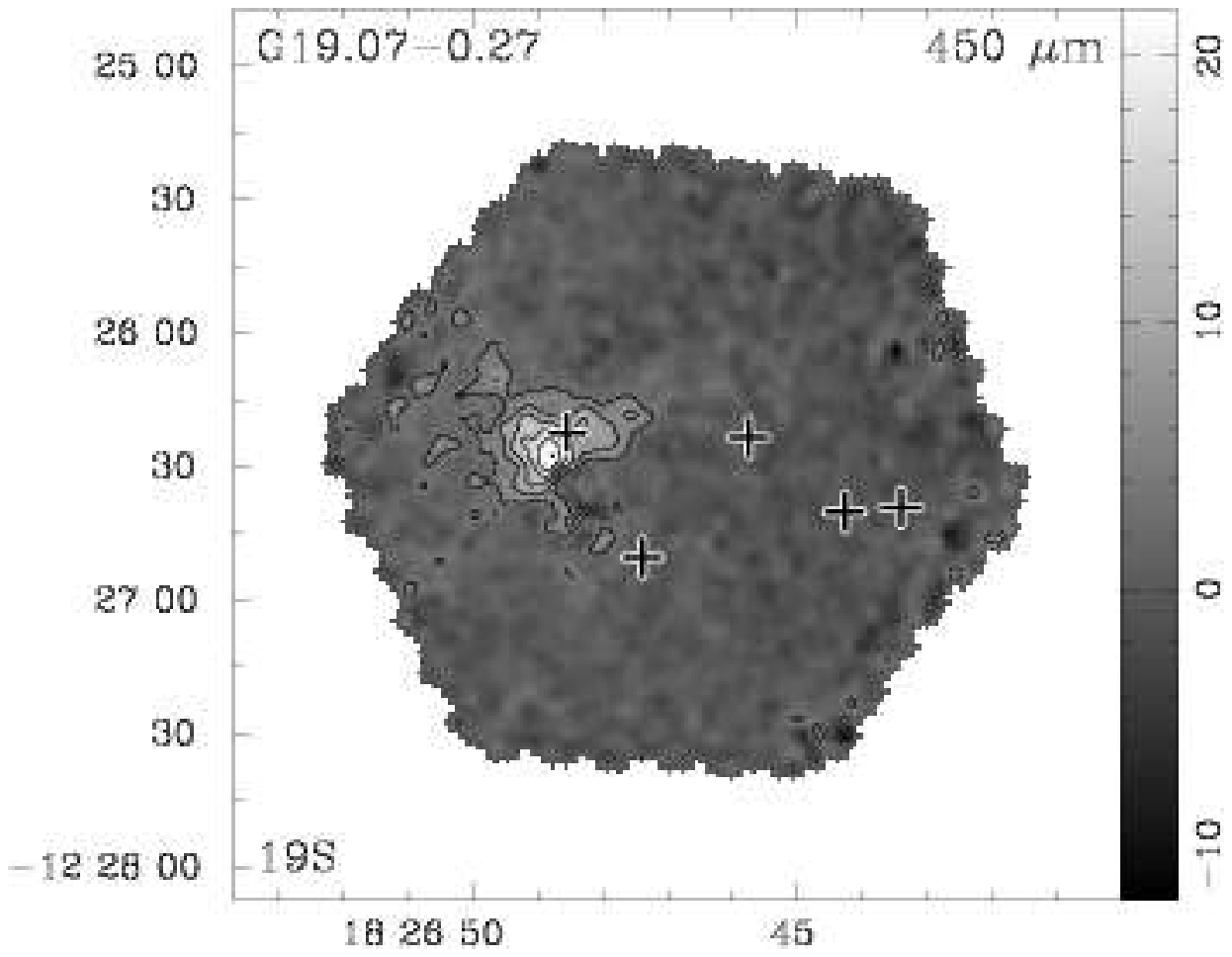}
 \includegraphics[scale=0.35,angle=-90]{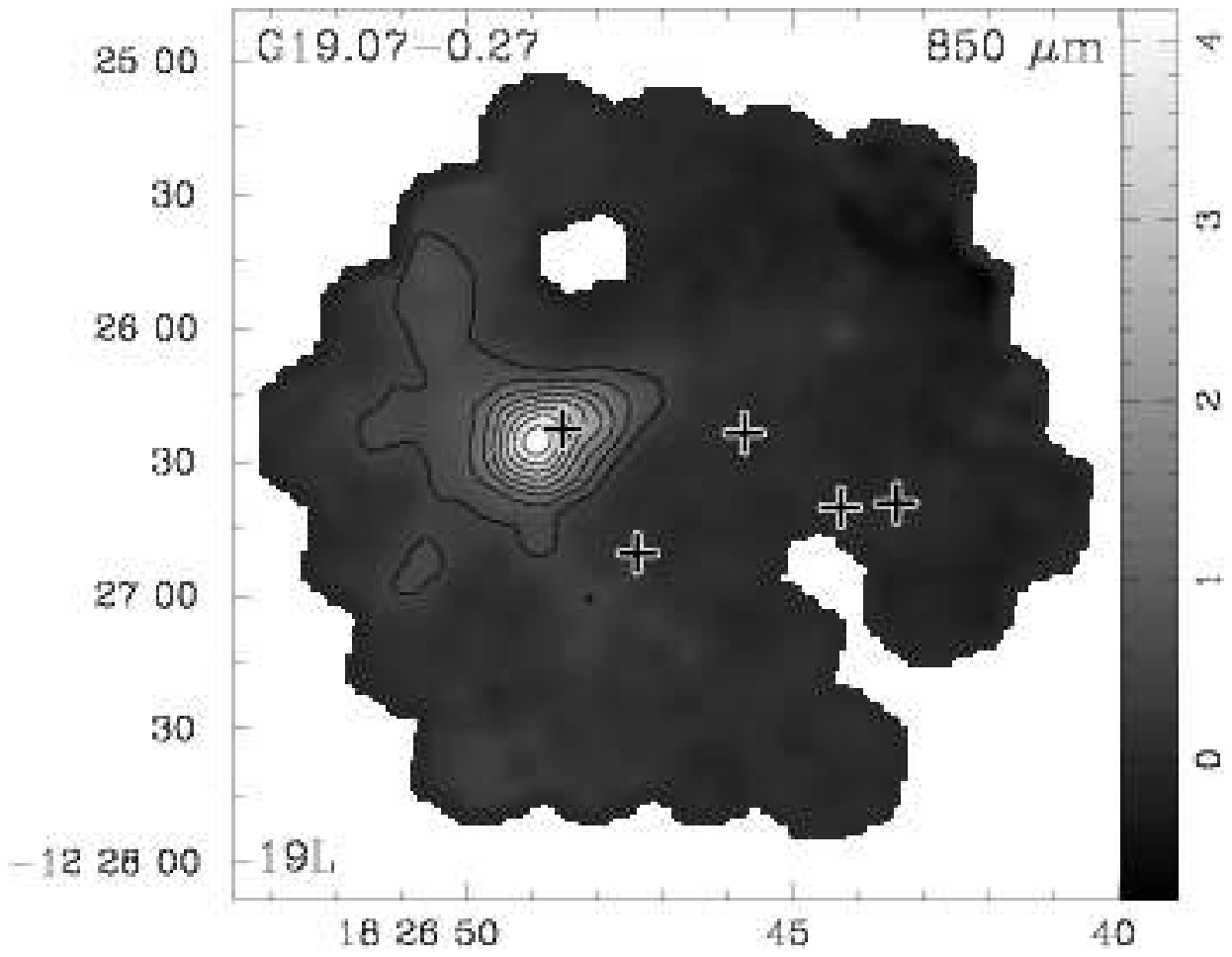} 
 \includegraphics[scale=0.35,angle=-90]{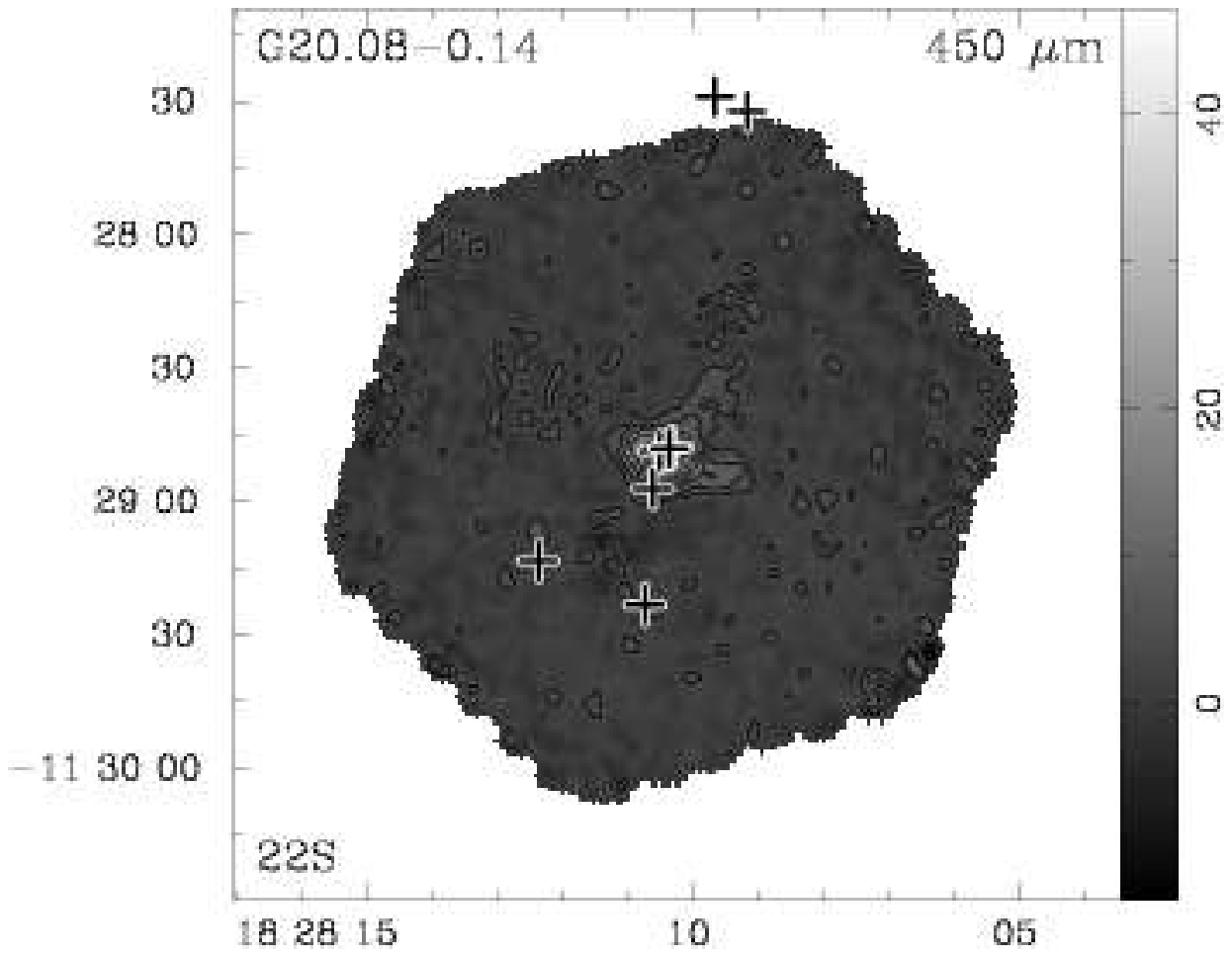}
 \includegraphics[scale=0.35,angle=-90]{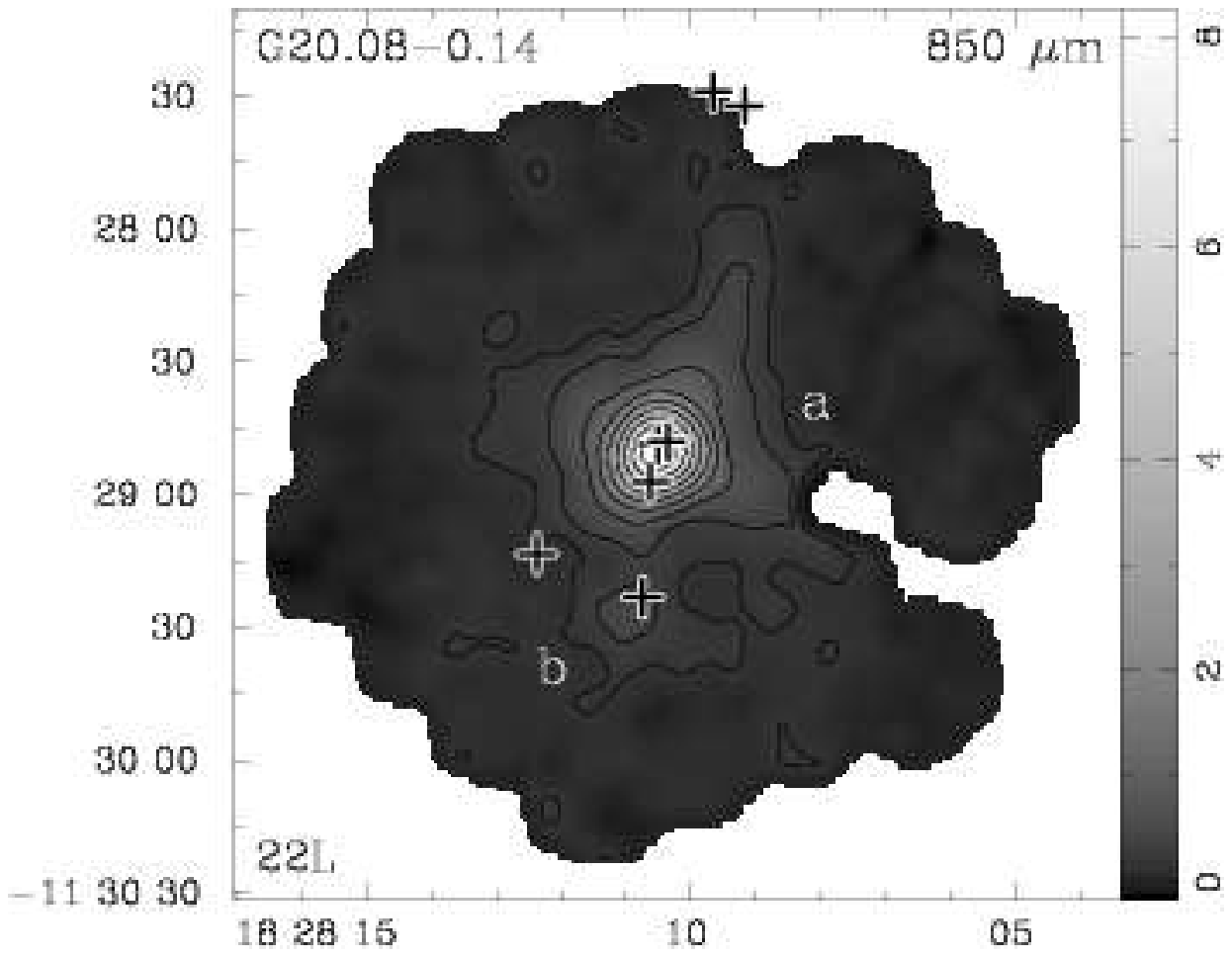}\\
 
 \includegraphics[scale=0.35,angle=-90]{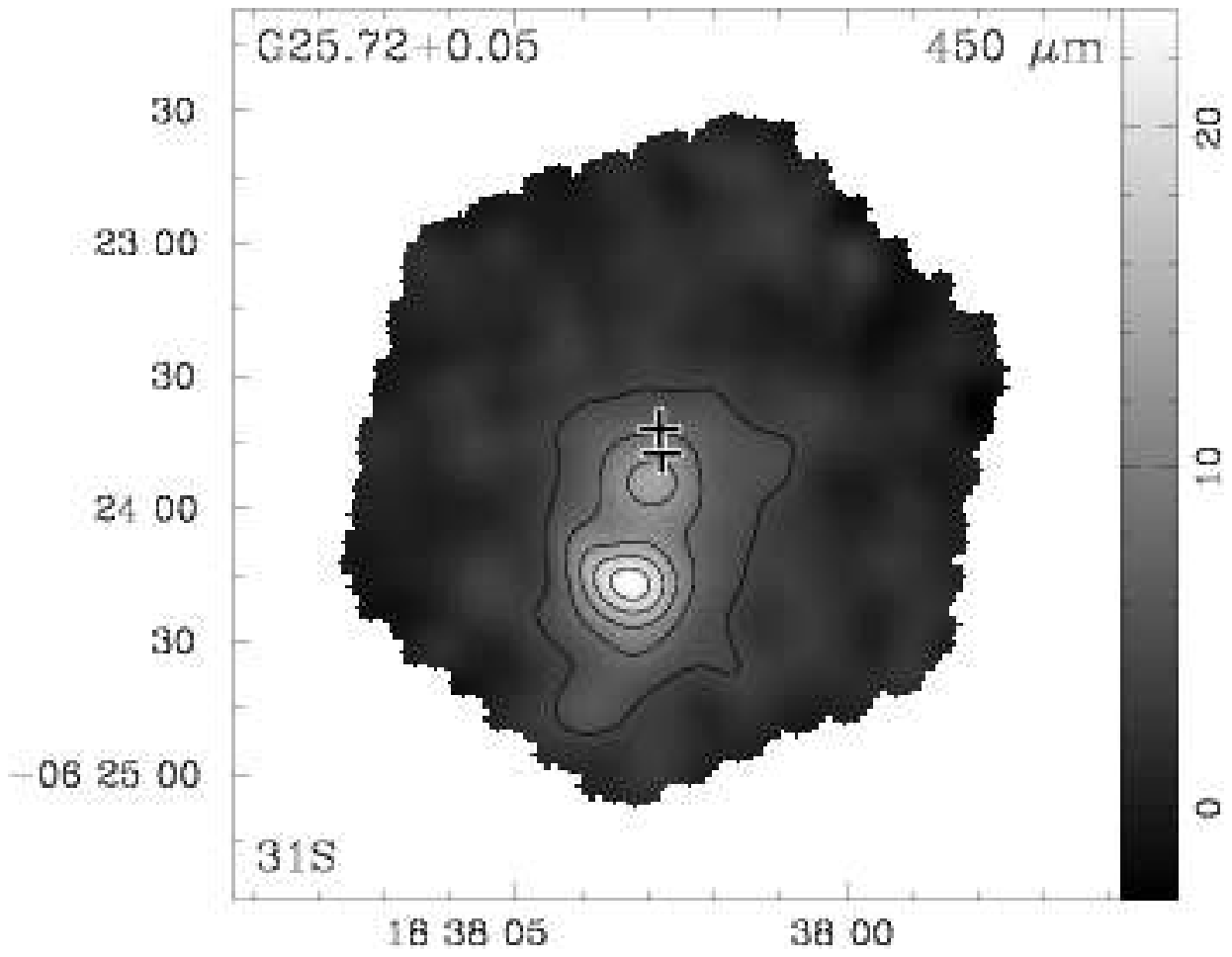}
 \includegraphics[scale=0.35,angle=-90]{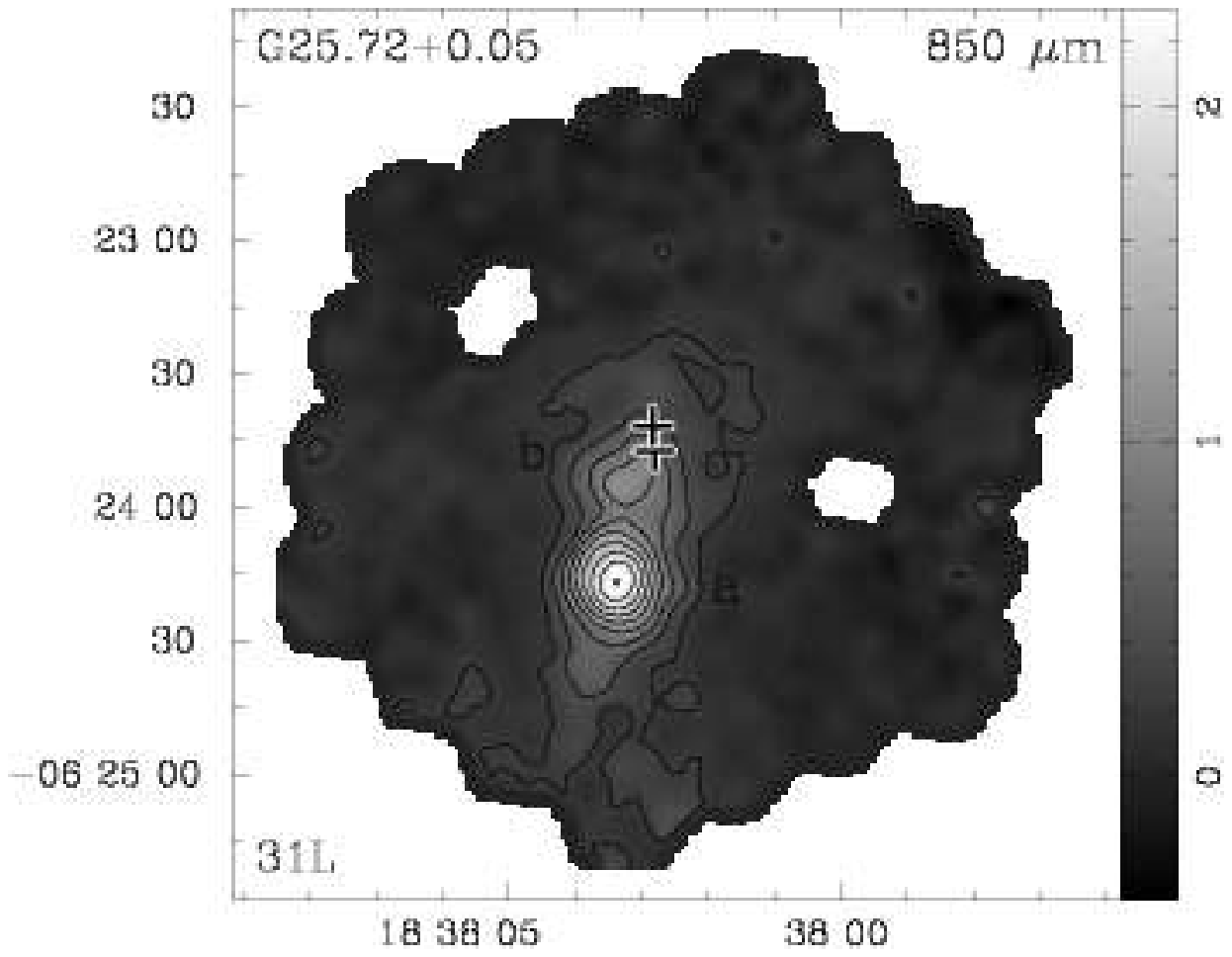} 
 \includegraphics[scale=0.35,angle=-90]{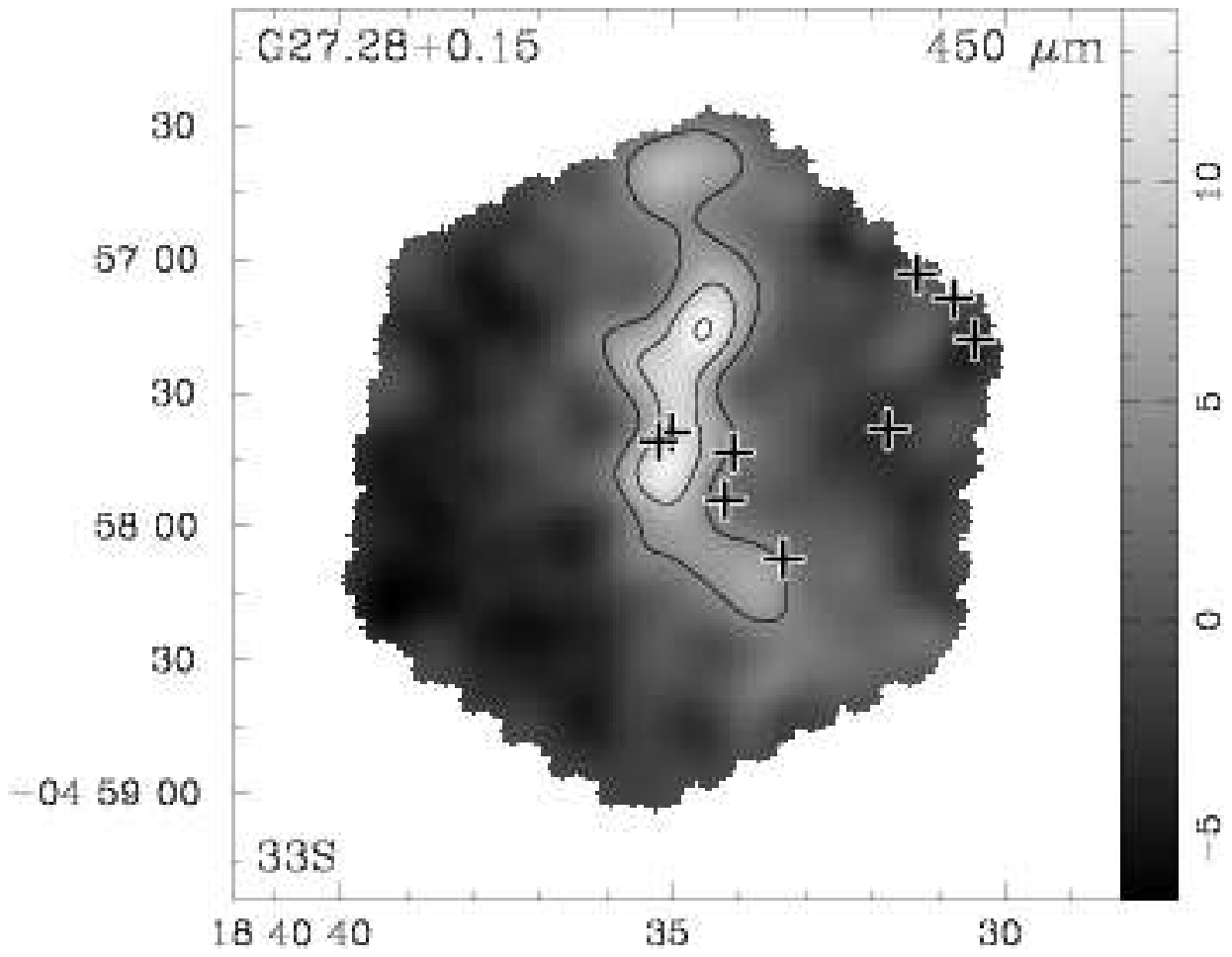}
 \includegraphics[scale=0.35,angle=-90]{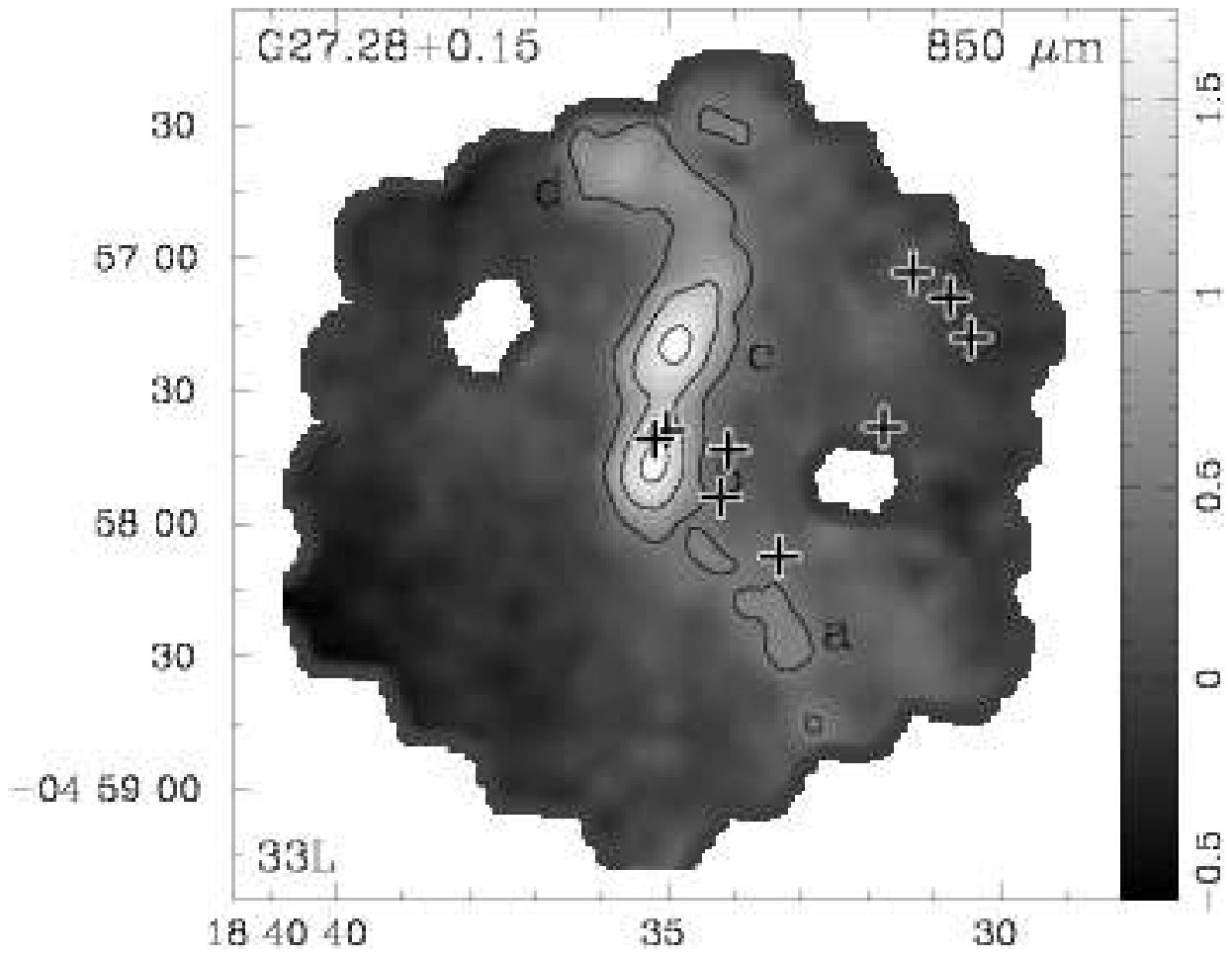}\\
 
 \includegraphics[scale=0.35,angle=-90]{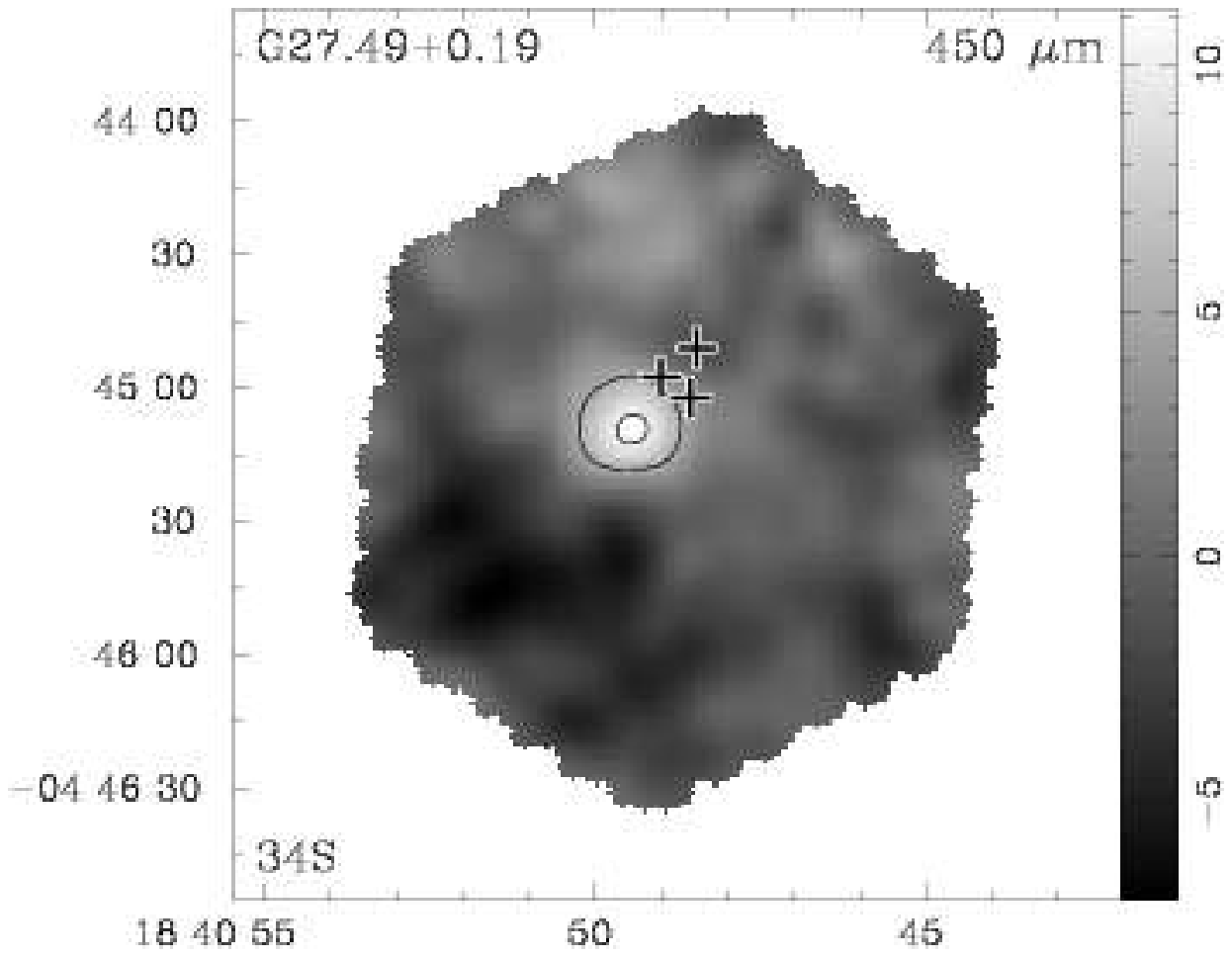}
 \includegraphics[scale=0.35,angle=-90]{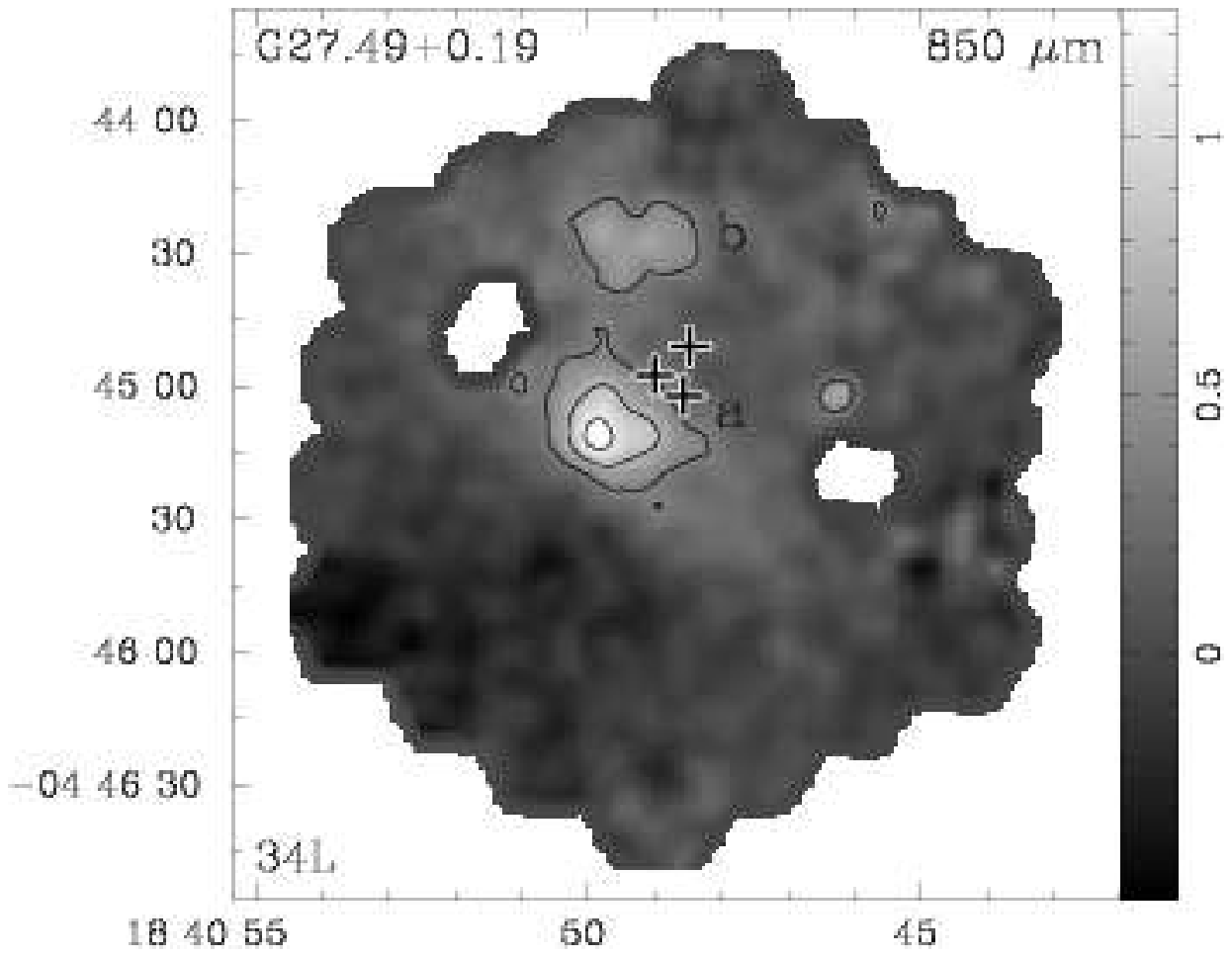} 
 \includegraphics[scale=0.35,angle=-90]{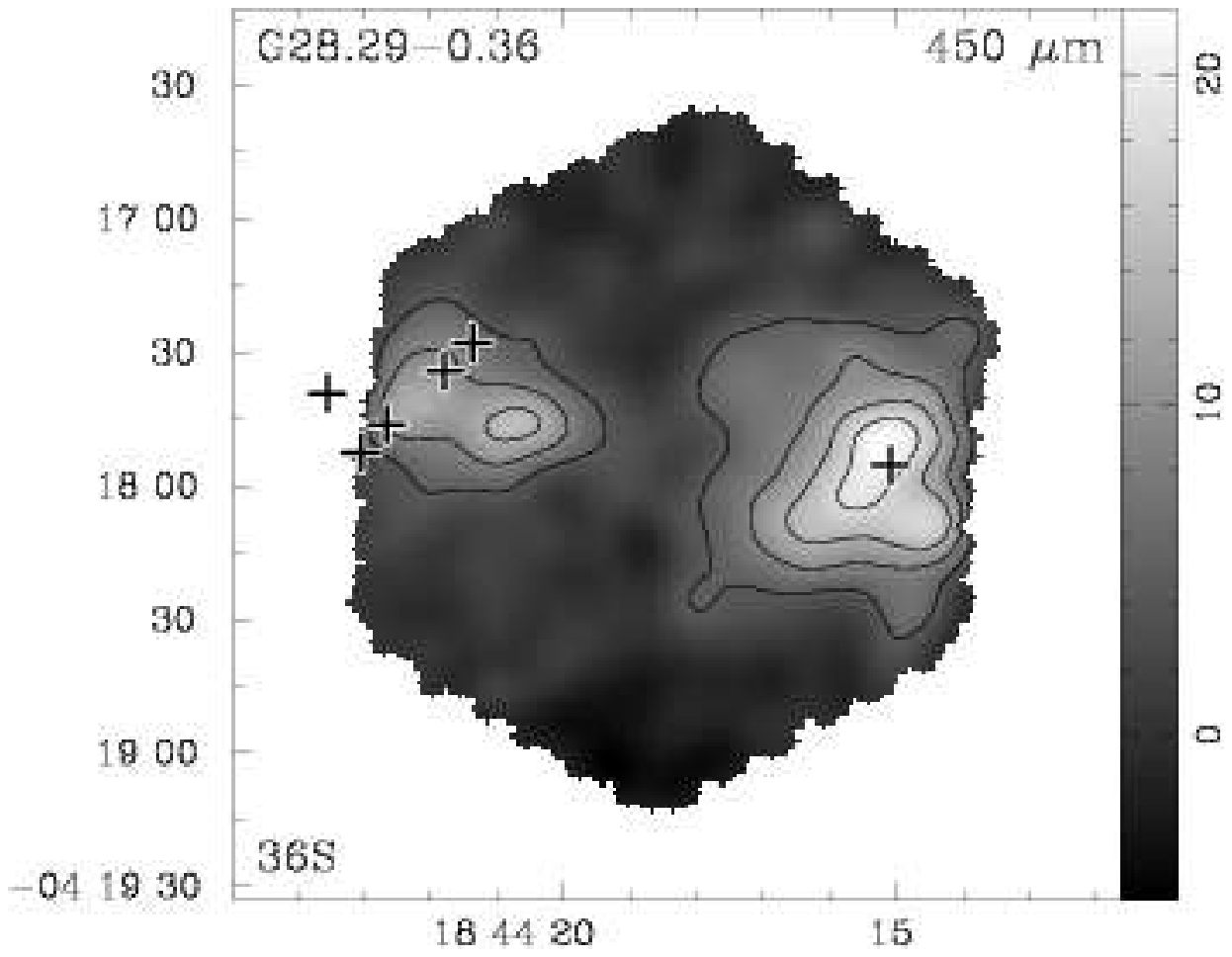}
 \includegraphics[scale=0.35,angle=-90]{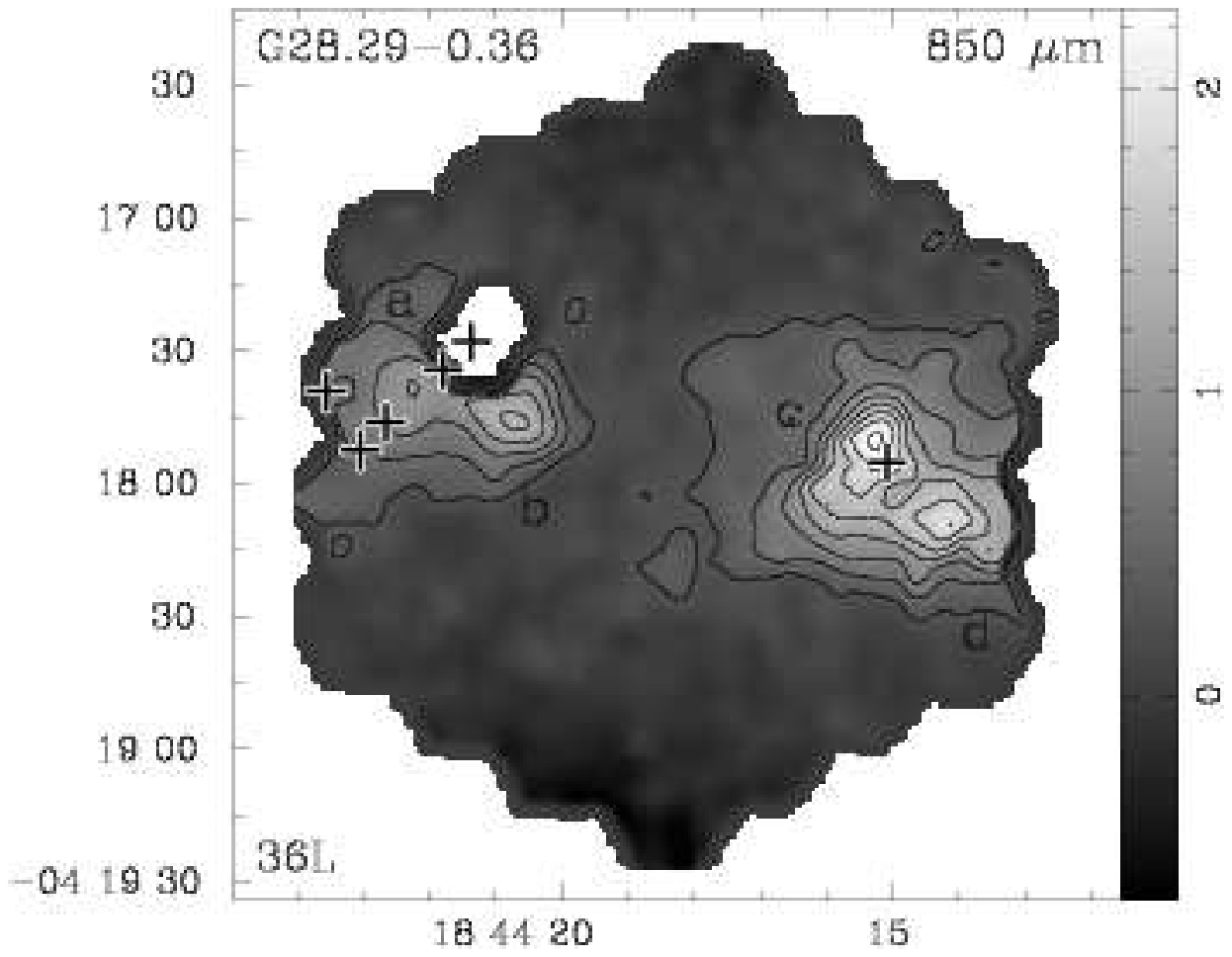} \\
 
 \includegraphics[scale=0.35,angle=-90]{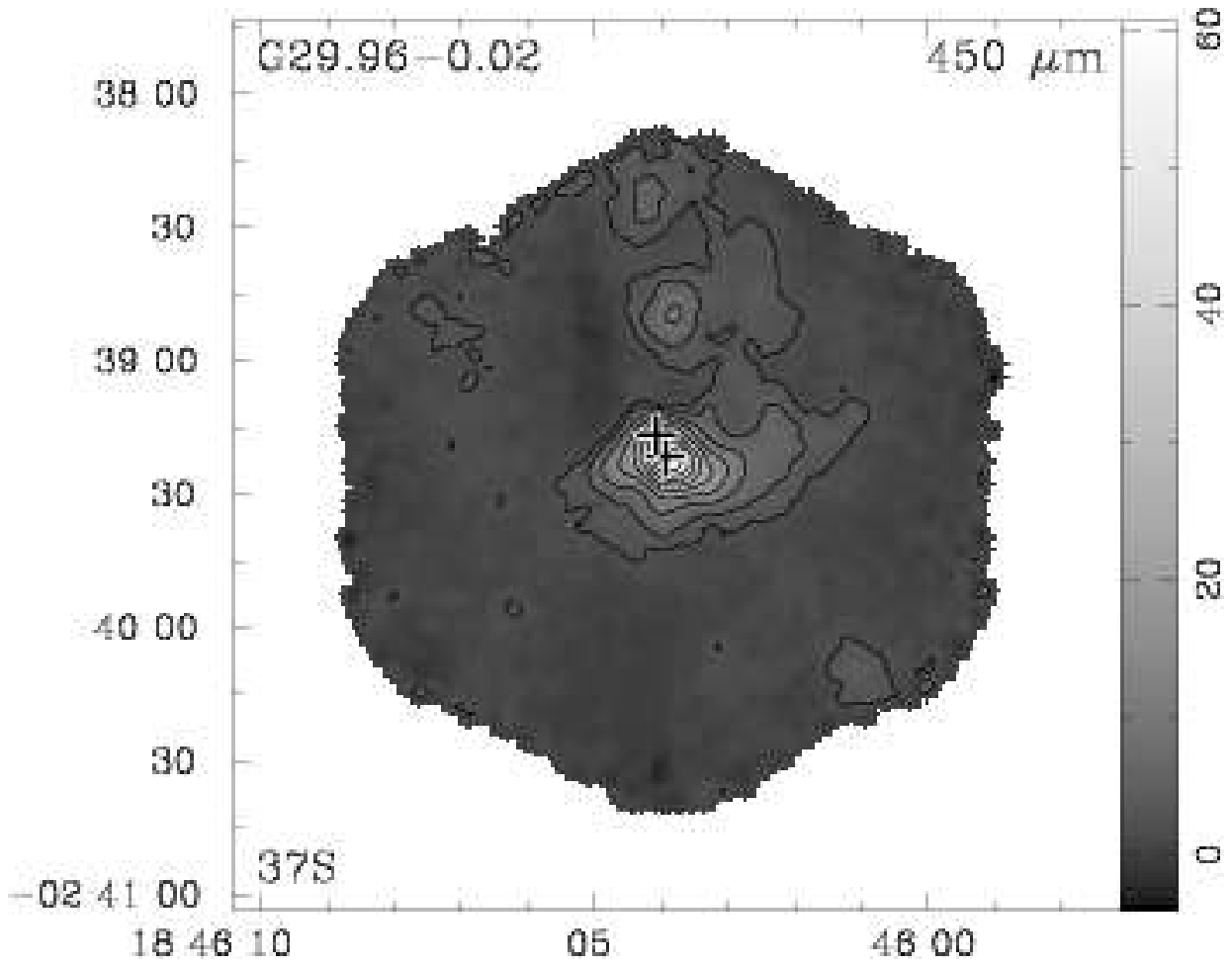}
 \includegraphics[scale=0.35,angle=-90]{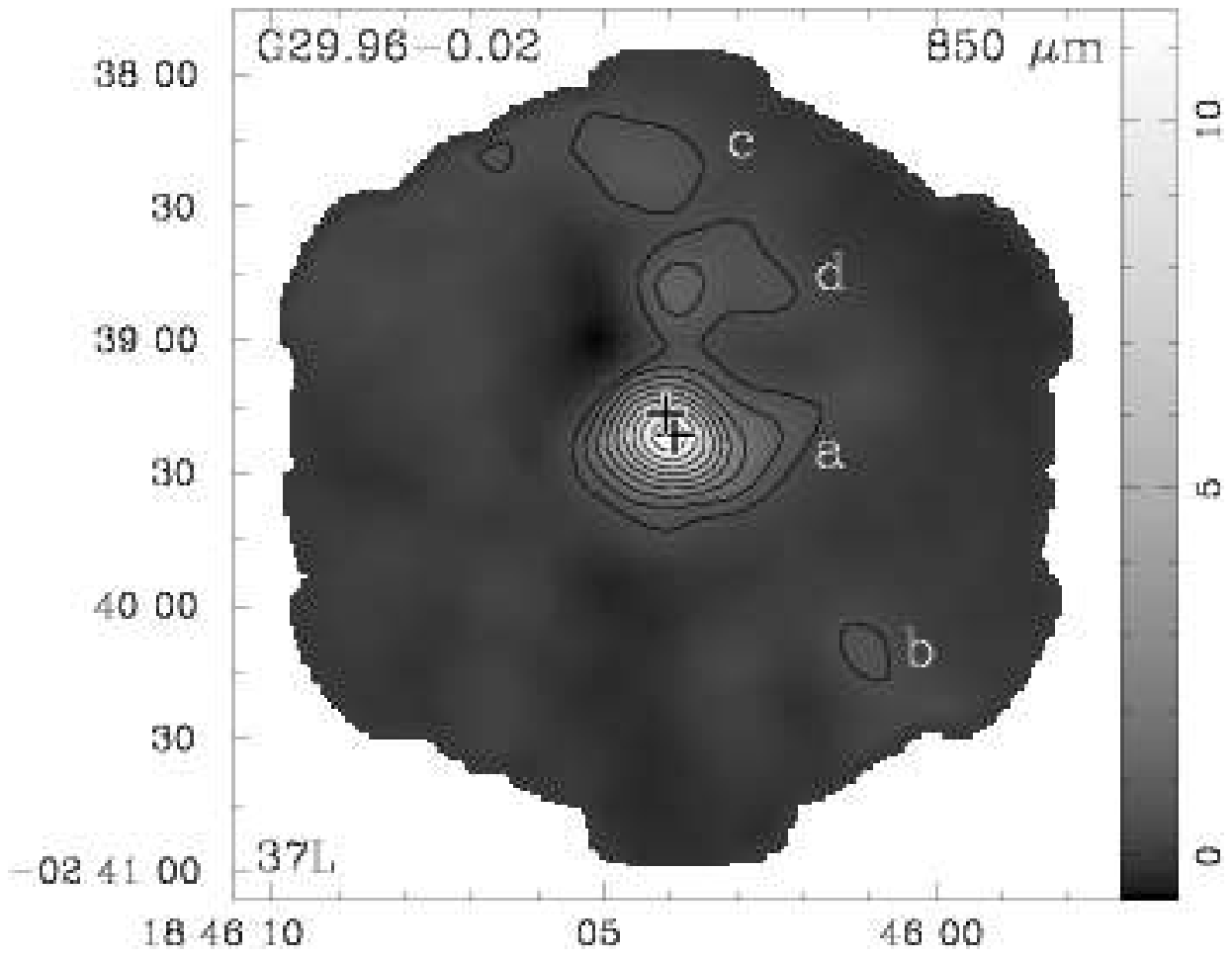} 
 \hspace*{0.11 cm}\includegraphics[scale=0.35,angle=-90]{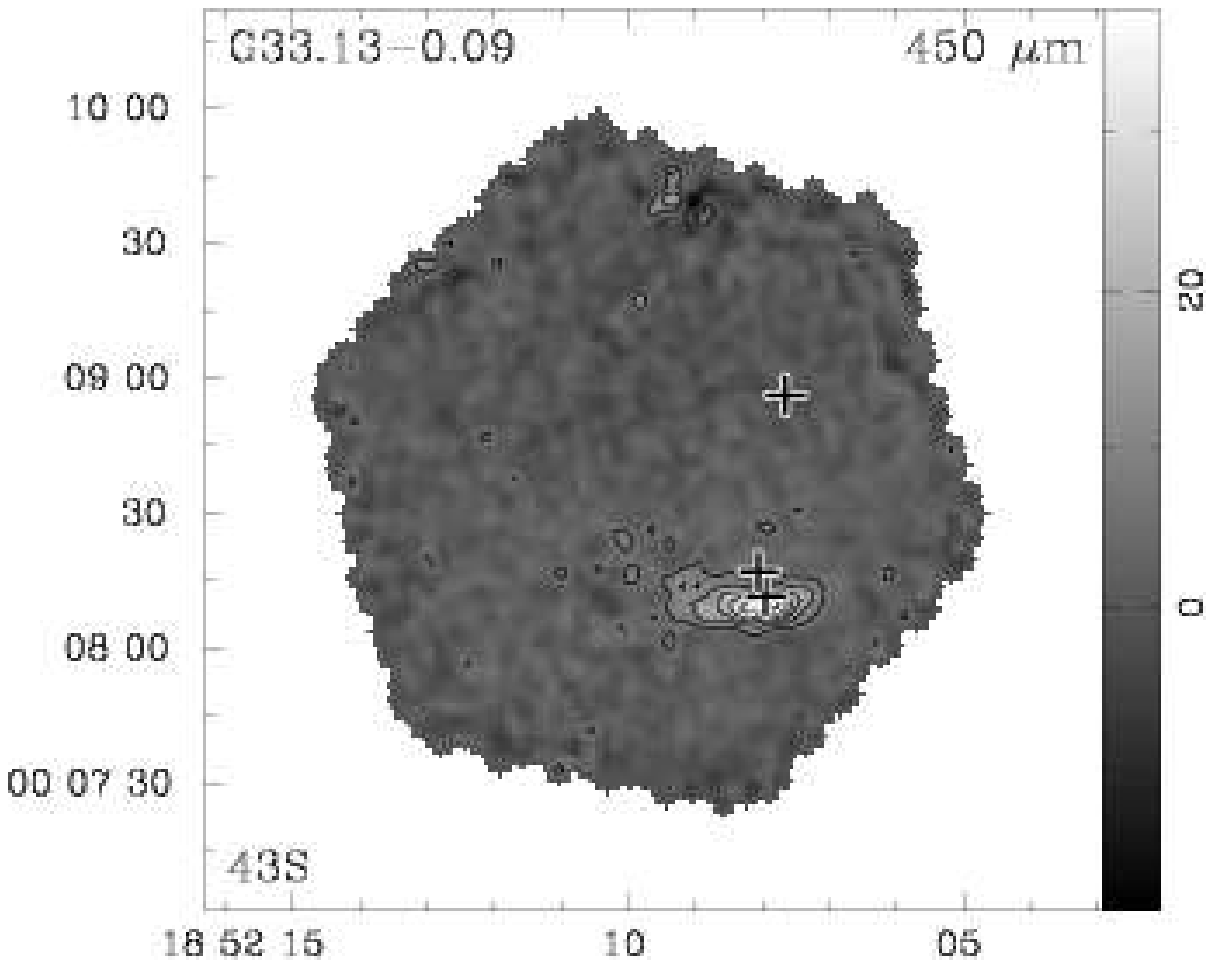}
 \hspace*{0.11 cm}\includegraphics[scale=0.35,angle=-90]{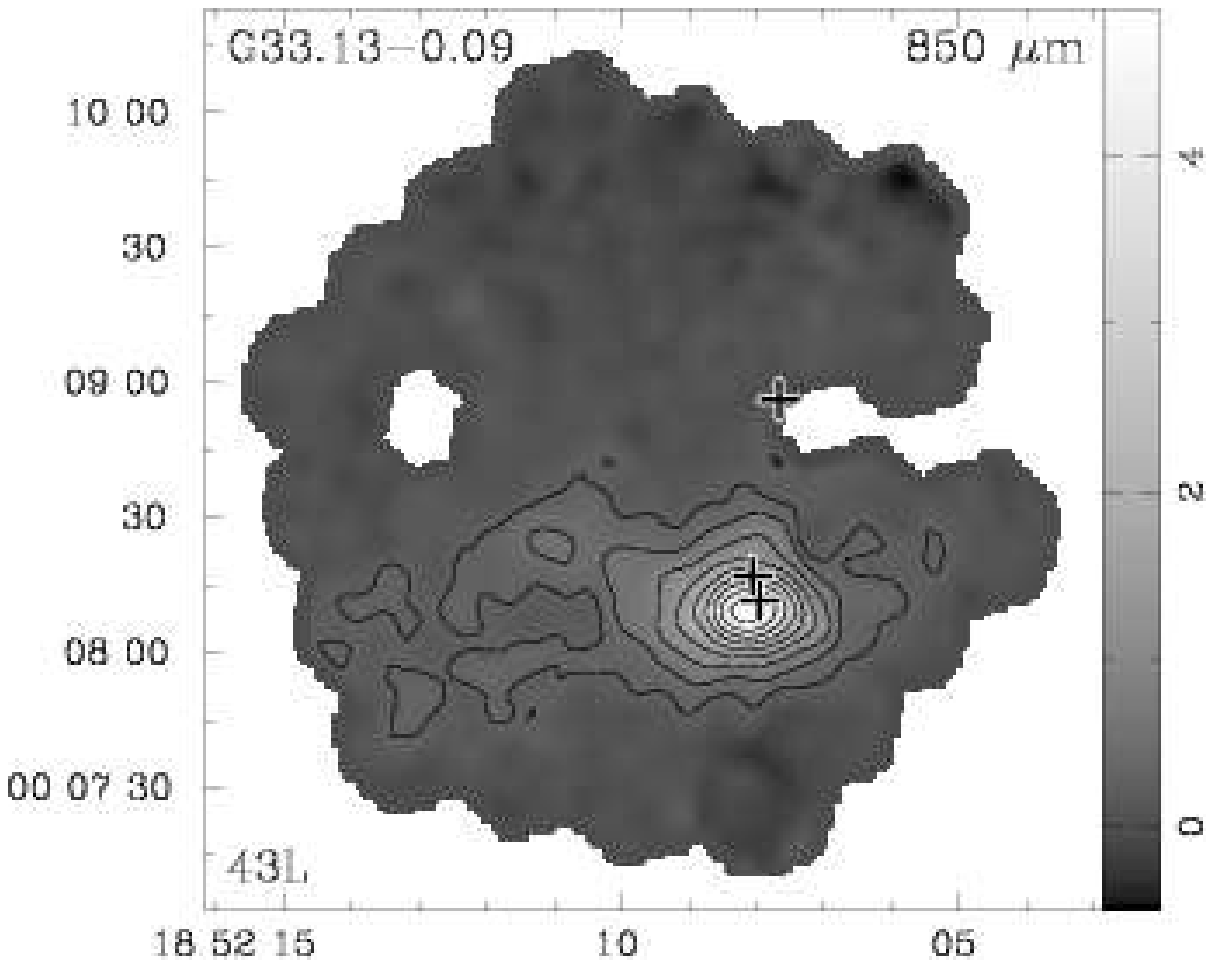}\\ 
 
 \hspace*{0.11 cm}\includegraphics[scale=0.35,angle=-90]{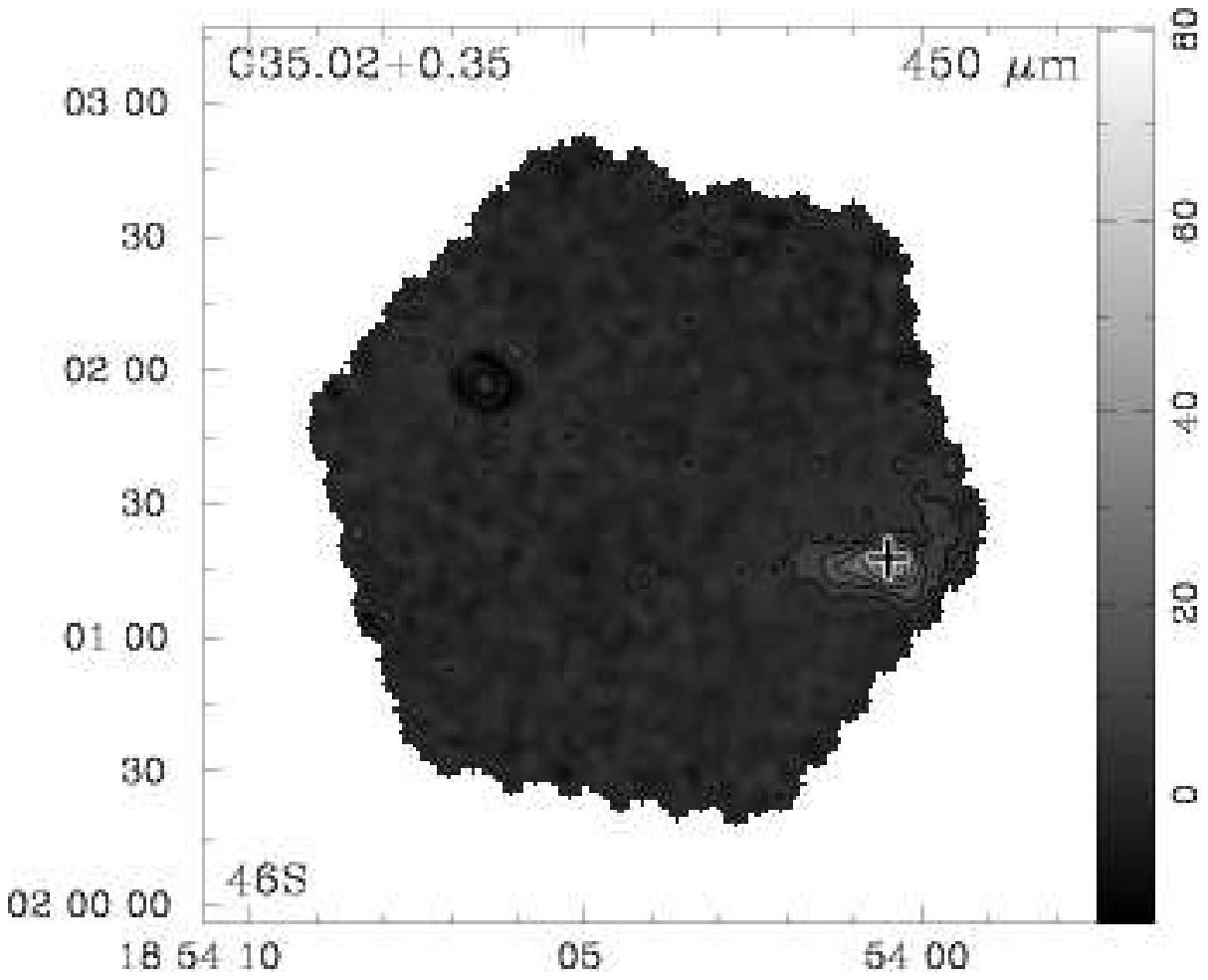}
 \hspace*{0.11 cm}\includegraphics[scale=0.35,angle=-90]{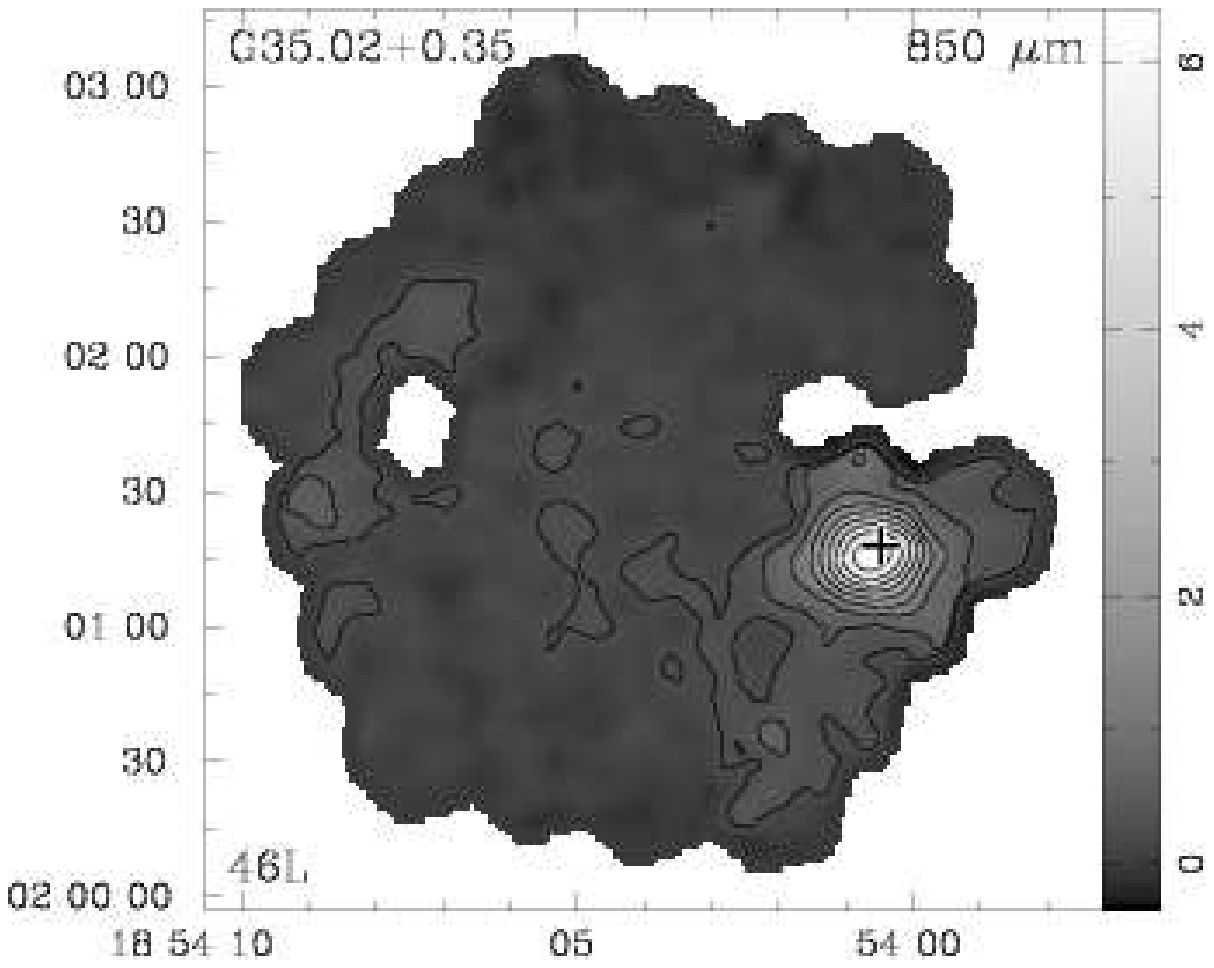}
 \hspace*{0.11 cm}\includegraphics[scale=0.35,angle=-90]{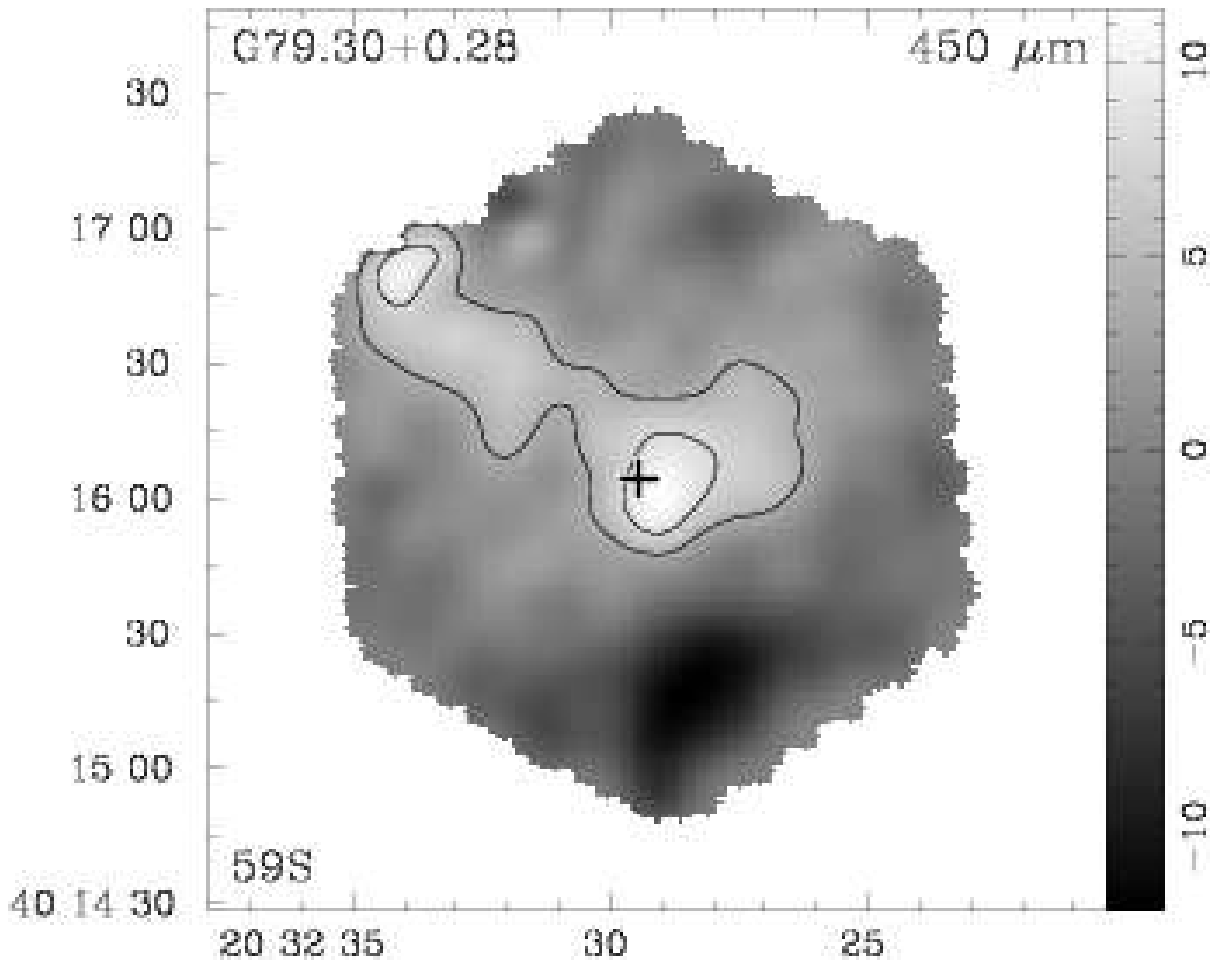}
 \hspace*{0.11 cm}\includegraphics[scale=0.35,angle=-90]{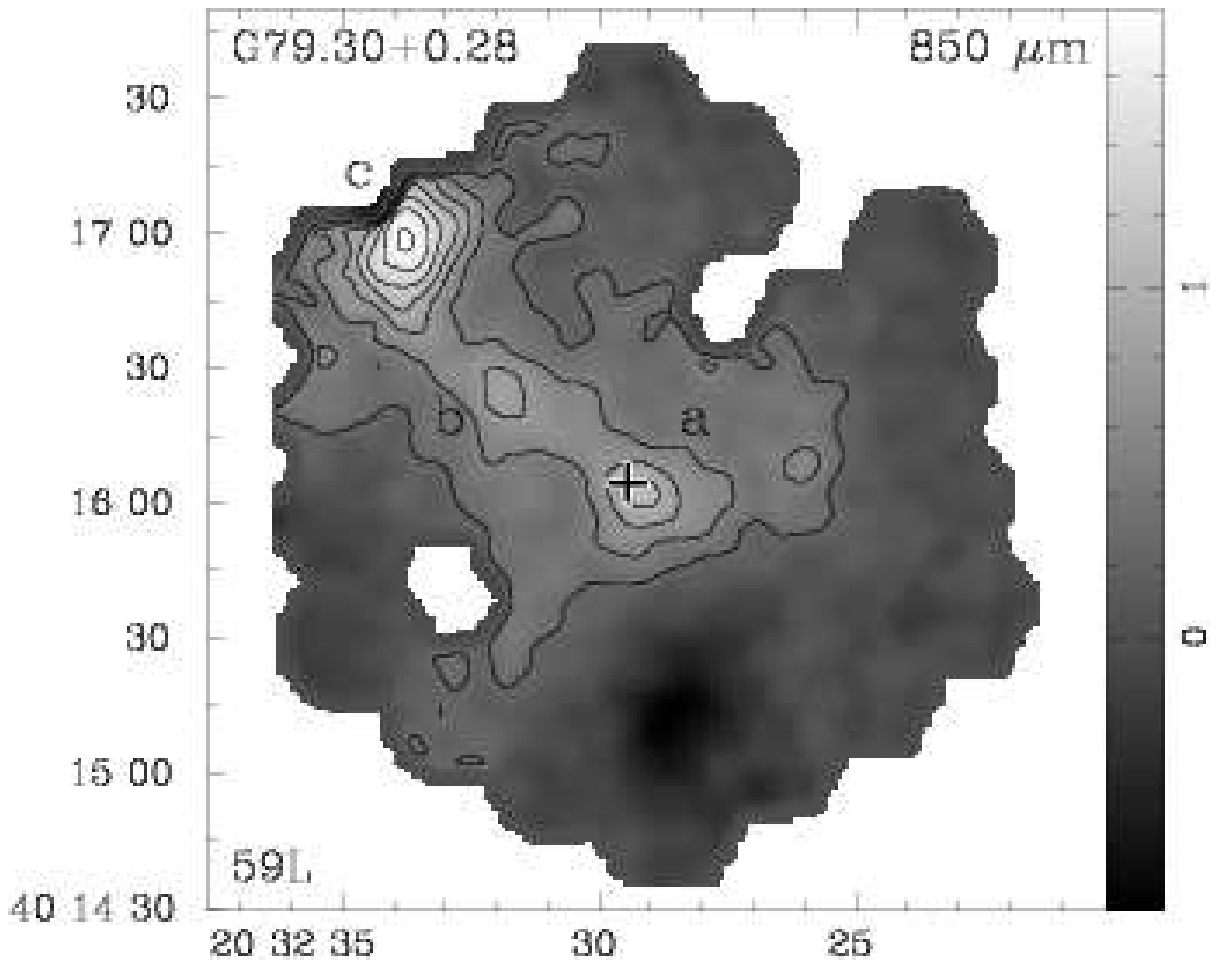} \\

 \caption{Selected SCUBA images from the survey with detections at both 450 and 850 $\mu$m. The images displayed here are those
 discussed further in the text of the paper. A full version of Fig.~\ref{fig:scuboth} can be found in the Online Supplement.
 Each UC H{\sc ii} region is represented by a pair of images at 450 $\mu$m (\emph{left image}) and 850$\mu$m (\emph{right
 image}). Coordinates are Right Ascension and Declination in the J2000 system.  Crosses indicate the positions of ultracompact
 H{\sc ii} regions from Wood \& Churchwell (\cite{wc89a}), Kurtz, Churchwell \& Wood (\cite{kcw94}), Becker et al.~(\cite{b94})
 or Giveon et al.~(\cite{gbh2005}). All images have been deconvolved with a model of the JCMT beam to remove the contribution
 from the error lobe and 450 $\mu$m images with limited signal-to-noise have been smoothed to the same resolution as the 850
 $\mu$m images to improve the source detections.}   
\label{fig:scuboth} 
\end{minipage} 
\end{figure*}

 \begin{figure*}[p]
 \begin{minipage}{\linewidth}
 \includegraphics[scale=0.35,angle=-90]{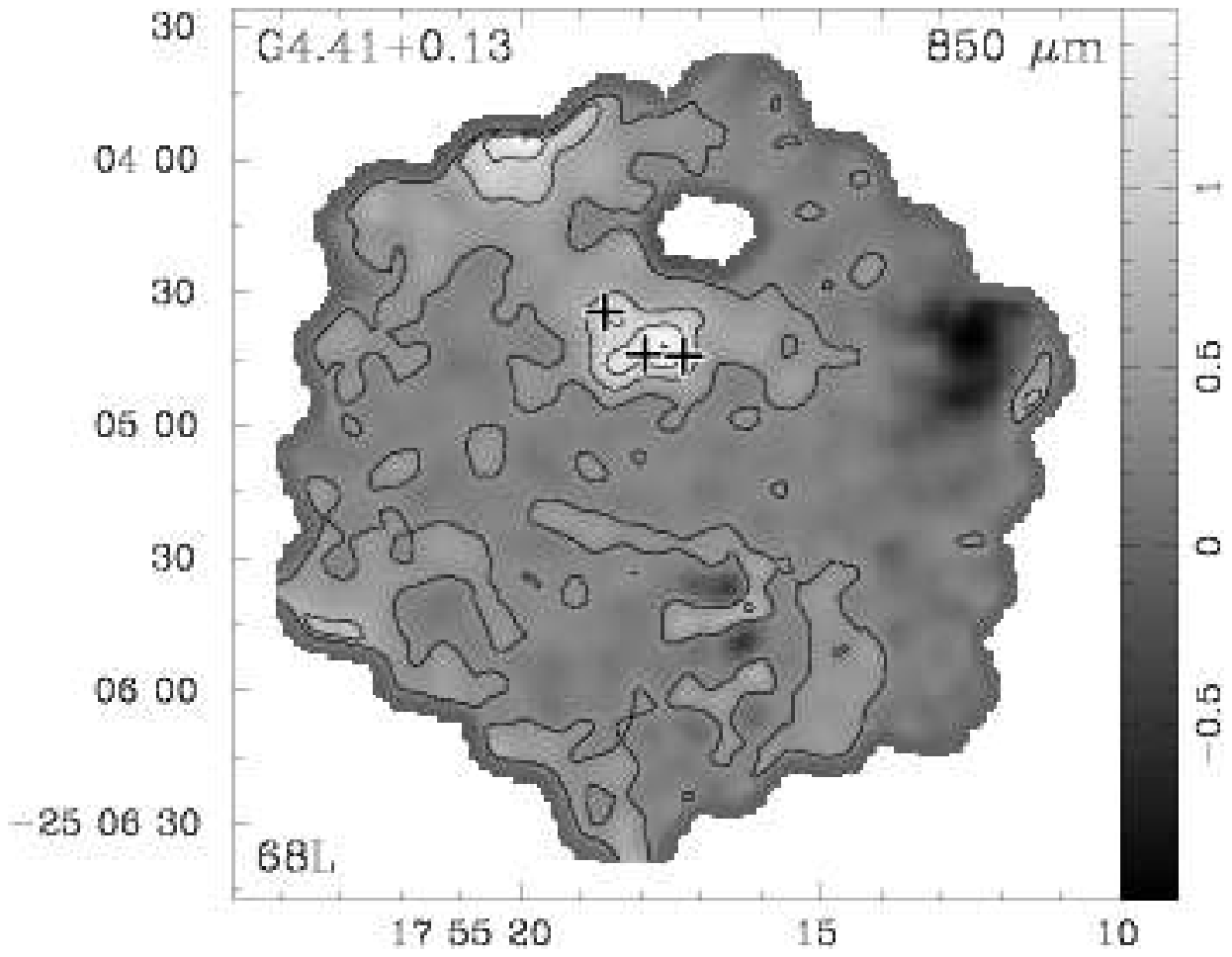} 
 \includegraphics[scale=0.35,angle=-90]{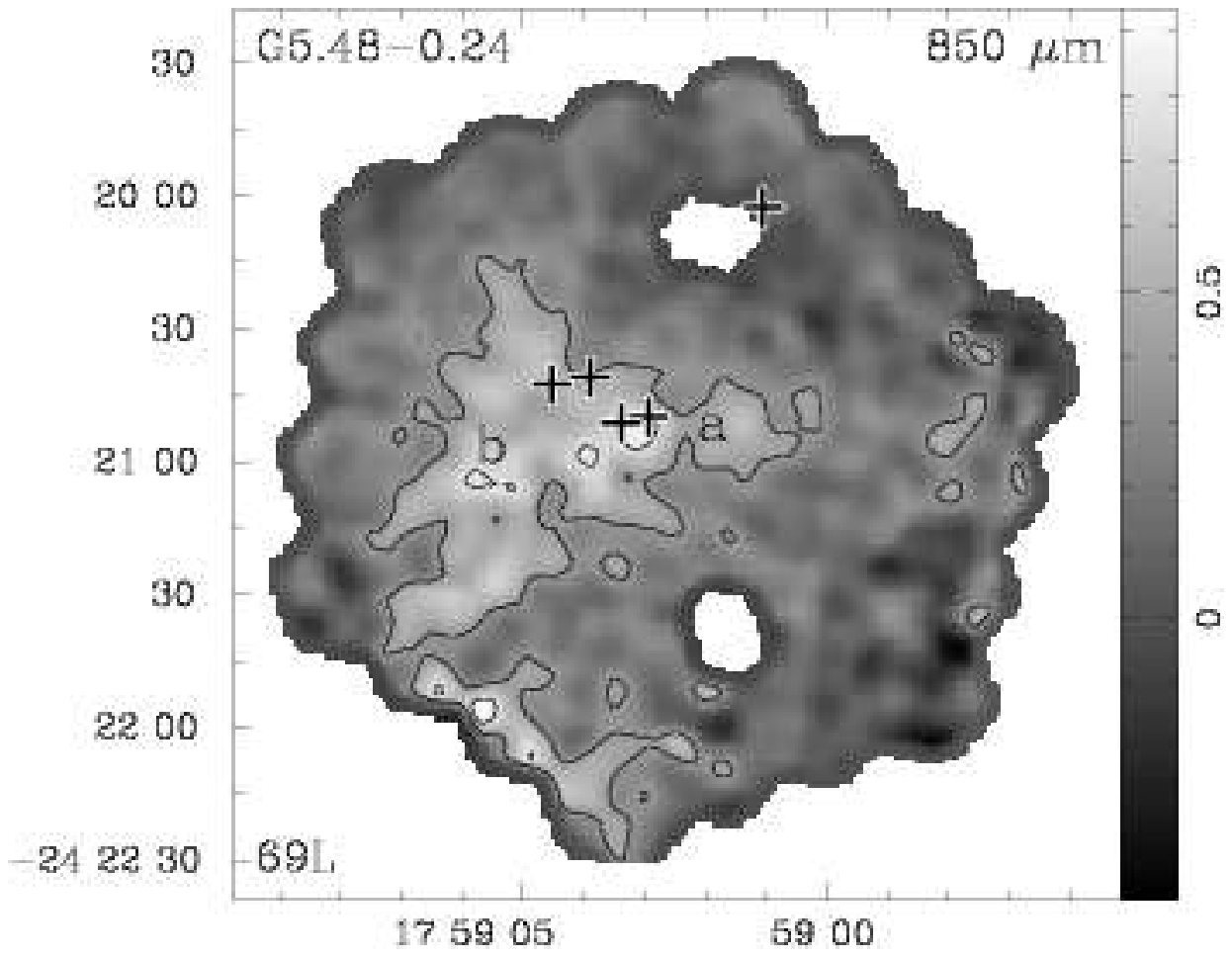} 
 \includegraphics[scale=0.35,angle=-90]{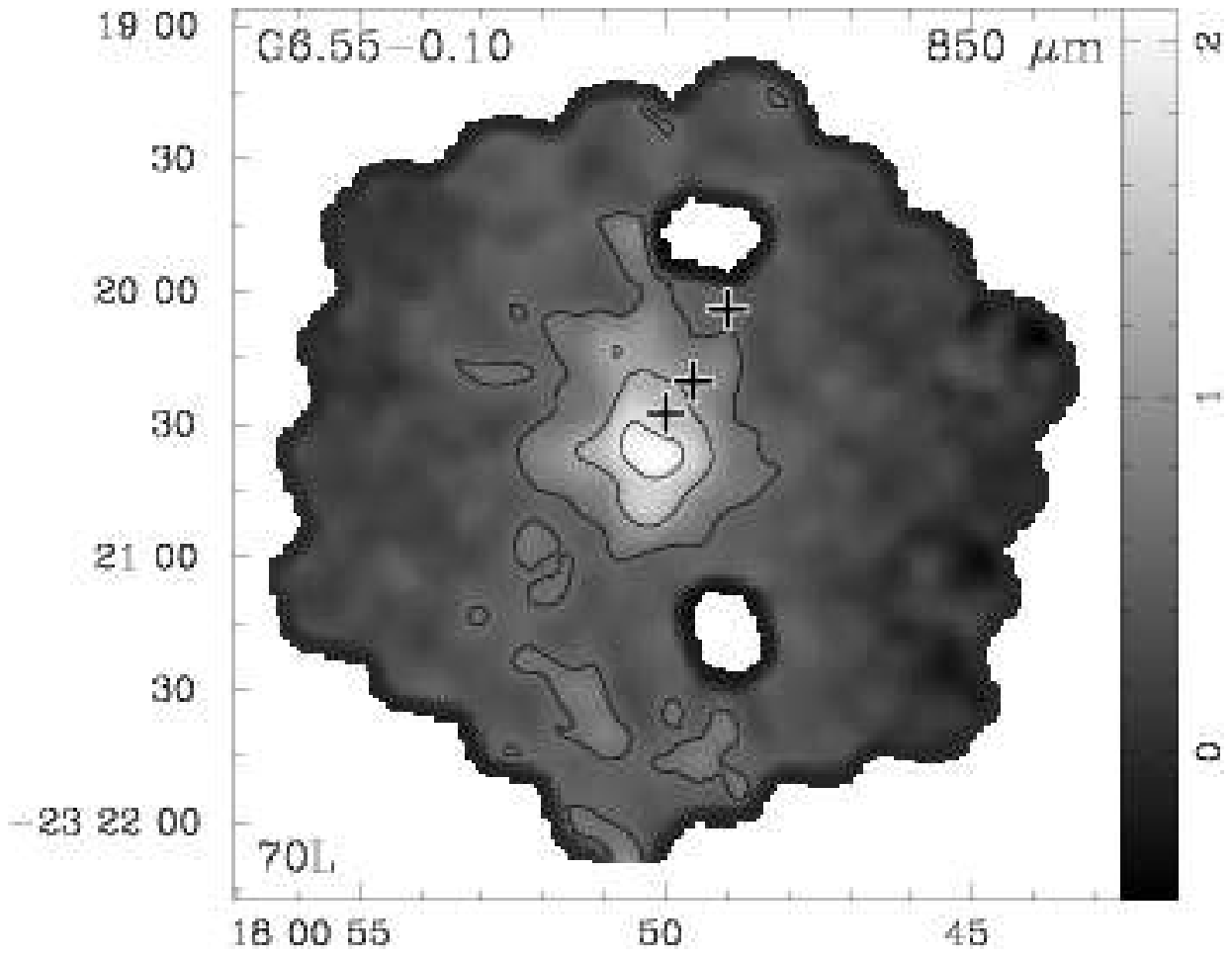} 
 \includegraphics[scale=0.35,angle=-90]{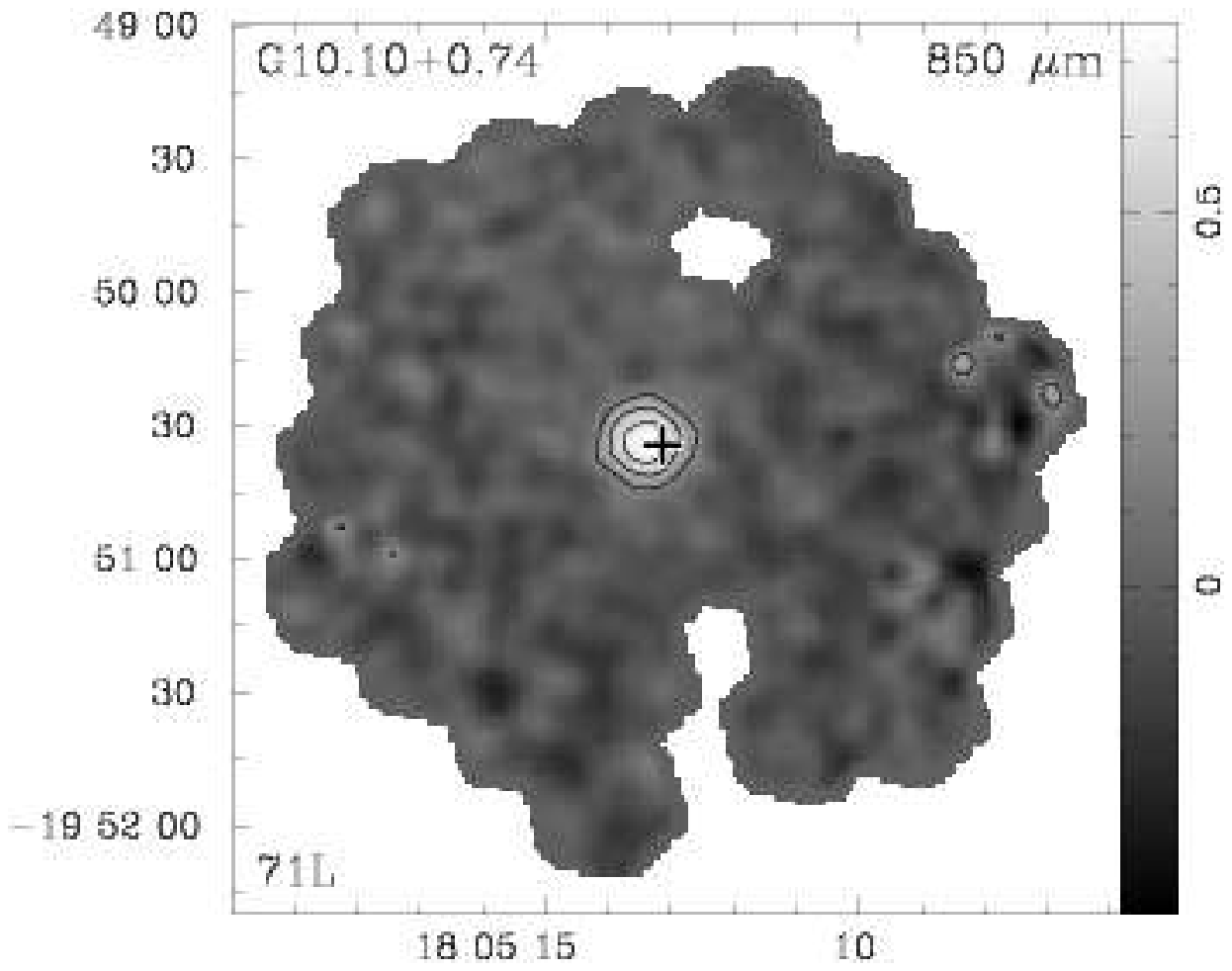}\\
 
 \includegraphics[scale=0.35,angle=-90]{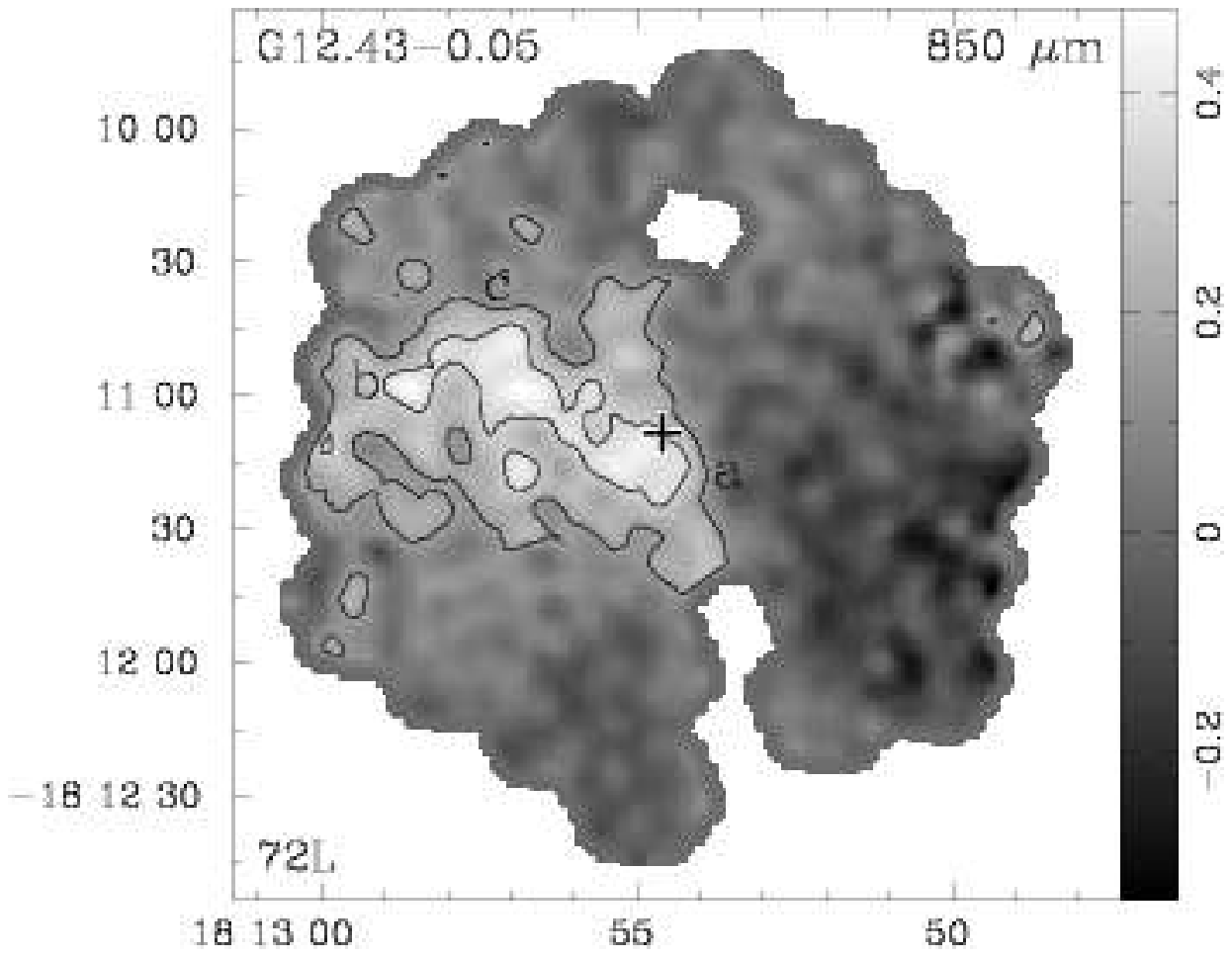} 
 \includegraphics[scale=0.35,angle=-90]{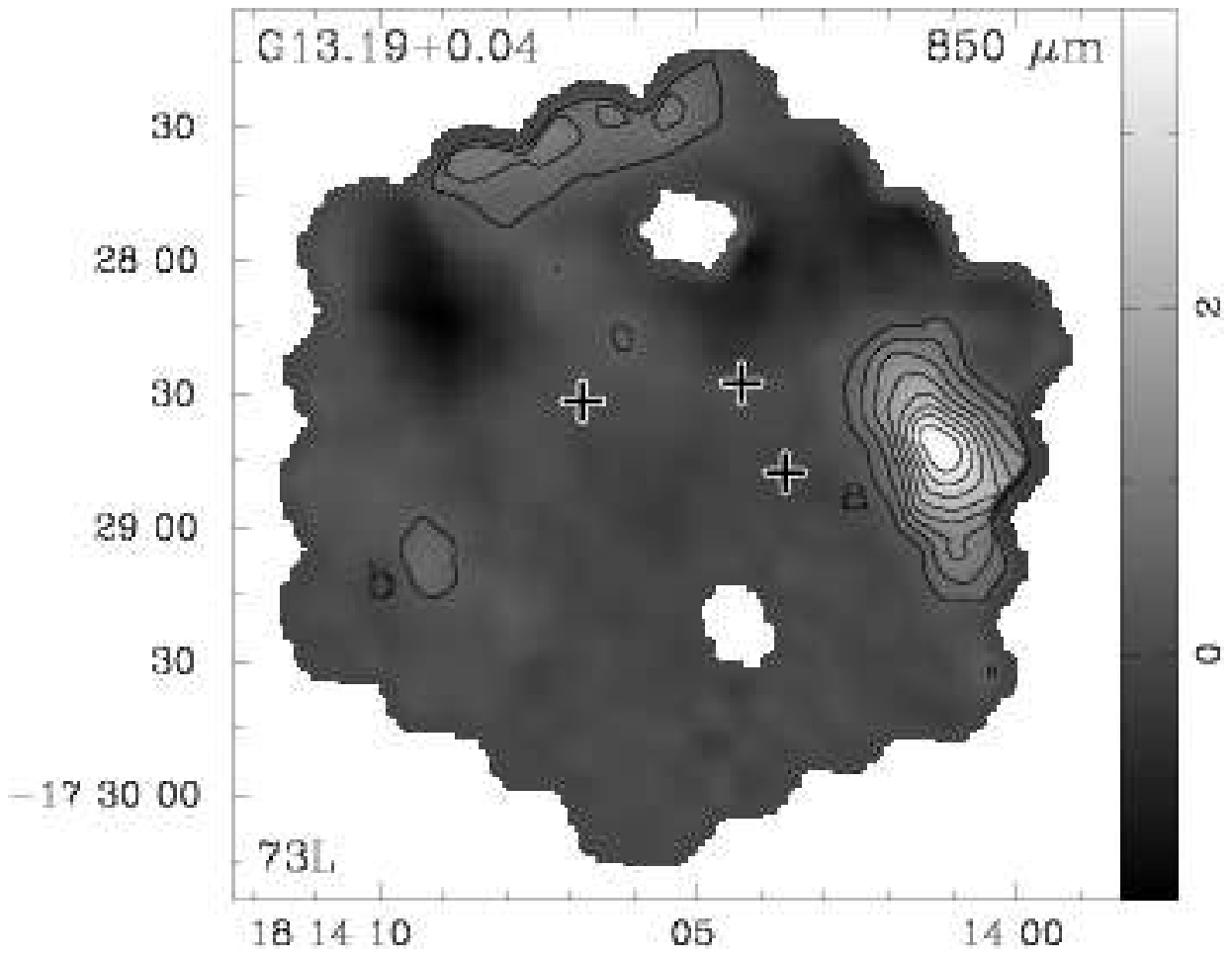} 
 \includegraphics[scale=0.35,angle=-90]{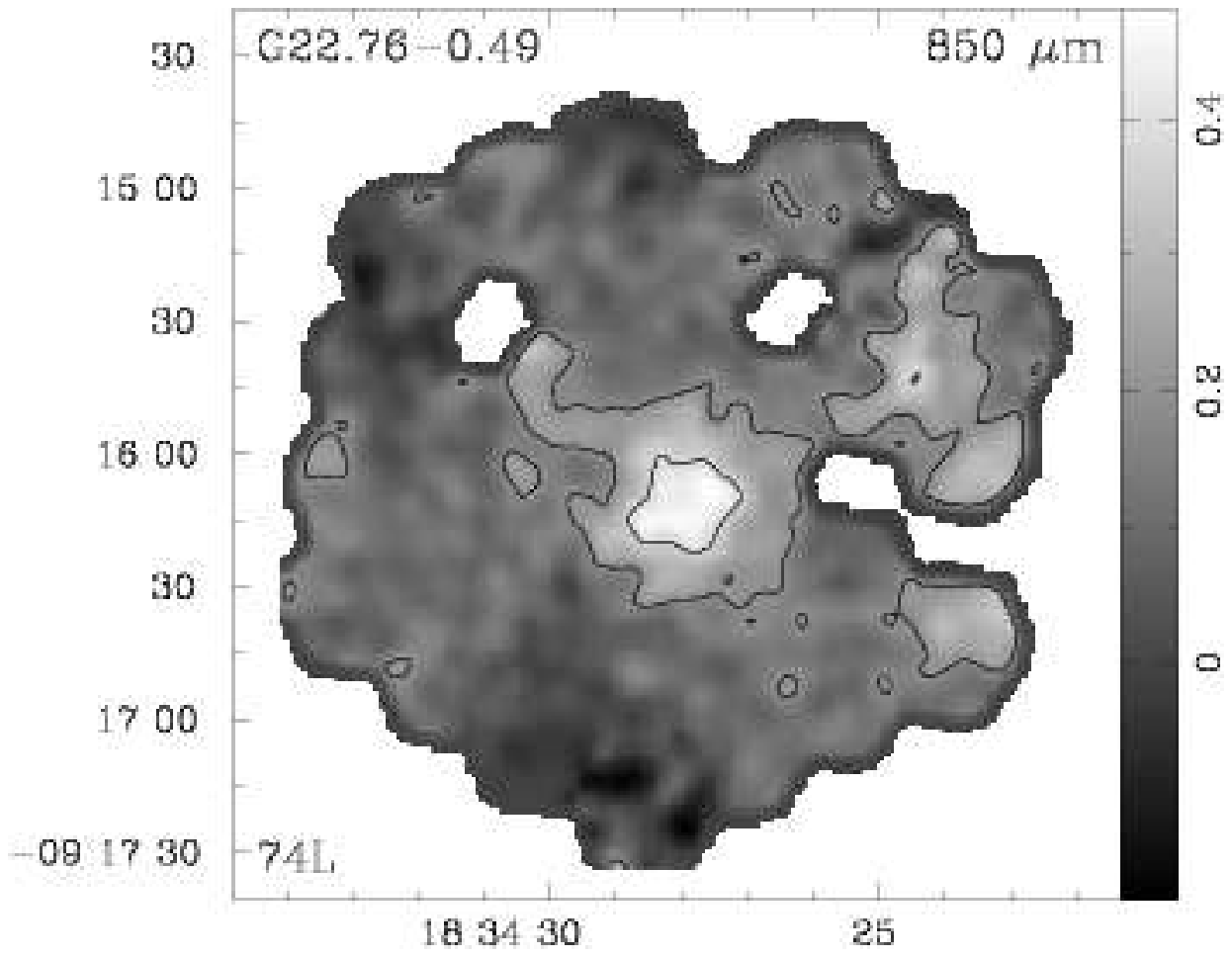} 
 \includegraphics[scale=0.35,angle=-90]{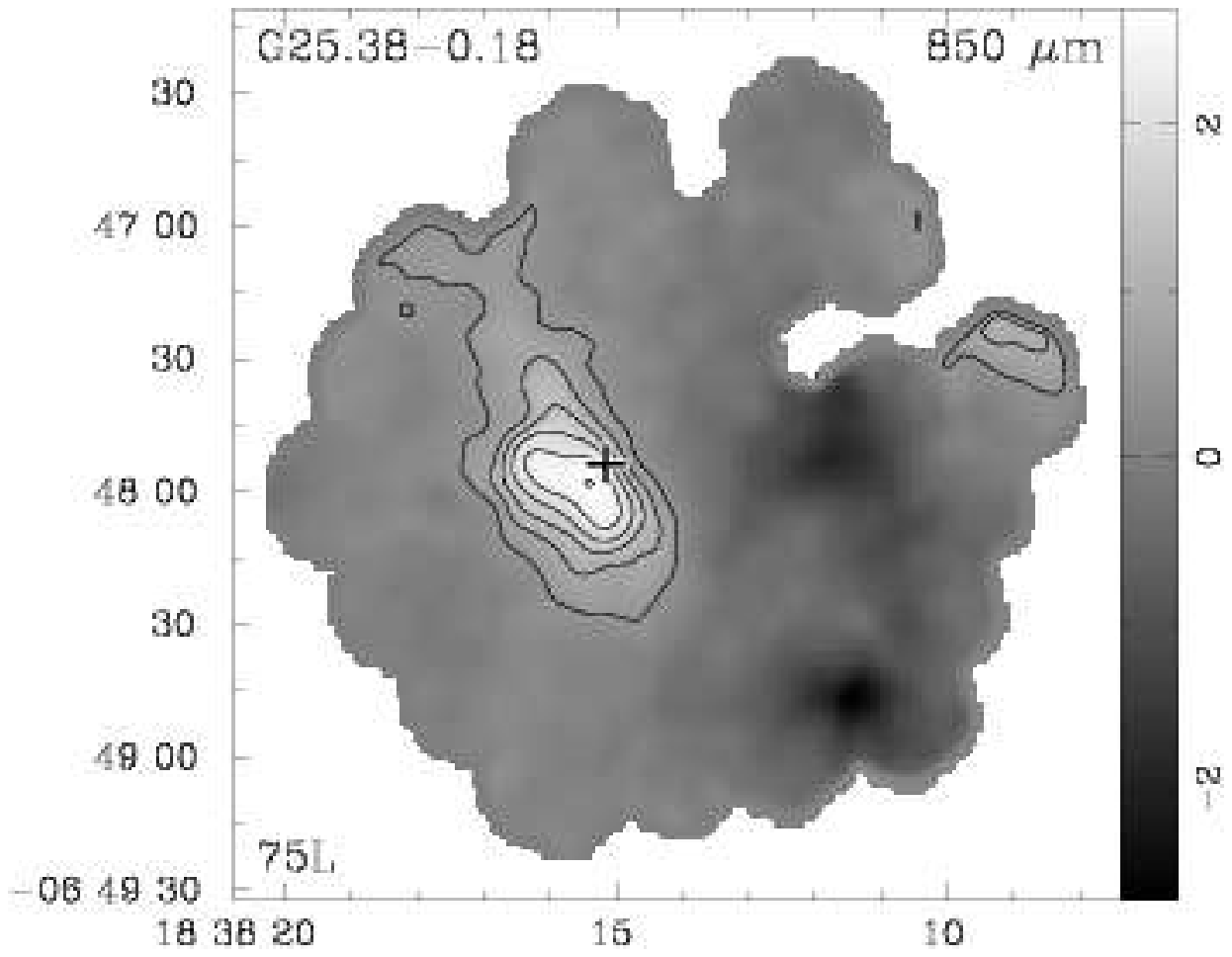}\\
 
 \includegraphics[scale=0.35,angle=-90]{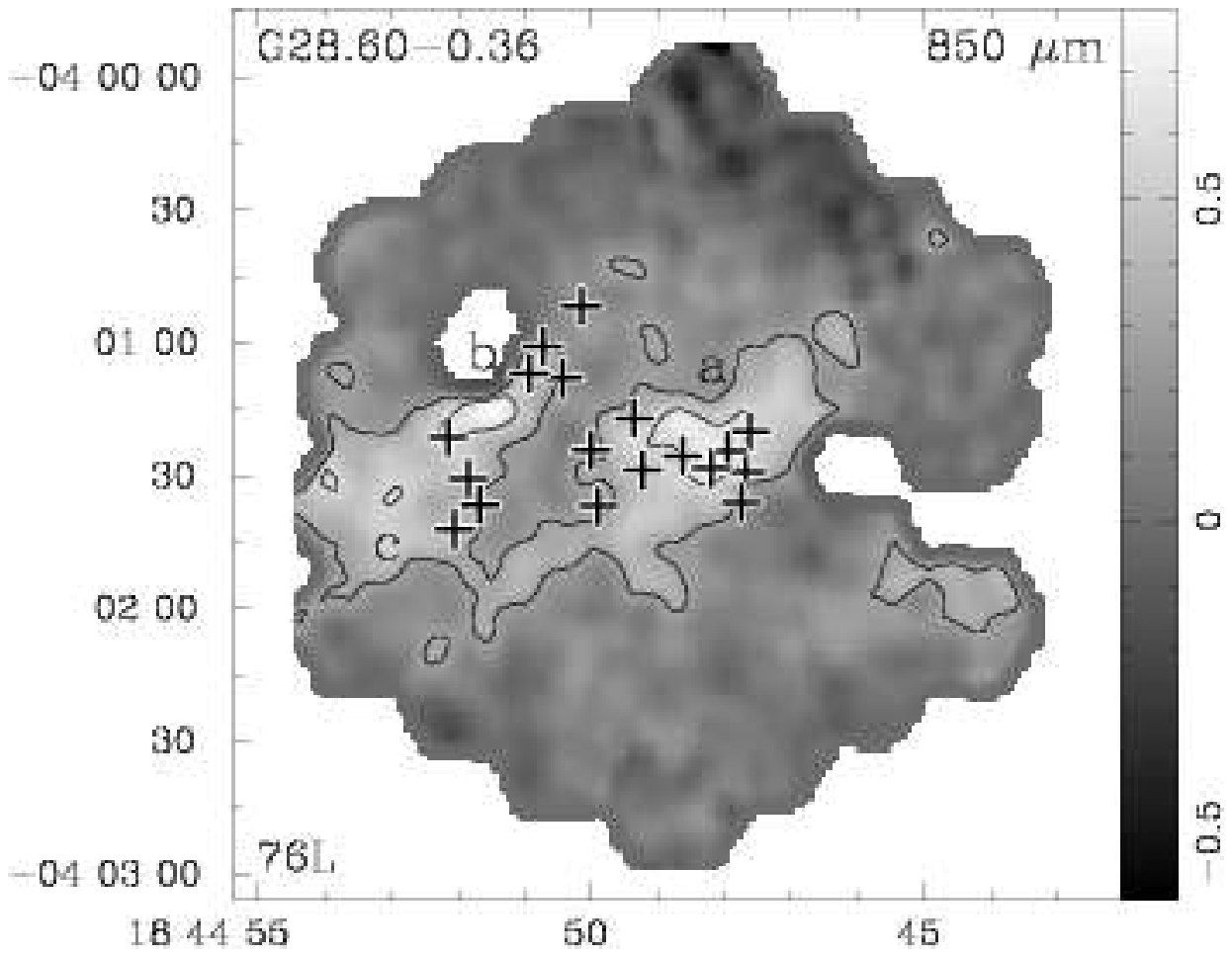} 
 \includegraphics[scale=0.35,angle=-90]{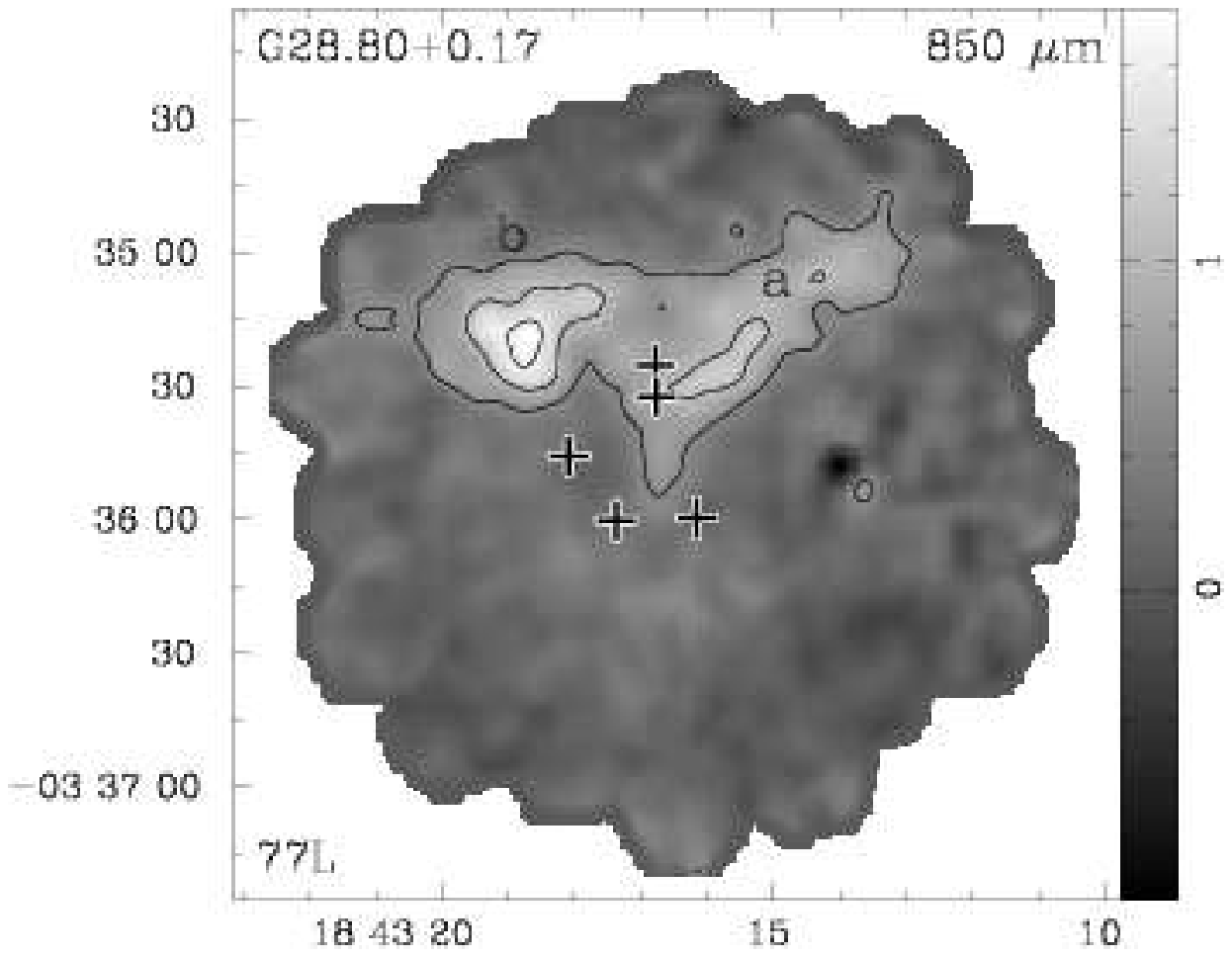}
 \includegraphics[scale=0.35,angle=-90]{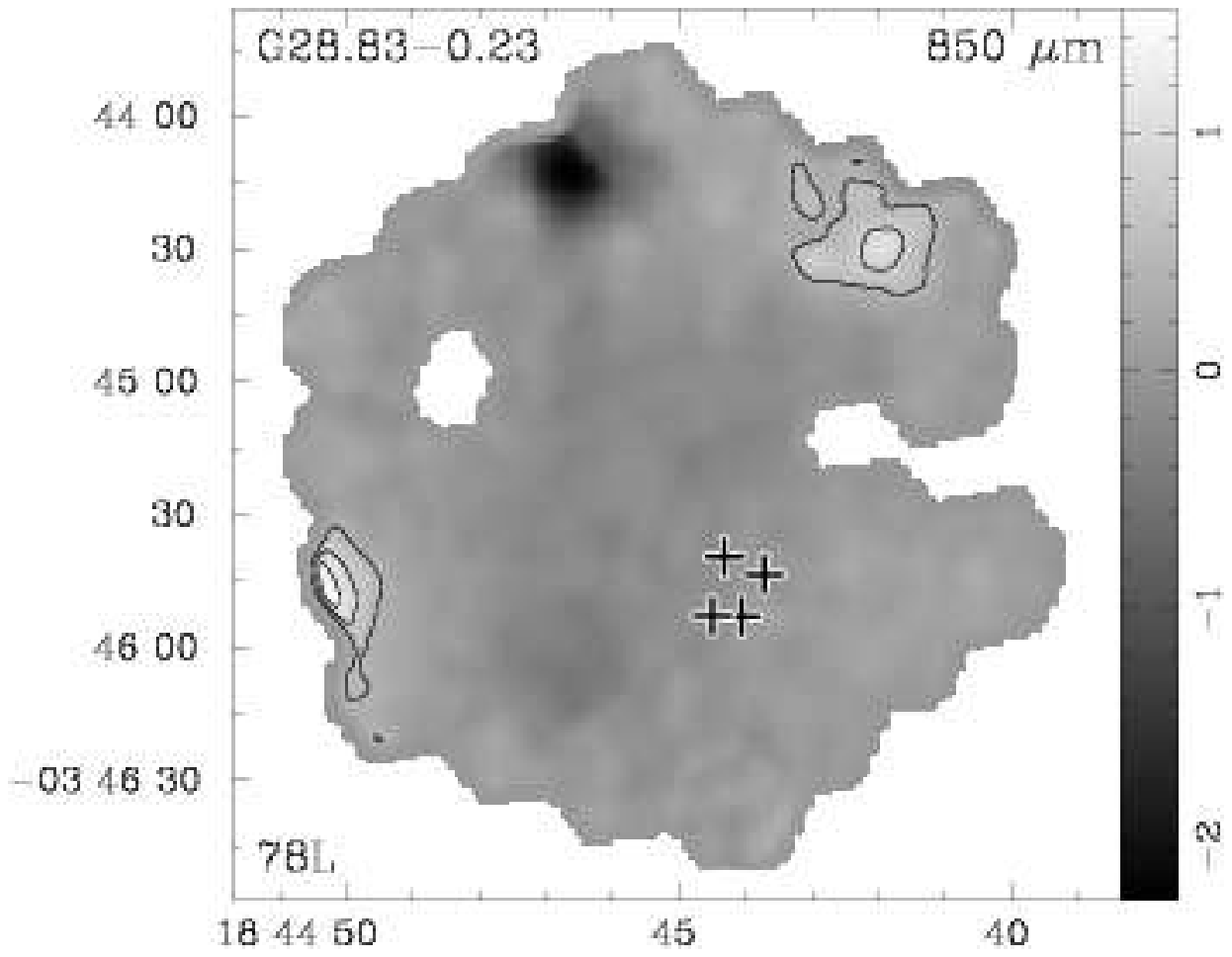} 
 \includegraphics[scale=0.35,angle=-90]{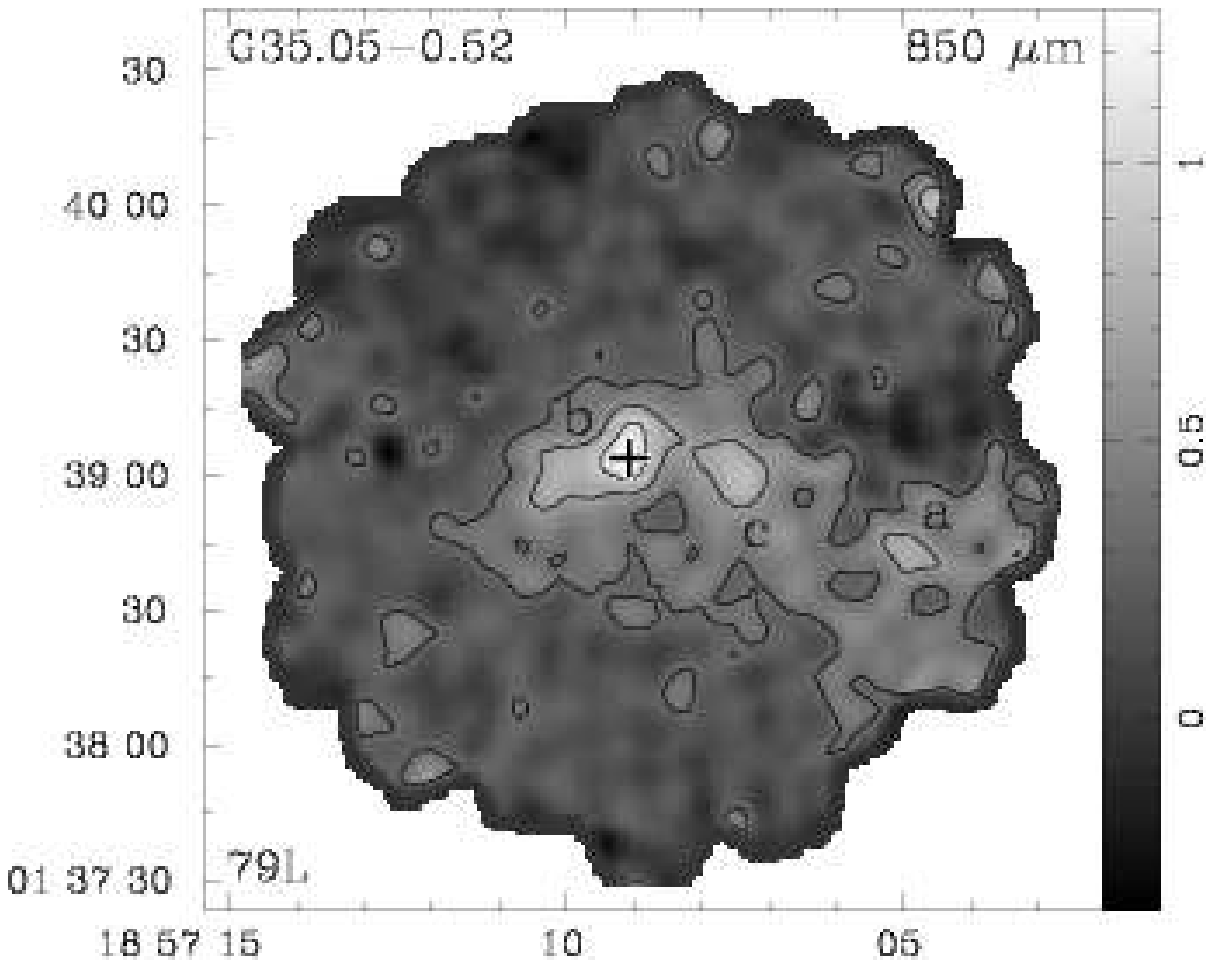}\\ 
 
 \hspace*{0.11cm}\includegraphics[scale=0.35,angle=-90]{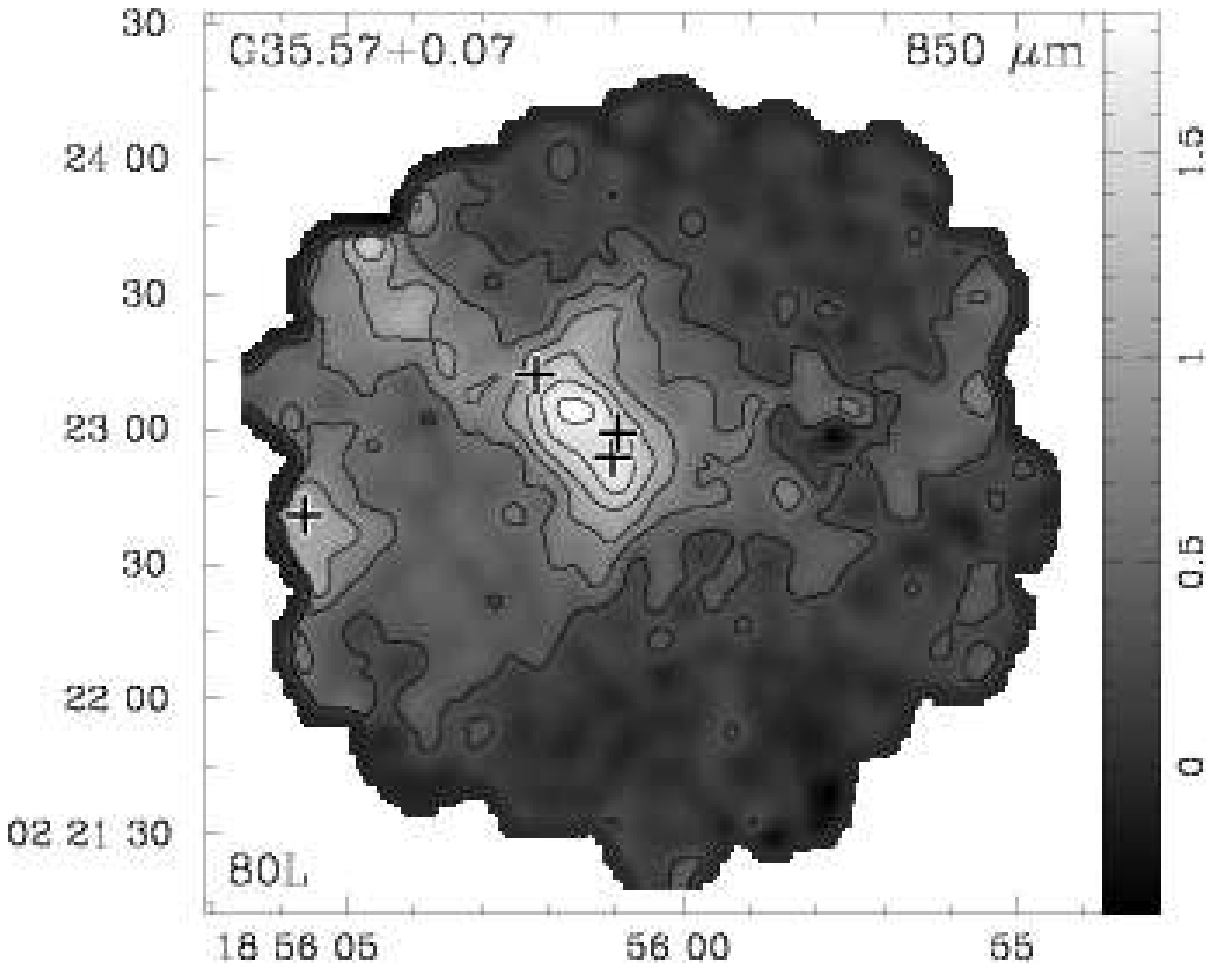}
 \hspace*{0.11cm}\includegraphics[scale=0.35,angle=-90]{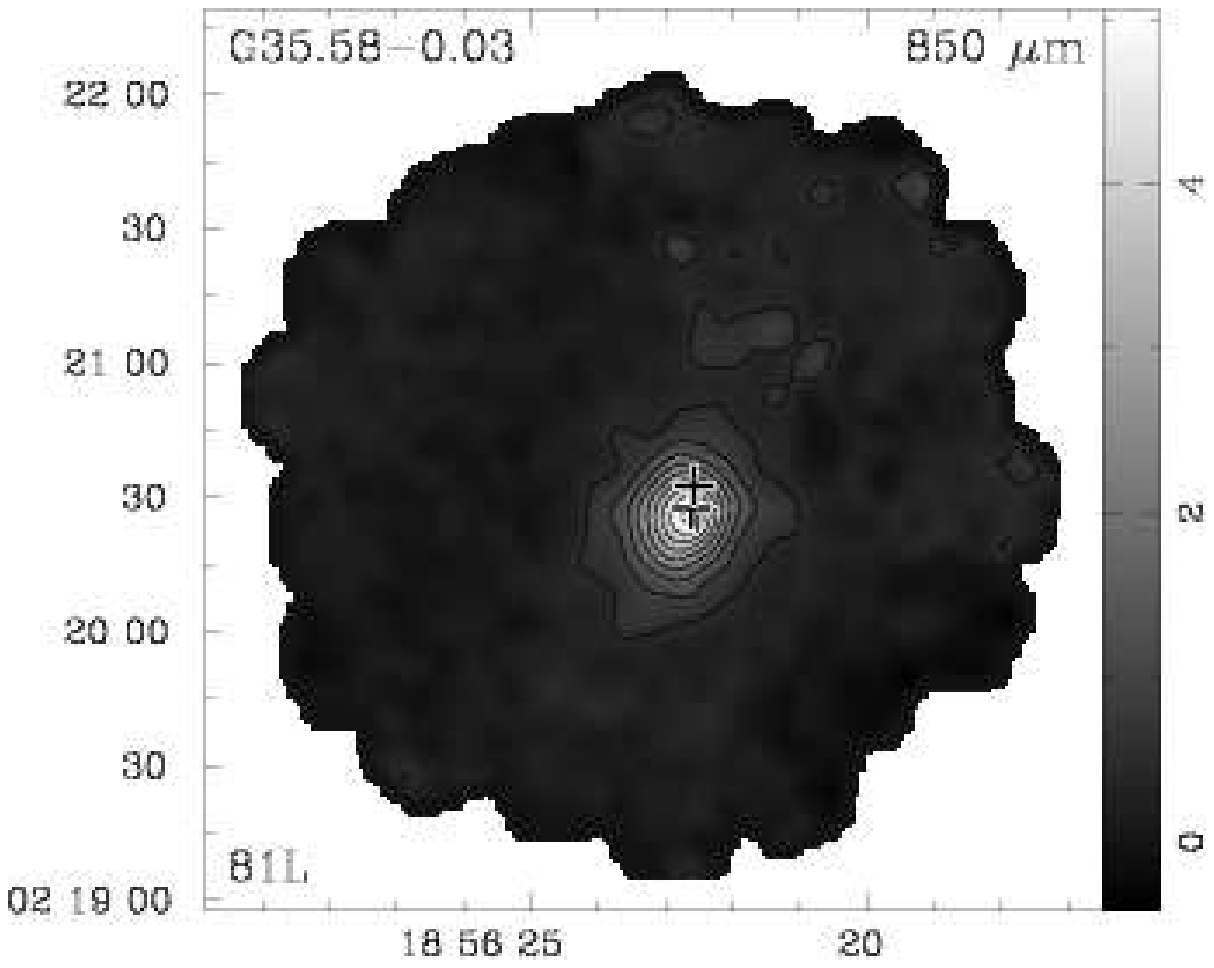} 
 \hspace*{0.11cm}\includegraphics[scale=0.35,angle=-90]{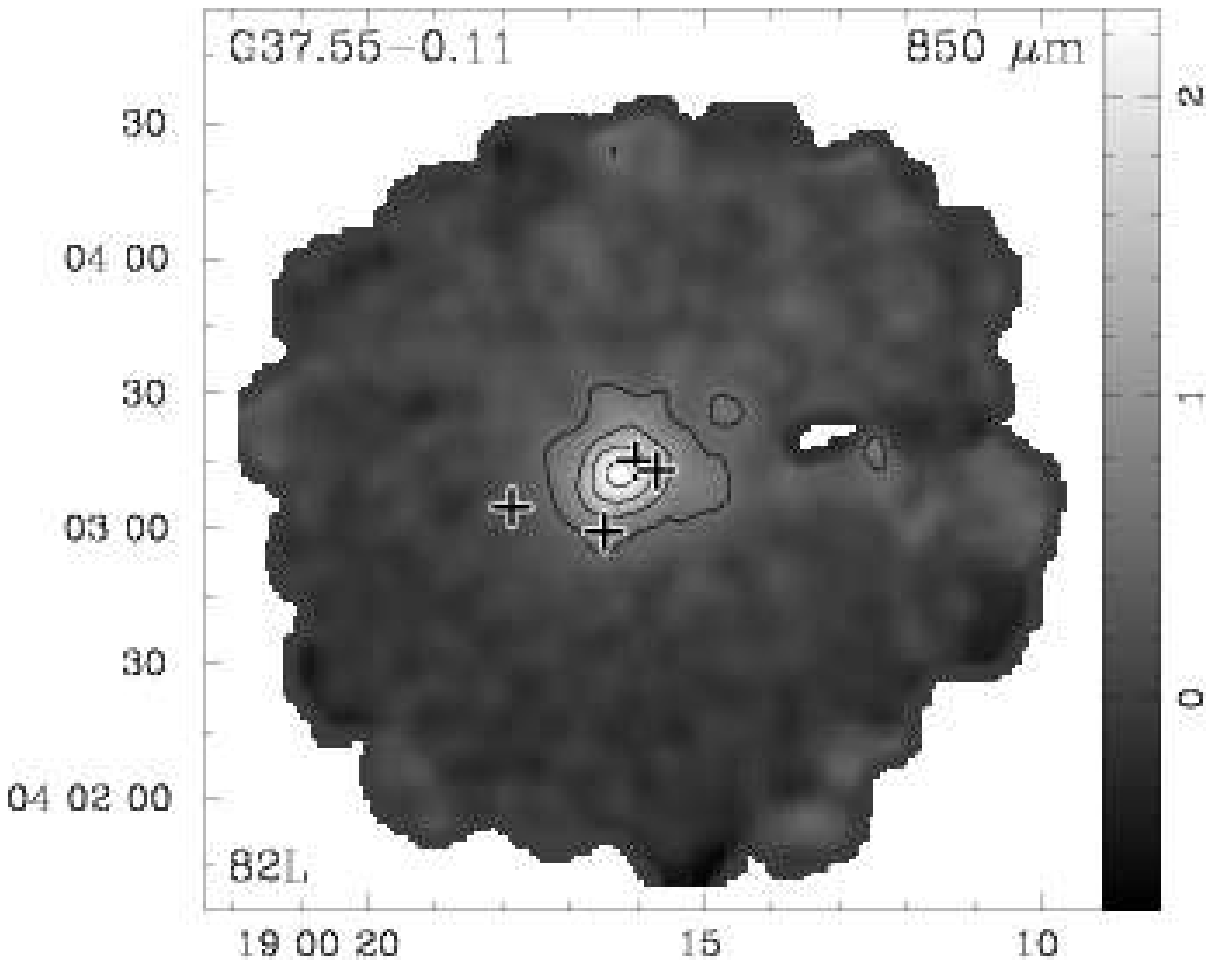}
 \hspace*{0.11cm}\includegraphics[scale=0.35,angle=-90]{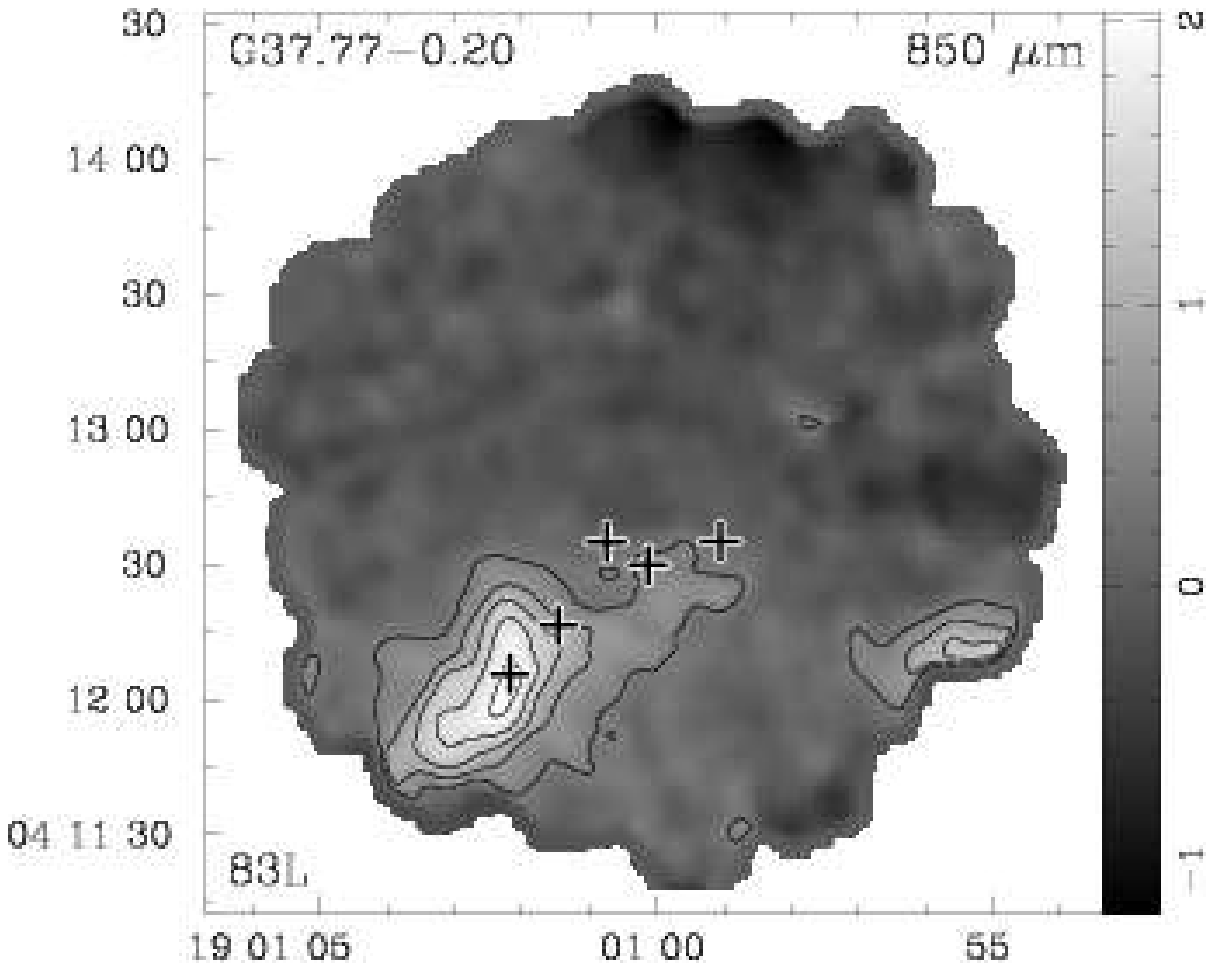}\\
 
 \hspace*{0.11cm}\includegraphics[scale=0.35,angle=-90]{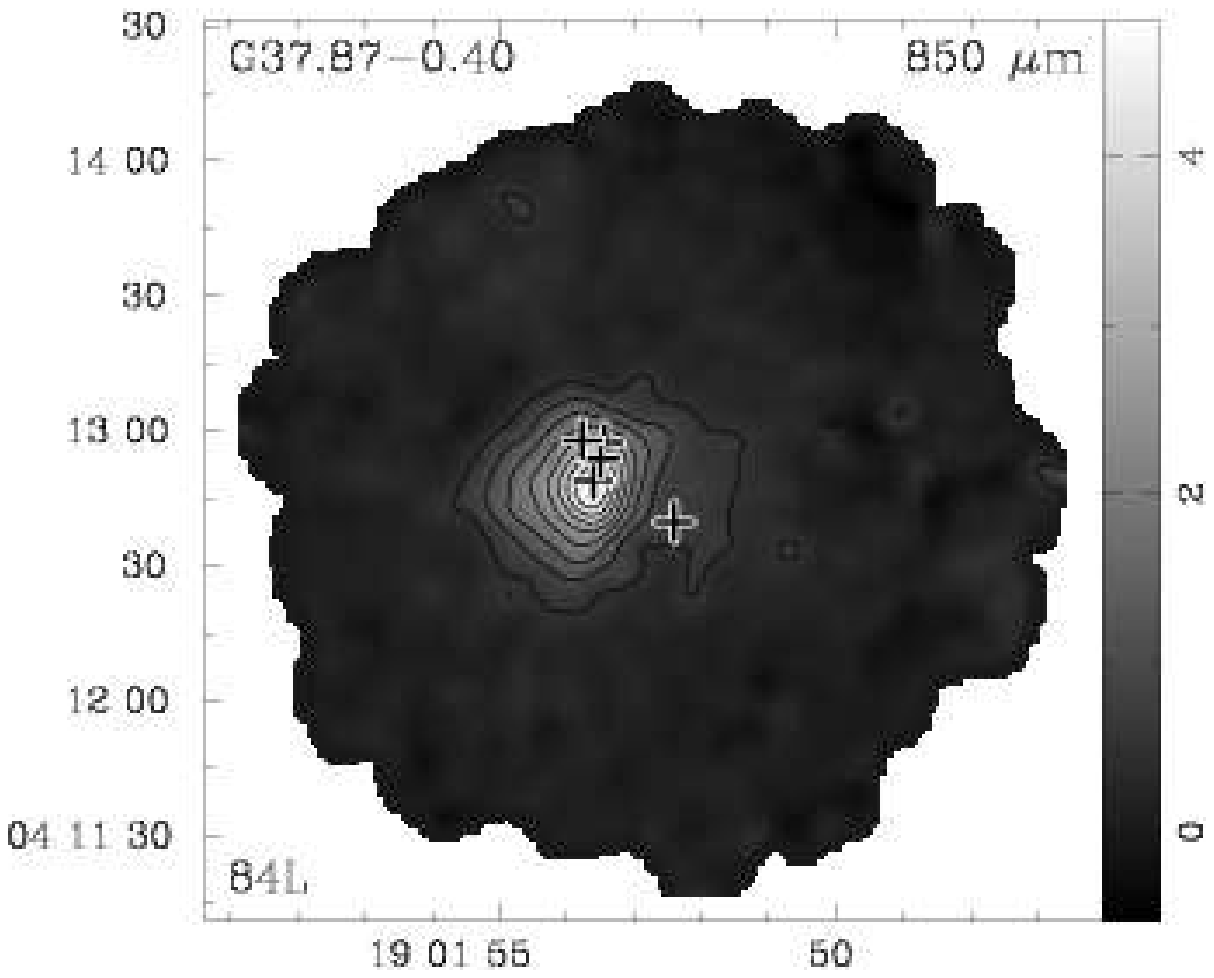}
 \hspace*{0.11cm}\includegraphics[scale=0.35,angle=-90]{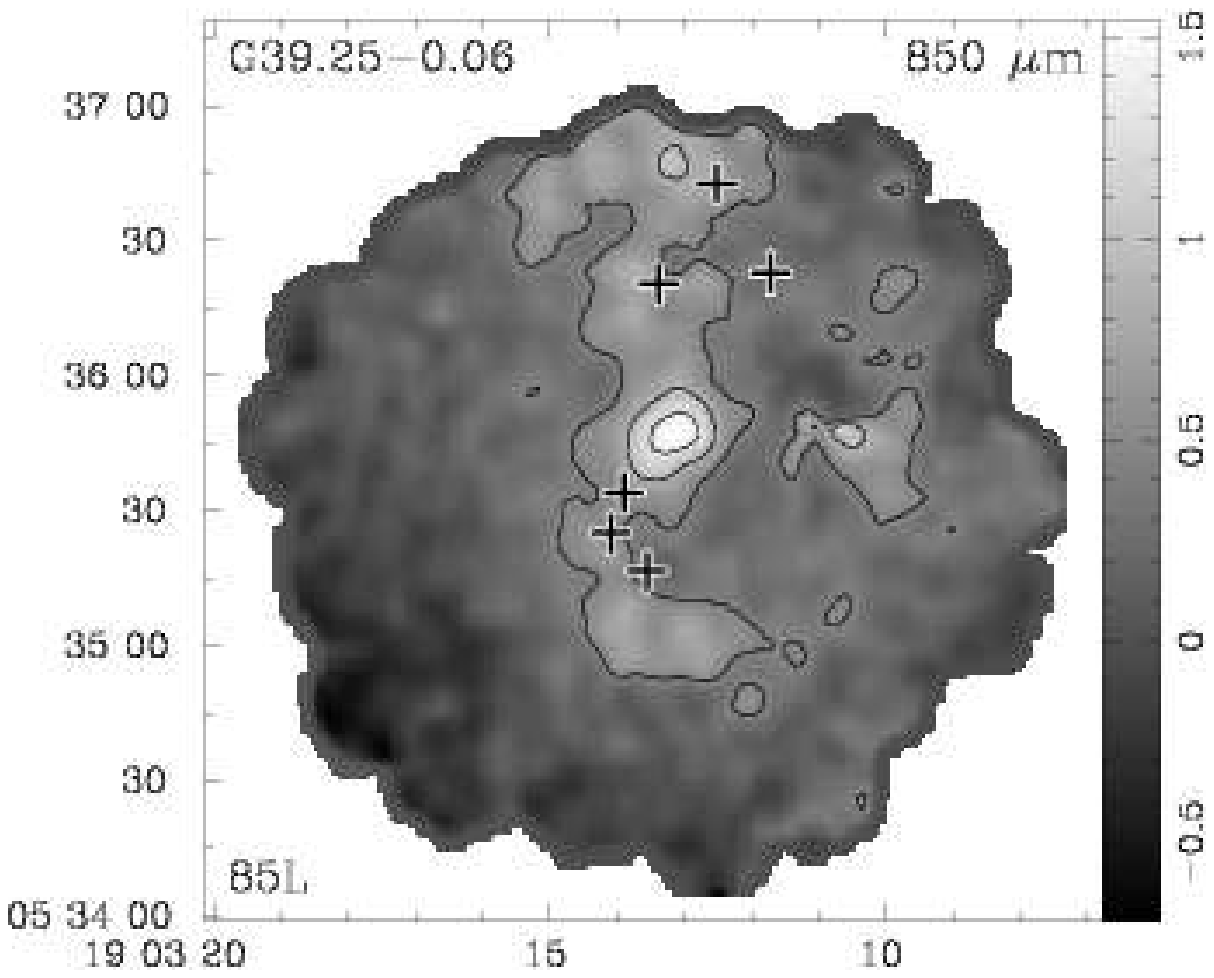}
 \hspace*{0.11cm}\includegraphics[scale=0.35,angle=-90]{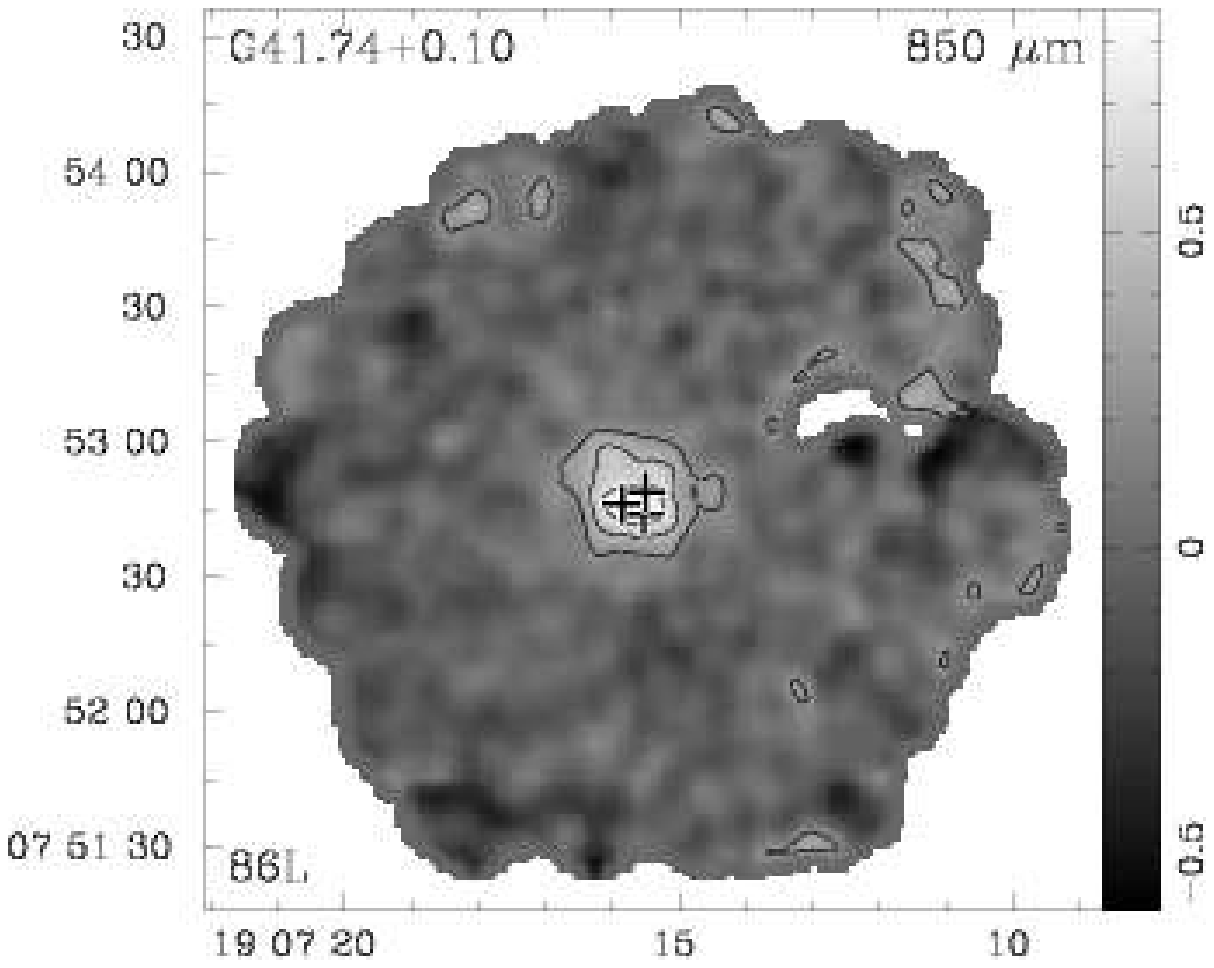}
 \hspace*{0.11cm}\includegraphics[scale=0.35,angle=-90]{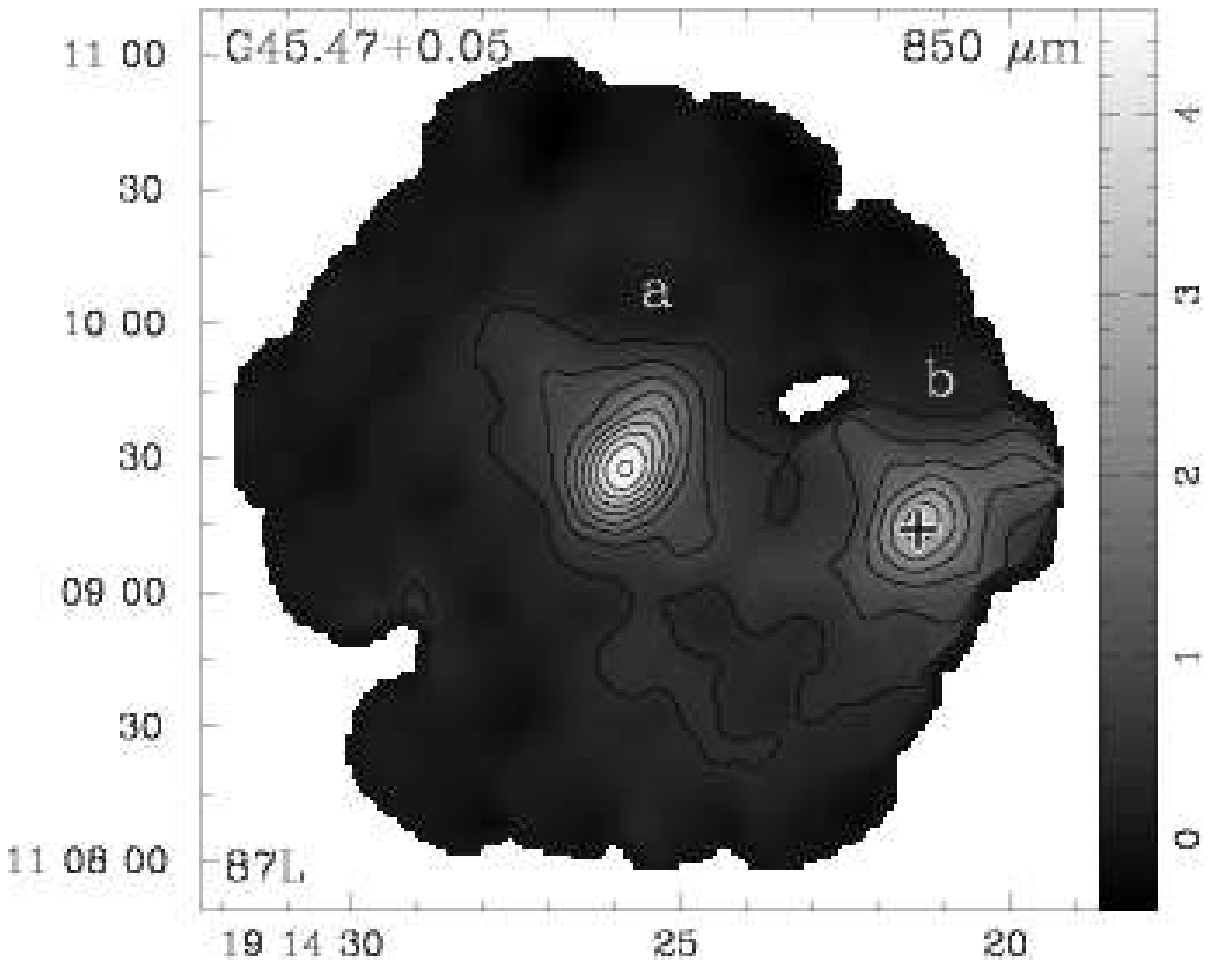}\\
 
 \hspace*{0.11cm}\includegraphics[scale=0.35,angle=-90]{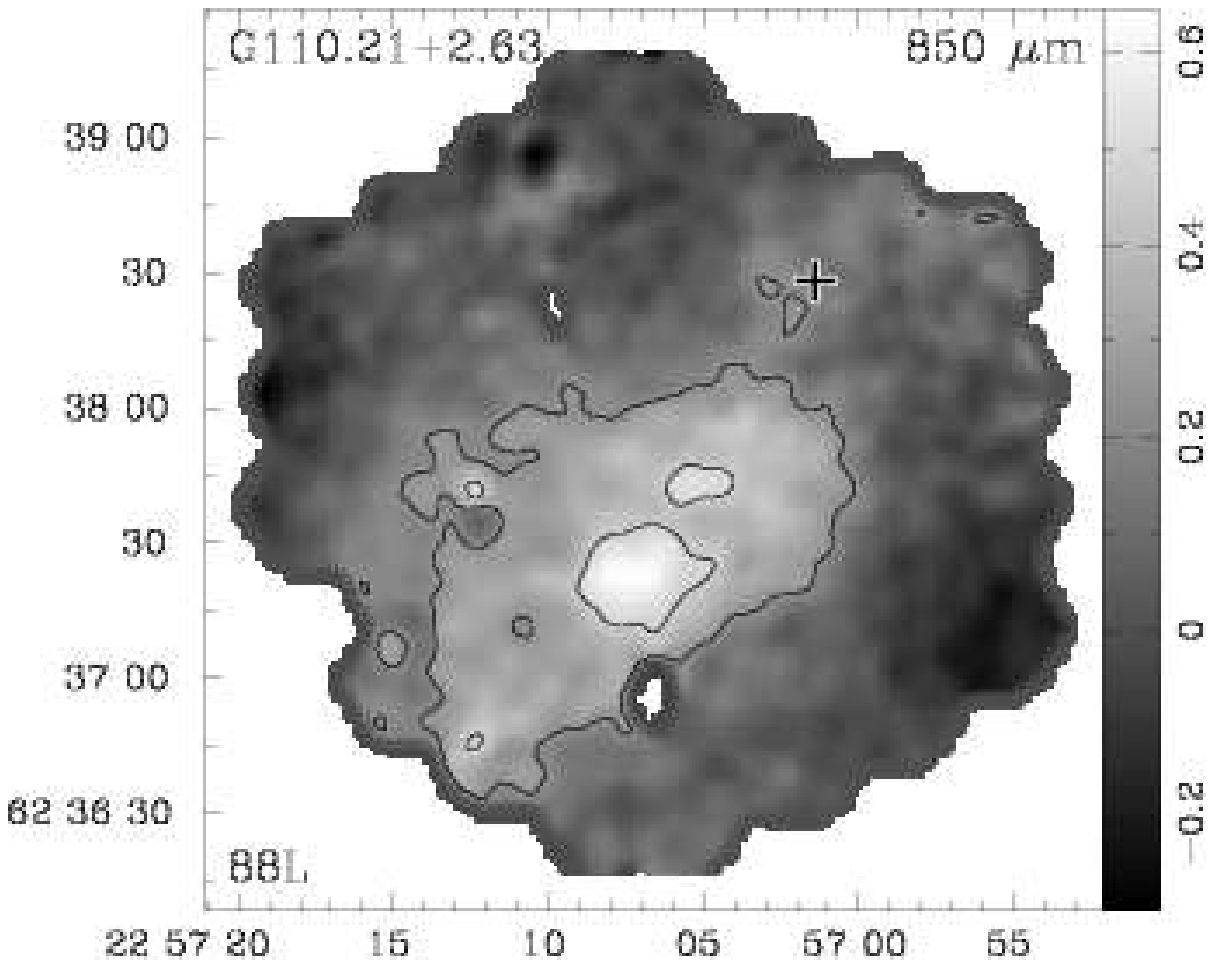}
 \hspace*{0.11cm}\includegraphics[scale=0.35,angle=-90]{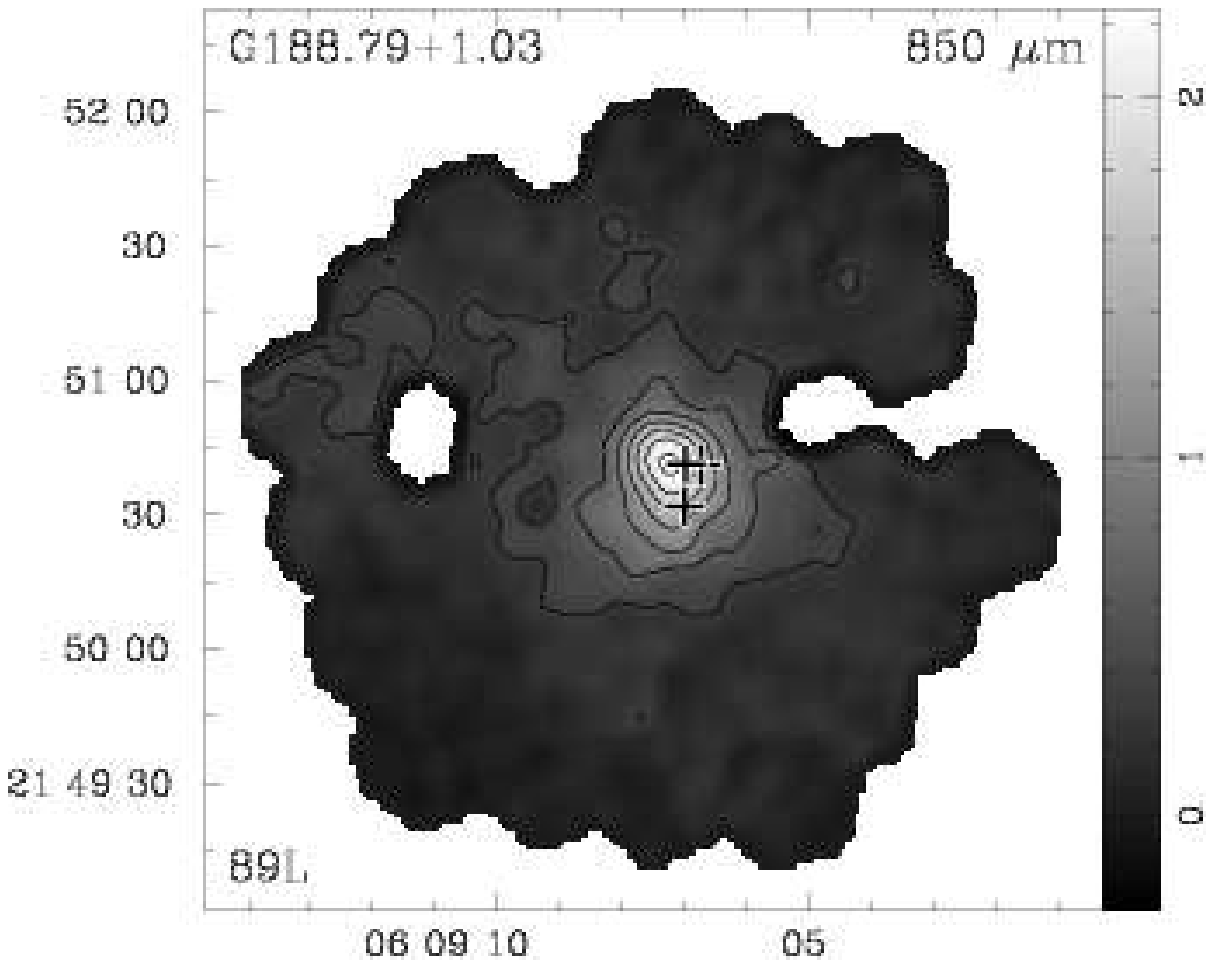}
 \hspace*{0.11cm}\includegraphics[scale=0.35,angle=-90]{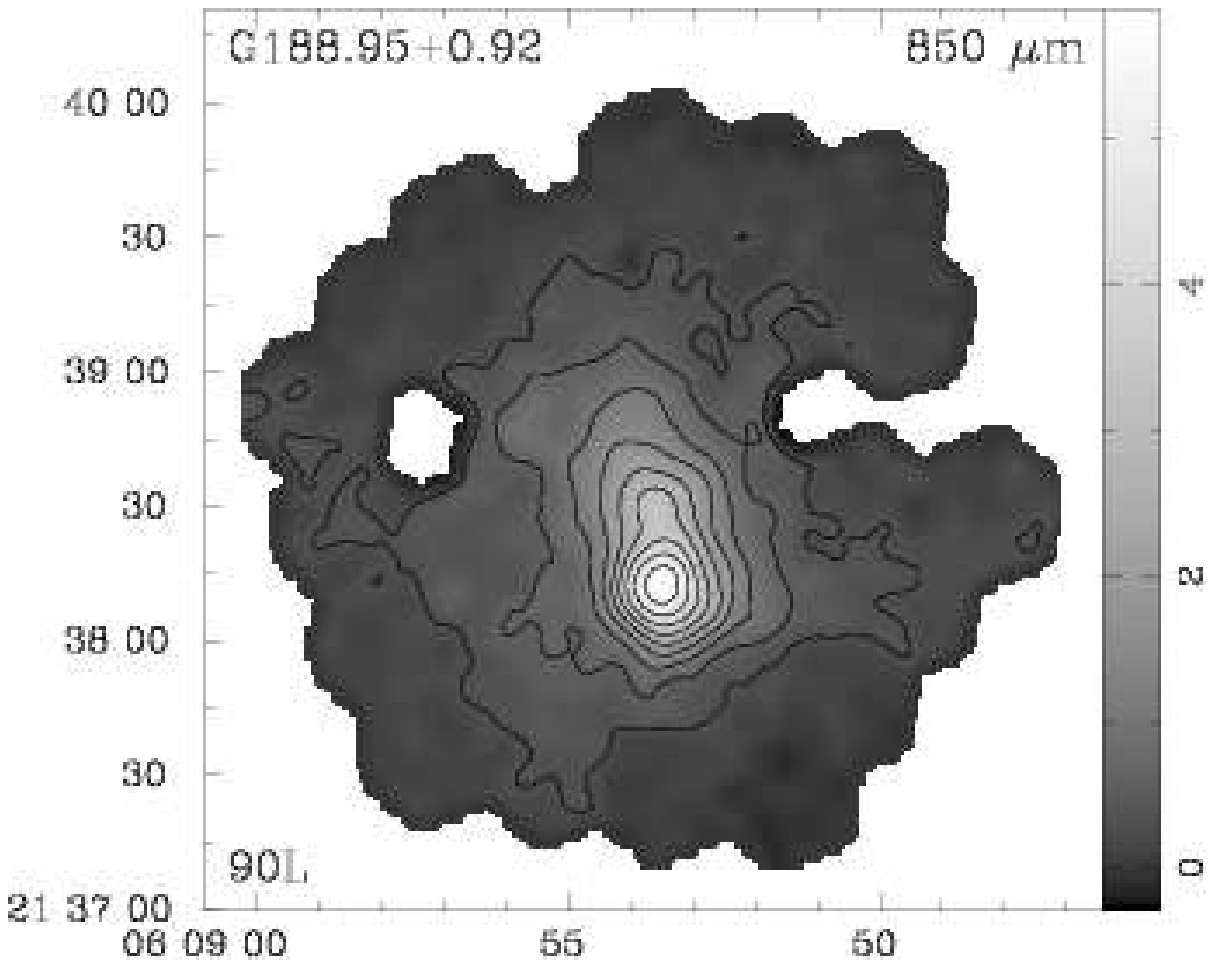}
 \includegraphics[scale=0.35,angle=-90]{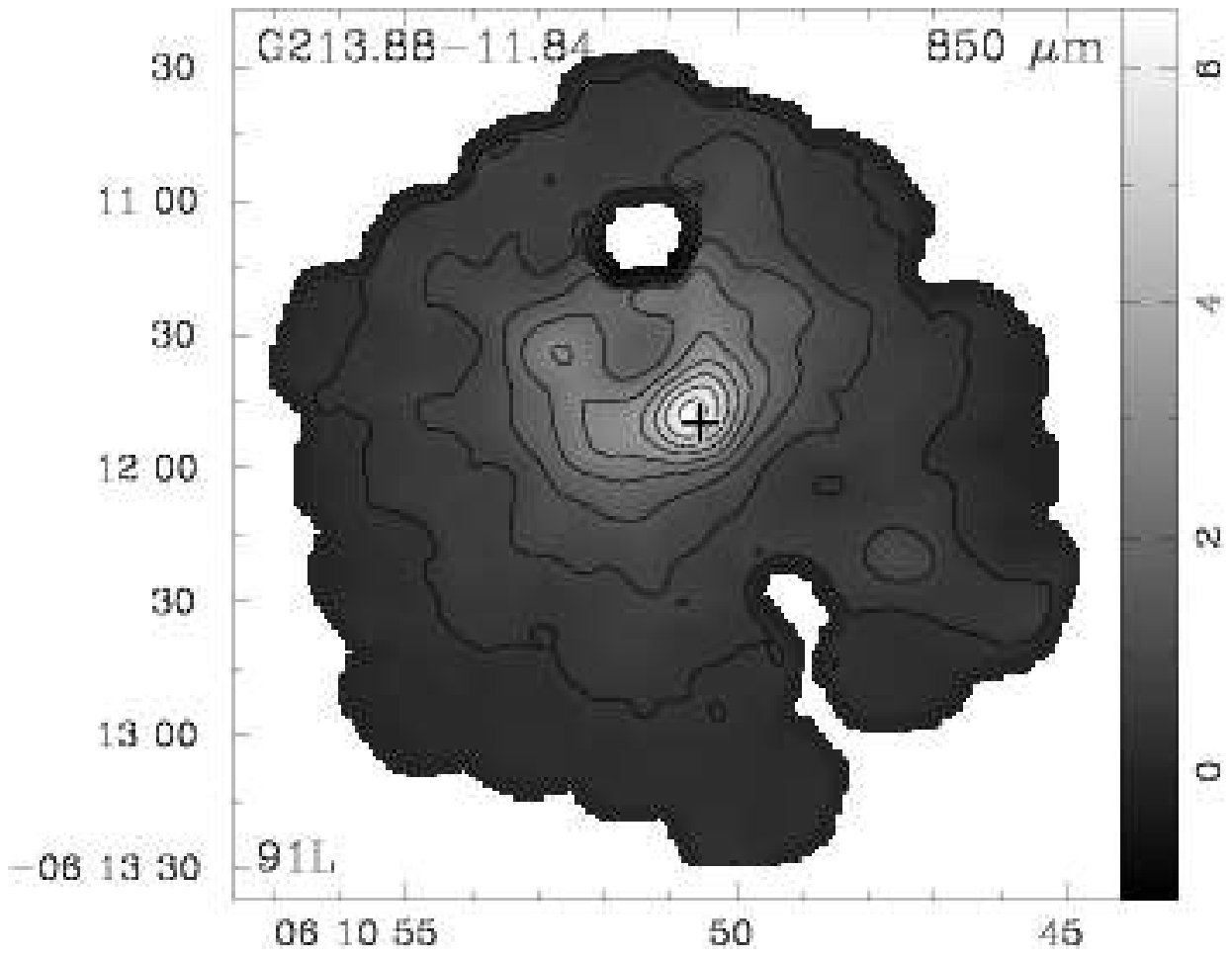}\\
 
 \setcounter{figure}{1}
 
 \caption{SCUBA images of the ultracompact H{\sc ii} regions in the survey with
 detections at 850 $\mu$m only. Each UC H{\sc ii} region is represented by a
 pair of images at 450 $\mu$m (\emph{left image}) and 850$\mu$m (\emph{right
 image}). Coordinates are Right Ascension and Declination in the J2000 system. 
 Crosses indicate the positions of ultracompact H{\sc ii} regions from Wood \&
 Churchwell (\cite{wc89a}), Kurtz, Churchwell \& Wood (\cite{kcw94}), Becker
 et al.~(\cite{b94}) or Giveon et al.~(\cite{gbh2005}). All images have been
 deconvolved with a model of the JCMT beam to remove the contribution from the
 error lobe.} 
 
  \label{fig:scu850}
 \end{minipage}
 \end{figure*}

The calibrated images were then converted into FITS format and deconvolved to remove the
contribution from the error beam. The deconvolution was performed using the \emph{clean}
task in MIRIAD (Sault, Teuben \& Wright \cite{miriad})  with a circularly symmetric
two-component Gaussian beam derived from azimuthal averages of the primary calibrator maps
in a manner similar to that described in Hogerheijde \& Sandell (\cite{hs00}).  Each image
was cleaned down to a cutoff level of twice the 1$\sigma$ r.m.s.~noise and then restored
back to a resolution appropriate for the wavelength (8\arcsec~for 450 $\mu$m and
14\arcsec~for 850 $\mu$m). The advantage of this technique is that the different error
beam contributions from each wavelength are removed, facilitating comparison of 450 $\mu$m
and 850 $\mu$m maps and allowing the integrated fluxes to be determined more accurately.

After deconvolution to remove the error beam the 450 $\mu$m images were inspected and those displaying
insufficient signal-to-noise to permit accurate source identification were smoothed from a native
resolution of 8\arcsec\ to the same resolution as that in the 850 $\mu$m images (14\arcsec).
Sacrificing the angular resolution of the 450 $\mu$m images  results in an improvement of the
signal-to-noise ratio for extended emission by approximately a factor of 3, which considerably aided
in the identification of faint 450 $\mu$m sources.


\section{Results}
\label{sect:results}

\subsection{Deconvolved images}
\label{sect:images}

We present the final deconvolved SCUBA images in Figs.~\ref{fig:scuboth} and \ref{fig:scu850},
marked  with the positions of the UC H{\sc ii} regions in each image. Fig.~\ref{fig:scuboth} contains
images of sources that were detected at 450 and 850 $\mu$m, whereas Fig.~\ref{fig:scu850} contains
images of those sources with data only at 850 $\mu$m, either because the 450 $\mu$m emission was
undetected or due to the 450 $\mu$m calibration problems discussed in the previous section. Only
those SCUBA fields with detected sub-mm sources are included in the figures. Sub-mm emission
was detected toward $\sim$ 90\% of the sample (92 out of 105 fields).  Here we outline a few
attributes of the SCUBA images in order to aid in their interpretation. 

``Holes'' in the 850 $\mu$m images result from the blanking of noisy bolometers in the data reduction
process. Typically between 1 and 3 bolometers were blanked from each 850 $\mu$m image and any flux from
sources found near the blanked image regions may be underestimated. Individual sources whose fluxes may
be affected are indicated in Table \ref{tbl:uchiis}. Noisy bolometers were also blanked from the 450
$\mu$m data, but as the short wavelength array in SCUBA has a higher spatial sampling, the blanking of
bolometers from the short-wavelength array does not cause gaps in the spatial coverage of the array.
Negative ``bowls'' of emission are also present in some images (e.g.~G13.19+0.04 in panel
73L\footnote{Note that the panel numbers  refer to individual images in  Figs.~\ref{fig:scuboth} and
\ref{fig:scu850}. L and S refer to the long (850 $\mu$m) and short (450 $\mu$m) wavelength channels of
SCUBA.}, and G25.38$-$0.18 in panel 75L) and these bowls are
caused by the presence of sub-mm emission in the chopping positions (i.e.~approximately 120\arcsec\ in
azimuth from the field centre). The morphology of a number of these bowls indicates that the emission is
compact and resembles that of the positive emission features seen in the sub-mm image. These compact
negative bowls may indicate the presence of other sub-mm cores located near the chopping positions,
possibly similar to those associated with the UC H{\sc ii} regions in the main image.

The flux contour levels on each image were determined by a dynamic range power-law fitting scheme
in order to emphasise both the low-level and bright emission. Logarithmic contour levels were
fitted to the dynamic range of the image $D$ (defined as the image peak divided by the 1$\sigma$
r.m.s.~noise) following the relation $D = 3 \times N^{i} +3$, where $N$ is the number of contours
(in this case 9) and $i$ is the contour power-law index. The minimum power-law index used was one,
which results in linear contours starting at a level of 3$\sigma$ and spaced by 3$\sigma$. This
dynamic power-law contouring scheme was found to give excellent results in both high and low
dynamic range images, concentrating the contours around low surface brightness features in the
images to adequately represent low-level structure in the images.

The SCUBA images show that the sub-mm emission associated with UC H{\sc ii} regions follows a range of
morphologies: from single, bright, centrally peaked compact cores (e.g.~G20.08$-$0.14 in panel 22L,
G35.02+0.35 in panel 46L); to more complex multiply-peaked regions with peaks both associated and
unassociated with known UC H{\sc ii} regions (e.g.~G28.29$-$0.36 in panel 36L); and ridge-shaped
structures with multiple condensations strung along the ridge (e.g. G27.28+0.15 in panel 33L,
G79.30+0.28 in panel 59L). Often the sub-mm cores that are \emph{not} associated with the UC H{\sc ii}
regions in an image are of comparable brightness or sometimes even brighter than the neighbouring
cores that are (e.g.~G8.14+0.23 in panel 4L, G25.72+0.05 in panel 31L).
The SCUBA images of the non-UC H{\sc ii} associated IRAS point sources (those indicated by a $^{\dag}$ in
Table \ref{tbl:samplelist}) also  reveal the presence of bright radio-quiet sub-mm cores similar in
characteristics and appearance to the cores that are known to be associated with UC H{\sc ii} regions
(e.g.~G13.19+0.04 in panel 73L). If these radio-quiet sub-mm cores are forming massive stars in a
similar manner to their radio-loud UC H{\sc ii}-associated neighbours, it is possible that the
radio-quiet cores represent an earlier evolutionary phase of star formation prior to the
development of an UC H{\sc ii} region as suggested by Garay et al.~(\cite{garay2004}). We will dwell
further on this hypothesis in Sect.~\ref{sect:analysis}.

\subsection{Source identification}
\label{subsect:sourceid}

Individual sources within the 450 and 850 $\mu$m images were identified using the source extraction
algorithm SEXTRACTOR (Bertin \& Arnouts \cite{ba96}). Whilst SEXTRACTOR was initially developed as
a source extraction routine for visible and infrared-wavelength images, it nevertheless provides
the means of separating the  often closely blended sub-mm cores in the SCUBA images and measuring
their individual fluxes in a consistent fashion. As SEXTRACTOR treats each pixel of the target
image in a statistically independent way, whereas each of our sub-mm images is gridded in pixels
smaller than the beam FWHM (and are thus not statistically independent), it was necessary to
account for the beam area in the final measured fluxes and their uncertainties. We ran SEXTRACTOR
on each of our images, setting the analysis threshold to be three times the r.m.s.~noise in each
image. The r.m.s.~noise in each image was determined by hand from the standard deviation of
source-free pixels, as it was found that SEXTRACTOR could not make an accurate automatic
determination of this quantity in the relatively small and crowded SCUBA images (as compared to
typical visible-wavelength data). 

A combined source catalogue was compiled from the individual SEXTRACTOR runs on each SCUBA image.
Beam-corrected 450 and 850 $\mu$m fluxes were determined for each source in the catalogue (using
the FLUX\_AUTO keyword) and were checked against a random sample of fluxes determined by manual
aperture photometry. In all cases the fluxes determined by SEXTRACTOR were consistent with those
measured by manual aperture photometry. We also compared the measured fluxes of those sources from
our catalogue that have been published in the literature. We share 11 fields with the methanol
maser survey of Walsh et al.~(\cite{walsh03}) and in all but two cases the integrated and peak
fluxes are consistent to within the uncertainties. An inspection of the two images (G15.04$-$0.68
\& G28.29$-$0.36) where the flux measurements are inconsistent reveals that the likely cause is due
to the confused nature of these regions and the  different chop positions used by ourselves and
Walsh et al.~(\cite{walsh03}). Our measured fluxes of the sub-mm cores associated with 
G76.38$-$0.62 , G25.65+1.05 and G106.80+5.31 (also known as S 106, RAFGL 7009S and S 140
respectively)  also agree very closely with the sub-mm measurements of Richer et
al.~(\cite{rpw93}), McCutcheon et al.~(\cite{msp95}) and Minchin, Ward-Thompson \& White
(\cite{mww95}), taking into account the slightly different filter bandwidths used in these
observations. 

The combined source catalogue with the centroid position of each sub-mm source, its peak flux
and integrated flux at both 450 \& 850 $\mu$m are given in Table \ref{tbl:uchiis}.  The
positions quoted for each source are usually taken from the 850 $\mu$m images (unless otherwise
stated) as the the 850 $\mu$m images have a considerably higher signal-to-noise ratio. Where it
was possible to deblend sources at 450 $\mu$m  but not at 850 $\mu$m images the centroid
positions are those measured from the 450 $\mu$m images. The errors quoted in Table
\ref{tbl:uchiis} include the absolute calibration errors of 30\% at 450 $\mu$m and 10\% at 850
$\mu$m combined in quadrature with the intrinsic measurement error derived from the image
r.m.s.~noise.

\begin{table}
\setcounter{table}{2}
\caption{SCUBA fields from the survey with  no detected sub-mm emission (to a level of
3$\sigma$). Fields indicated with a dagger ($^{\dag}$) show no evidence for UC H{\sc ii} regions in the
WC89a and KCW94 surveys, or in the literature search described in Sect.~\ref{subsect:sample}.}
\label{tbl:uchii_nondets}
\begin{tabular}{llc}\hline\hline
 UC H{\sc ii} & IRAS PSC& 850 $\mu$m flux upper limit \\ 
 field name & identifier &(Jy/beam) \\\hline
G24.39+0.07$^{\dag}$ & 18328-0735 & 0.3 \\
G26.10$-$0.07 & 18365-0609 & 0.4 \\
G32.96+0.28$^{\dag}$ & 18481-0014 & 0.2 \\
G37.36$-$0.23 & 18578+0345 & 0.4 \\
G41.13$-$0.32$^{\dag}$ & 19051+0704 & 0.2 \\
G41.52+0.04 & 19046+0734 & 0.2 \\
G41.71+0.11 & 19048+0748 & 0.2 \\
G42.90+0.57 & 19054+0901 & 0.1 \\
G43.26$-$0.18$^{\dag}$ & 19087+0900 & 0.1 \\
G43.82+0.39$^{\dag}$ & 19077+0946 & 0.2 \\
G44.26+0.10 & 19096+1001 & 0.1 \\
G45.20+0.74$^{\dag}$ & 19090+1109 & 0.2 \\
G45.40$-$0.72$^{\dag}$ & 19147+1040 & 0.3 \\
G188.77+1.07 & 06061+2151 & 0.2 \\ \hline
\end{tabular}
\end{table}

Out of the total 105 SCUBA fields observed as part of the survey we have identified 155 sub-mm
sources at either 450 or 850 $\mu$m. Approximately 10\% of the fields observed (13 in total) did
not display any sub-mm emission down to a level of 3$\sigma$. Approximately half of the
non-detected SCUBA fields (7/13) are associated with ultracompact centimetre-wave radio
components. As these objects do not display significant sub-mm emission, which indicates a low
column density of dust, it is unlikely that the centimetre-wave radio emission traces embedded
UC H{\sc ii} regions unless the distance to these objects is large. We will discuss these objects in
more detail in Sect.~\ref{sect:analysis}. The non-detected SCUBA fields and their upper 850 $\mu$m flux
limits are are indicated in Table \ref{tbl:uchii_nondets}.

\subsection{Notes on individual sources}

In this section we describe selected fields from the survey in further detail to aid in the
interpretation of their sub-mm emission. We briefly describe their morphology in the sub-mm,
their associated UC H{\sc ii} regions (or lack of them) and whether the fields are better known by
another name than their galactic coordinates. 

\paragraph{G1.13$-$0.11: panels 1S \& 1L} This source is also known as \object{Sgr D} (Mehringer et al.~\cite{mgl98}) and
is a cluster of compact H{\sc ii} regions mostly less than 10\arcsec\ in diameter. The cluster is
surrounded by a low-intensity extended cm-wave continuum component and is   dominated by a
single compact source that is consistent with excitation by an O5.5 star  (Mehringer et
al.~\cite{mgl98}). We identify one sub-mm clump associated with G1.13$-$0.11, which takes the
form of an extended ridge running  roughly NE-SW across the SCUBA FOV. This morphology is
consistent with the 350 $\mu$m map of this source obtained by Hunter et al.~(\cite{hcwcb00}).
The central region of the ridge possesses two small peaks which may indicate the presence of
more than one sub-mm clump, however we are unable to separate the emission into two
discrete cores.  The dominating  compact H{\sc ii} region is located close to the central peak,
suggesting that it is embedded within the dust ridge. The presence of H$_{2}$CO absorption lines
at the molecular cloud velocity  towards this compact H{\sc ii} region strengthens this hypothesis
(Mehringer et al.~\cite{mgl98}). The large-scale radio and sub-mm morphology of this region
suggest that the extended cm-wave  emission represents a blister H{\sc ii} region. Because G1.13$-$0.11
is located so near to the Galactic Centre on the sky it is difficult to estimate an accurate
distance. Following Mehringer et al.~(\cite{mgl98}) we assume that the distance to G1.13$-$0.11
is the same as that to the Galactic Centre (8.5 kpc), although we caution that the true distance
is likely to be greater than this estimate (see Mehringer et al.~\cite{mgl98} for a more detailed
argument).

\paragraph{G5.89$-$0.39: panels 2S \& 2L} We identify one sub-mm core in this region (G5.887$-$0.394SMM) which is
also the well-known outflow source \object{OH 5.89$-$0.39} (Zijlstra et al.~\cite{zpe90}) or \object{W28 A2} (Harvey
\& Forveille \cite{hf88}). G5.89$-$0.39 is a well-known luminous massive star-forming region,
displaying compact cm-wave emission from an UC H{\sc ii} region (WC89a), H$_{2}$O, OH and CH$_{3}$OH
maser emission (Hofner \& Churchwell \cite{hc96}; Palagi et al.~\cite{pcc93}; Argon, Reid \&
Menten \cite{arm00}; Val'tts et al.~\cite{ves2000}) and thermal emission from a variety of
molecular tracers (e.g.~Fontani et al.~\cite{fcco02}; Thompson et al.~\cite{tm99}; Gomez et
al.~\cite{grgm91}). G5.89$-$0.39 has also been observed comprehensively in the infrared and
millimetre-wave continuum (Feldt et al.~\cite{fsh99}) and in 350 $\mu$m continuum emission 
(Hunter et al.~\cite{hcwcb00}; Mueller et al.~(\cite{msej02}). The morphologies of the millimetre
and 350 $\mu$m maps agree well with our 450 \& 850 $\mu$m maps. The sub-mm core appears to
possess a clumpy substructure, with the UC H{\sc ii} region slightly offset from the peak of the sub-mm
emission. This may indicate that the sub-mm core contains multiple sub-components, which is
consistent with the scenario proposed by Feldt et al.~(\cite{fsh99}). For further discussion of
the structure of the sub-mm clump associated with G5.89$-$0.39 the reader is referred to the
models of Hatchell \& van der Tak (\cite{hvdt03}) and Feldt et al.~(\cite{fsh99}).

\paragraph{G5.97$-$1.17: panels 3S \& 3L} This object is also the well-known Hourglass region of the Lagoon Nebula
and is home to the O7 star Herschel 36. The Lagoon Nebula has been extensively investigated by
Tothill et al.~(\cite{twmmmk02}) and the interested reader is referred to this paper and its
companion (White et al.~\cite{wtm97}) for an overview of the star formation within the Lagoon
Nebula. We identify two sub-mm clumps in our SCUBA images: G5.972$-$1.159SMM (which corresponds
to clump WC1 in Tothill et al.~\cite{twmmmk02}) and G5.974$-$1.175SMM (clump
HG in  Tothill et al.~\cite{twmmmk02}). We note that the sub-mm fluxes measured in this paper are
roughly a factor of 2 greater than those measured by Tothill et al.~(\cite{twmmmk02}). This is
more than likely due to the different procedures used to measure the sub-mm fluxes: Tothill et
al.~use a Gaussian fitting procedure to extract the clumps from local background emission, whereas our
aperture photometry is likely to include this extended component. 

\paragraph{G6.55$-$0.10: panel 70L} This object is also known as \object{W28 A1} and exhibits maser emission from both
H$_{2}$O and CH$_{3}$OH masers (Churchwell, Walmsley \& Cesaroni \cite{cwc90}; Walsh et al.~\cite{wbh98}), plus
thermal NH$_{3}$ emission (Churchwell, Walmsley \& Cesaroni \cite{cwc90}). Kim \& Koo (\cite{kk01}) show that the
WC89a UC H{\sc ii} region represents the peak of a compact H{\sc ii} region approximately 30\arcsec\ in diameter.
The sub-mm images reveal a single diffuse core in the centre of the field, which is positionally associated with
both the compact H{\sc ii} region and the H$_{2}$O \& CH$_{3}$OH masers. 

\paragraph{G8.14+0.23: panels 4S \& 4L} The sub-mm emission in this fields follows a ridge-like morphology, with
four identified sub-mm clumps (G8.143+0.219SMM, G8.140+0.224SMM, G8.135+0.245SMM,
G8.136+0.212SMM). This object has also been mapped by Walsh et al.~(\cite{walsh03}) and   our
450 \& 850 $\mu$m images agree closely with those of Walsh et al. Kim \& Koo (\cite{kk03})
present wide-field maps of G8.14+0.23 in the $^{13}$CO J=1--0 line, which show a ridge of
molecular emission in close correspondence with the ridge seen in our SCUBA images. The
star-formation activity in this region appears restricted to the clump G8.140+0.224SMM, which is
positionally associated with two UC H{\sc ii} regions (WC89a; Becker et al.~\cite{b94}) and an H$_{2}$O
maser (Palagi et al.~\cite{pcc93}). There is also a CH$_{3}$OH maser detected toward this region
(Walsh et al.~\cite{wbh98}), which is located at the Western edge of clump G8.143+0.219SMM.

\paragraph{G8.67$-$0.36: panels 5S \& 5L} There are two sub-mm clumps identified in the SCUBA images 
(G8.683$-$0.368SMM \& G8.671$-$0.357SMM). Both clumps are associated with CH$_{3}$OH masers
(Caswell et al.~\cite{caswell95}), whereas only the brightest sub-mm clump G8.671$-$0.357SMM is
associated with an H$_{2}$O maser (Palagi et al.~\cite{pcc93}) and an UC H{\sc ii} region (WC89a). 
Mueller et al.~(\cite{msej02}) detected only a single clump in their 350 $\mu$m map of
G8.67$-$0.36, as the edge of their mapped field truncates G8.683$-$0.368SMM.

\paragraph{G9.62+0.20: panels 6S \& 6L} This object is the well studied massive star-forming region
\object{G9.62+0.20},
which is host to several UC H{\sc ii} regions, CH$_{3}$OH \& H$_{2}$O masers and a hot molecular core (Persi et
al.~\cite{ptr03}; Testi et al.~\cite{thkr00}; Hofner et al.~\cite{hkcwc96}). We detect a single
sub-mm clump in this field that appears to be elongated along the NW-SE axis, in agreement with
the SCUBA map of Walsh et al.~\cite{walsh03}) and the 350 $\mu$m maps of Hunter et
al.~(\cite{hcwcb00}) \& Mueller et al.~(\cite{msej02}). For a wider overview of the
star-formation in this region the reader is directed to the excellent summaries by Hofner et
al.~(\cite{hkcwc96}) and Persi et al.~(\cite{ptr03}).

\paragraph{G10.15$-$0.34/G10.30$-$0.15: panels 8S \& 8L and 9S \&9L} These two complexes of UC H{\sc ii} regions identified  by
WC89a lie in the \object{W31} Giant Molecular Cloud. Kim \& Koo (\cite{kk01}) present wide-field 20 cm
maps of both these regions which reveal that the WC89a UC H{\sc ii} regions are associated with
extended radio emission (a shell morphology in the case of G10.15$-$0.34 and a bipolar H{\sc ii}
region morphology for G10.30$-$0.15). The sub-mm ridge observed in G10.30$-$0.15 aligns well with
the centre of the bipolar H{\sc ii} region observed by Kim \& Koo (\cite{kk01}) and the elongated CS
emssion mapped by Kim \& Koo (\cite{kk02}). This suggests that the stars exciting the H{\sc ii} region
are still embedded in a dense flattened molecular clump seen edge-on, but their UV photons are
escaping from the top and bottom of the clump to ionise the surrounding material (as in, for
example, S 106; Richer et al.~\cite{rpw93}). G10.30$-$0.15 is also associated with CH$_{3}$OH
maser emission (Walsh et al.~\cite{wbh98}; Goedhart, van der Walt \& Schutte \cite{gvs00}). 

\paragraph{G10.84$-$2.59: panels 11S \& 11L} This source is also known as \object{GGD 27}, which comprises  an infrared
reflection nebula (Yamashita et al.~\cite{ysn87}), a cluster of potentially massive YSOs (Aspin
\& Geballe \cite{ag92}; Bica, Dutra \& Barbuy \cite{bdb2003}), the HH 80--81 thermal radio jet
(Marti, Rodriguez \& Reipurth \cite{mrr93}) an H$_{2}$O maser (G\'omez, Rodriguez \& Marti
\cite{grm95}) and a CH$_{3}$OH maser (Val'tts et al.~\cite{ves2000}). G\'omez et al.~(\cite{gomez2003})
present high resolution NH$_{3}$ and HCN images of the GGD 27 region, showing that the peak of the
molecular emission is closely associated with the H$_{2}$O maser. Although at the limit of our SCUBA
positional accuracy, we note that the sub-mm peak identified in this field (G10.842-2.594SMM) is
displaced by $\sim$6\arcsec\ from that of the H$_{2}$O maser. It is likely that the UC
H{\sc ii} region identified by KCW94 is actually the central source of the  HH 80--81 thermal jet. The
luminosity of the UC H{\sc ii} region suggests that a B0-B1 star is responsible for its excitation
(KCW94).

\paragraph{G11.94$-$0.62: panels 13S \& 13L} Two sub-mm sources are identified in this field (G11.936$-$0.618SMM,
G11.918$-$0.615SMM). The southernmost source (G11.918$-$0.615SMM) is associated with an H$_{2}$O
maser (Hofner \& Churchwell \cite{hc96}) whereas the brighter central source
(G11.936$-$0.618SMM) is associated with an UC H{\sc ii} region (WC89a), H$_{2}$O (Hofner \& Churchwell
\cite{hc96}) and CH$_{3}$OH masers (Val'tts et al.~\cite{ves2000}). No cm-wave radio emission is
found to be associated with the southernmost sub-mm clump G11.918$-$0.615SMM. Given the presence
of an H$_{2}$O maser in this core the lack of cm-wave radio continuum emission suggets that
G11.918$-$0.615SMM may be at an earlier phase of massive star formation than
G11.936$-$0.618SMM.  These two sub-mm sources were also observed by Walsh et
al.~(\cite{walsh03}).

\paragraph{G12.90$-$0.25: panels 14S \& 14L} This source is the well-studied star-forming region \object{W33A} (e.g.~Jaffe
et al.~\cite{jhk84}). W33A is a deeply embedded luminous infrared star-forming region exhibiting
several molecular absorption lines (Grim et al.~\cite{gbg91}; Palumbo, Tielens \& Tokunaga
\cite{ptt95}). We have identified two sub-mm clumps within our SCUBA images of which the
brightest (G12.908$-$0.261SMM) is associated with W33A. Numerous masers are associated with this
sub-mm clump; OH (Argon, Reid \& Menten \cite{arm00}), H$_{2}$O (Braz \& Epchtein \cite{be83})
and CH$_{3}$OH (Szymczak, Hrynek \& Kus \cite{shk2000}; Caswell et al.~\cite{caswell95}). Van
der Tak et al.~(\cite{vve2000}) have also mapped the W33A clump in millimetre-wave continuum  at
high angular resolution, which reveals two embedded compact millimetre cores. One of these cores
(W33A MM 1) is associated with a weak centimetre continuum  source that is consistent with an
ionised stellar wind (Rengarajan \& Ho \cite{rh96}). In contrast the remaining sub-mm clump in
this field (G12.908$-$0.261SMM) displays no known cm-wave continuum or maser emission. This
clump was also undetected by either Walsh et al.~(\cite{walsh03}) using SCUBA or at 350 $\mu$m by
van der Tak et al.~(\cite{vve2000}), as it lies outside both their mapped regions.

\paragraph{G13.19+0.04: panel 73L} This field potentially contains three or more sub-mm components; a
bright clump to the west (G13.177+0.059SMM), a fainter source to the east (G13.186+0.028SMM) and
a bright ``ridge'' at the northern edge of the SCUBA FOV. This latter emission is not classified
as a source in the catalogue due to its truncation by the edge of the SCUBA FOV, which means
that the centroid and flux of this source cannot be constrained. Whilst the field contains two
compact radio components from the Becker et al.~(\cite{b94}) 5 GHz survey, neither are
positionally associated with these sub-mm clumps. Braz \& Epchtein (\cite{be83}) and Szymczak,
Hrynek \& Kus (\cite{shk2000}) report the detection of an H$_{2}$O and CH$_{3}$OH maser
respectively toward this region, although the large positional accuracy in their single-dish
observations means that the masers cannot be associated with individual sub-mm clumps.

\paragraph{G15.04$-$0.68: panels 16S \& 16L} This region is the well known nebula \object{M17} (also known as the Omega
Nebula), an H{\sc ii} region/molecular cloud complex containing a number of early O stars (Nielbock et
al.~\cite{ncjm01}; Hanson, Howarth \& Conti \cite{hhc97}). The strongly negative emission
features seen in our SCUBA maps (and those of Walsh et al.~\cite{walsh03}) imply that the
chopping positions were contaminated by sub-mm emission and thus the fluxes measured in this
SCUBA field are strictly lower limits to the true flux. The H{\sc ii} region in M17 appears to be
viewed as an edge-on blister to the molecular cloud complex, which is in turn comprised of two
molecular clouds A \& B (Wilson, Hanson \& Muders \cite{whm03}). The UC H{\sc ii} region identified by
WC89a lies in the region M17SW found in cloud B, which is also host to a number of H$_{2}$O
masers (Braz \& Epchtein \cite{be83}).

\paragraph{G19.61$-$0.23: panels 21S \& 21L} This field contains a single centrally peaked sub-mm core, which is
known to harbour three embedded compact (3--8\arcsec\ FWHM) ammonia cores and a complex of UC
H{\sc ii} regions (Garay et al.~\cite{gmr98}).

\paragraph{G20.99+0.09: panels 23S \& 23L} We identify two sub-mm clumps in this field, G20.990+0.089SMM and
G20.981+0.094SMM, which comprise a tadpole-like morphology at 850 $\mu$m  with a brighter head
(G20.981+0.094SMM) and fainter tail.(G20.990+0.089SMM). The head of the tadpole is unfortunately
affected by blanked bolometers and the flux of this component is thus a lower limit to the true
flux. This field was a non-detection in the WC89a survey, but Becker et al.~(\cite{b94})
detected three ultracompact radio components at 5 GHz which are found to be mainly associated
with the fainter tail of the tadpole. The tail was not detected at 450 $\mu$m.

\paragraph{G23.46$-$0.20: panels 24S \& 24L} This field comprises a faint central sub-mm source,
G23.454$-$0.203SMM, and also brighter emission to the SW that is truncated by the edge of the
field. G23.454$-$0.203SMM is  positionally coincident with the UC H{\sc ii} region identified by
WC89a. The remaining  bright  source is not included in our catalogue, in view of the
impossibility of determining a sensible centroid or flux of this source. Walsh et
al.~(\cite{walsh03}) observed the 450 \& 850 $\mu$m emission from this source (identified in
their study as G23.44$-$0.18), as it is known to contain two CH$_{3}$OH masers (Walsh et
al.~\cite{wbh98}). Wider-field $^{13}$CO maps of this region reveal the presence of a further
molecular clump to the south of G23.44$-$0.18 (Kim \& Koo \cite{kk03}).

\paragraph{G23.87$-$0.12: panels 26S \& 26L} Four sub-mm clumps are identified in this region, of which only one
(G23.861$-$0.125SMM) is associated with an UC H{\sc ii} region. A further ultracompact 5 GHz  radio
source was detected in this field by Becker et al.~(\cite{b94}), which lies mid-way  between the
two sources G23.872$-$0.122SMM \& G23.866$-$0.114SMM  and is positionally associated with
neither.

\paragraph{G23.96+0.15: panels 27S \& 27L} We have identified four sub-mm clumps in this field  (G23.955+0.148SMM,
G23.952+0.152SMM, G23.944+0.159SMM, G23.949+0.163SMM). The first two of these clumps form a
bright peanut-shaped complex in the centre of the image, whilst the latter two are truncated by
two blanked bolometers in the final image and may in fact be a single object. This field has also
been mapped at 350 $\mu$m by Hunter et al.~(\cite{hcwcb00}), who found a very similar
morphology. The two bright clumps in the peanut-shaped complex are designated IRAS 18917$-$0757
SMM1 and SMM2 by Hunter et al., which corresponds to our SCUBA sources G23.955+0.148SMM and 
G23.952+0.152SMM respectively. G23.955+0.148SMM is known to be associated with an UC H{\sc ii} region
from the WC89a catalogue. Palagi et al.~(\cite{pcc93}) report the presence of an H$_{2}$O maser
associated with this region, but without confirmation at higher angular resolution it is not
possible to associate this single-dish maser detection to an individual sub-mm clump. 

\paragraph{G25.38$-$0.18: panel 75L} We identify one sub-mm clump in this field, although the two negative
``bowls'' seen in the image suggest the presence of other nearby clumps in the chopping
positions. G25.38$-$0.18 is also known as the \object{W42} giant H{\sc ii} region and has recently been found to
harbour a massive stellar cluster containing a number of YSOs and at least one O5.5 star (Blum,
Conti \& Damineli \cite{bcd00}). The position of this O5.5 star (W42 1) is close to the peak of
the sub-mm clump G25.382$-$0.184SMM identified in our 850 $\mu$m image. As the K-band spectra of
W42 1 (Blum, Conti \& Damineli \cite{bcd00}) show that this star has largely cleared away its
birth cocoon, this suggests that W42 1 lies in the foreground of G25.382$-$0.184SMM.

\paragraph{G25.65+1.05: panels 30S \& 30L} This object is also known as \object{RAFGL 7009S} and is a well-known embedded
massive YSO associated with an UC H{\sc ii} region and displaying strong mid-infrared ice absorption
features and a powerful molecular outflow (Zavagno et al.~\cite{zdnc02}; Dartois, Gerin \&
d'Hendecourt \cite{dgh2000}; Shepherd \& Churchwell \cite{sc96}; McCutcheon et
al.~\cite{msp95}). We identify a single strongly peaked  sub-mm clump in the SCUBA 450 \& 850
$\mu$m images, positionally coincident with the UC H{\sc ii} region of KCW94 and with flux levels
consistent with those quoted in McCutcheon et al.~(\cite{msp95}), allowing for the different
filter bandwidths used.

\paragraph{G27.28+0.15: panels 33S \& 33L} The sub-mm emission from this region takes the form of an elongated
filament with 4 associated sub-mm clumps (G27.267+0.146SMM, G27.280+0.144SMM, G27.296+0.151SMM
and G27.285+0.149SMM). Only one clump, G27.280+0.144SMM, is positionally associated with an UC
H{\sc ii} region from the WC89a survey. This clump is also associated with a supernova remnant  (\object{SNR
027.3+00.1}) that shows OH maser emission (Green et al.~\cite{gfg97}). CH$_{3}$OH maser emission
is also detected toward this region by Szycmzak et al.~\cite{shk2000}) but with the large
positional uncertainties involved in the single-dish detection it is not possible to localise
the maser emission within the field. Kim \& Koo (\cite{kk01}, \cite{kk03}) present wider-field
maps of this region at 21 cm and in the $^{13}$CO J=1--0 line respectively. The 21cm radio
continuum reveals an extended region of emission elongated to the NW, which in all likelihood
represents the SNR. The $^{13}$CO map shows an elongated filament of molecular emission extended
north-south, consistent with our SCUBA images.

\paragraph{G29.96$-$0.02: panels 37S \& 37L} We detected four sub-mm clumps in the SCUBA images of this object,
found approximately along a north-south axis. The brightest of these clumps, G29.956$-$0.017SMM,
is associated with the well-known UC H{\sc ii} region and massive star-forming complex G29.96$-$0.02.
This UC H{\sc ii} region has been extensively studied and was the first ultracompact H{\sc ii} region to
have its exciting star spectroscopically identified (Watson \& Hanson \cite{wh97}). The UC H{\sc ii}
region has a cometary morphology (WC89a), with a dense, hot molecular core located at the head
of the ``comet''. This hot molecular core has large abundances of NH$_{3}$ and other saturated
molecules (Hatchell et al.~\cite{htmm98}; Cesaroni et al.~\cite{cch94}), is positionally
associated with H$_{2}$O masers (Hofner \& Churchwell \cite{hc96}), a compact mid-infrared source (De Buizer et
al.~\cite{dbw02}) and a possible outflow (Pratap et al.~\cite{pmb99}). There is thus strong
evidence that this sub-mm clump  is an active massive star-forming cluster. There is, so far, no
evidence that massive star formation tracers are associated with the other sub-mm clumps in the
SCUBA image. 

\paragraph{G30.78$-$0.02: panels 39S \& 39L} This region is also known as the H{\sc ii} region/molecular cloud complex
\object{W43}, which has recently been extensively mapped by Motte, Schilke \& Lis (\cite{msl03}) at 1.3
mm and 350 $\mu$m. The two sub-mm sources that we detect towards this region, G30.785$-$0.022SMM
and  G30.786$-$0.028SMM, correspond to the two cores W43 MM6 and W43 MM8 respectively. For a
thorough overview of the star formation within W43 the interested reader is referred to Motte,
Schilke \& Lis (\cite{msl03}).

\paragraph{G33.92+0.11: panels 44S \& 44L} Although we have identified only a single sub-mm clump in this region
(G33.914+0.109SMM), it is clear from the 450 $\mu$m \& 850 $\mu$m images of this region that the
clump has an extension to the west. Higher
resolution C$^{18}$O maps of G33.92+0.11 by  Watt \& Mundy (\cite{wm99}) show that 
G33.914+0.109SMM is indeed comprised of two molecular cores, the fainter of which (Core B) is
coincident with the extension seen in the 450 \& 850 $\mu$m images (Fig.~\ref{fig:scuboth}). As
it is not possible to separate the cores in our SCUBA images we include this object as a single
sub-mm source in our catalogue.

\paragraph{G34.26+0.15: panels 45S \& 45L} This region is arguably one of the most well-studied massive
star-forming regions in the Galaxy. The \object{G34.26+0.15} region is a molecular cloud/H{\sc ii} region
complex comprising a large blister H{\sc ii} region to the SE of a kidney-shaped molecular clump
approximately 3.5 $\times$ 4 pc in size (Carral \& Welch \cite{cw92}; Heaton et
al.~\cite{hmld85}). Embedded within this molecular clump are a hot molecular core that exhibits
strong H$_{2}$O maser emission (Hofner \& Churchwell \cite{hc96}) and large abundances of
saturated molecules (Macdonald et al.~\cite{mghm96}), two unresolved UC H{\sc ii} regions and a
spectacular cometary H{\sc ii} region with a clearly defined tail stretching for at least 20\arcsec\ 
outwards from the centre of the molecular clump to the west (Gaume et al.~\cite{gfc94}).
Surprisingly the sub-mm morphology follows the appeareance of the cometary H{\sc ii} region, although
at a much larger scale. The sub-mm emission has a broad cometary morphology with little emission
to the west of the main core, suggesting that the cometary H{\sc ii} region may be expanding into this
low-density region. We identify three sub-mm sources in the SCUBA images; the central clump 
detailed above (G34.257+0.152SMM), an elongated clump to the northwest, G34.257+0.164SMM, that
is positionally coincident with an H$_{2}$O maser (Fey et al.~\cite{fgnc94}) and a further clump
to the southeast, G34.244+0.133SMM, previously discovered at 350 $\mu$m by Hunter et
al.~(\cite{hnb98}) and identified as a candidate embedded proto-B star.

\paragraph{G43.80$-$0.13: panels 50S \& 50L} This object is a well-known OH maser source known as \object{OH
43.8$-$0.1}
(Braz \& Epchtein \cite{be83}), which also harbours variable H$_{2}$O maser spots (Lehkt
\cite{lehkt00}). The SCUBA images reveal that the sub-mm clump is elongated roughly north-south,
which is not consistent with the recent CS J=5--4 map of Shirley et al.~(\cite{seykj03}), in
which the CS clump has a moderate east-west extension.

\paragraph{G45.07+0.13/G45.12+0.13: panels 51S \& 51L and 52S \& 52L} The molecular environment of both these UC H{\sc ii} regions
was the focus of a thorough study by Hunter, Phillips \& Menten (\cite{hpm97}), who found that
both of these UC H{\sc ii} regions are embedded in molecular clumps massing several thousand
M$_{\odot}$ and are associated with bipolar outflows. Our sub-mm images agree closely with those
obtained by Hunter, Phillips \& Menten (\cite{hpm97}), including the east-west extension seen in
the 450 $\mu$m image of G45.07+0.13.

\paragraph{G45.47+0.05: panel 87L} This field contains two UC H{\sc ii} regions, G45.47+0.05 and
G45.45+0.06,
which are both associated with compact 850 $\mu$m emission (Fig.~\ref{fig:scu850}). This latter
UC H{\sc ii} region is also associated with a cluster of massive YSOs (Feldt et al.~\cite{fsh98}).

\paragraph{G69.54$-$0.98: panels 53S \& 53L} This region is also known as \object{ON 1} (Winnberg \cite{winnberg70}) 
or OH 69.5$-$1.0 and is associated with a well-known OH maser (Braz \& Epchtein \cite{be83}) and
H$_{2}$O maser emission (Palagi et al.~\cite{pcc93}).

\paragraph{G70.29+1.60: panels 54S \& 54L} This region is also known as \object{K3-50A} and is a well-known UC H{\sc ii} region
residing within the star-forming complex W58. The ionised gas comprising \object{K3-50A} displays a 
strong velocity gradient that is suggestive of a bipolar ionised flow (de Pree et
al.~\cite{dpg94}). This ionised flow is surrounded by a massive molecular torus $\sim$ 0.5 pc 
in radius, in turn embedded within a larger lower-density molecular clump some 1.1 pc in radius
(Howard et al.~\cite{hkp97}). High resolution mid-infrared spectroscopic imaging has recently
revealed that the UC H{\sc ii} region is in fact ionised by multiple O stars (Okamoto et
al.~\cite{oky03}).

\paragraph{G76.38$-$0.62: panels 56S \& 56L} This object is the well-known bipolar H{\sc ii} region \object{S
106} (Sharpless
\cite{sharpless59}, an hourglass shaped H{\sc ii} region oriented north-south with a dark dust lane
running from east to west across the centre of the H{\sc ii} region (e.g.~Smith et al.~\cite{sjg01}).
The H{\sc ii} region is extended over roughly 3 $\times$ 2\arcmin\ and it is likely that the UC H{\sc ii}
region identified toward this source by KCW94 represents stellar wind emission from the central ionising
star. The SCUBA maps of G76.38$-$0.62 in Fig.~\ref{fig:scuboth} clearly show the
presence of the dark dust lane across the centre of S 106 and closely resemble those of 
Richer et al.~(\cite{rpw93}), to which the reader is referred for a fuller discussion of the
nature of S106.

\paragraph{G77.97$-$0.01: panels 57S \& 57L} The sub-mm emission in this field takes the form of two
diffuse clumps, of which the clump at the field centre (G77.962$-$0.008SMM) is associated with the KCW94
UC H{\sc ii} region  G77.965$-$0.006 and an infrared stellar cluster (Bica,  Dutra \& Barbuy
\cite{bdb2003}). The compact VLA configuration observations of Kurtz et al.~(\cite{kwho99}) reveal that there is an extended
diffuse component of 3.6 and 20 cm emission, peaking on both sub-mm clumps and with an extended more diffuse
tail stretching to the east.

\paragraph{G106.80+5.31: panels 62S \& 62L} This region is the well-known \object{S 140} H{\sc ii} region/molecular cloud complex
which is known to harbour at least three embedded infrared protostars, multiple outflows and a
variable H$_{2}$O maser (e.g.~Bally et al.~\cite{brwa02};  Minchin, Ward-Thompson \& White
\cite{mww95}; Brand et al.~\cite{bcf03}). G106.80+5.31 was mapped in the sub-mm by  Minchin,
Ward-Thompson \& White (\cite{mww95}) and our SCUBA images agree closely with these maps,
although we could not deblend the two sources S106 SMM1 and SMM3. These sources are represented
by our single sub-mm clump G106.794+5.313SMM. The remaining sub-mm clump in the field,
G106.800+5.314SMM corresponds to the source S 106 SMM2 from  Minchin, Ward-Thompson \& White
(\cite{mww95}).

\paragraph{G109.87+2.11: panels 63S \& 63L} This region is also known as \object{Cep A} and is an extremely well-studied
molecular cloud (Sargent \cite{sargent77}, \cite{sargent79}) exhibiting vigorous signs of star
formation including embedded UC H{\sc ii} regions, YSOs, protostars, outflows and H$_{2}$O masers
(e.g.~Richardson et al.~\cite{rwaw87}; Moriarty-Schieven, Snell \& Hughes \cite{msh91};
Torrelles et al.~\cite{tvh93}; Rodriguez et al.~\cite{rgc94}; G{\' o}mez et al.~\cite{gat99};
Hughes \cite{hughes2001}). We detect a single sub-mm clump in this field with a bright central
core associated with lower brightness extended emission along a NE-SW axis. The extended sub-mm
emission is elongated along almost the same axis as the string of UC H{\sc ii} regions observed toward
Cep A (Hughes \& Wouterloot \cite{hw84}). The fluxes measured from our SCUBA images are
consistent with those measured by Moriarty-Schieven, Snell \& Hughes (\cite{msh91}).

\paragraph{G110.21+2.63: panel 88L} A single diffuse clump of sub-mm emission is detected in this field
(G110.211+2.609SMM) at 850 $\mu$m only. This region is also known as the \object{Cep B} molecular cloud,
which is found along with Cep A in the Cep OB3 molecular cloud complex (Sargent
\cite{sargent77}, \cite{sargent79}). Cep B is found to be associated with a cluster of T Tauri
candidates (Moreno-Corral et al.~\cite{mcc93}). Testi et al.~(\cite{toh95}) identify three H{\sc ii}
regions toward Cep B, a blister H{\sc ii} region some 45 $\times$ 25\arcsec in extent (\object{[TOF95] A}) and two
ultracompact H{\sc ii} regions (\object{[TOF95] B} and C). Component B is coincident with the UC H{\sc ii} region
\object{G110.209+2.630} identified by KCW94. All of these H{\sc ii} regions are displaced from the sub-mm peak 
of G110.211+2.609SMM.

\paragraph{G111.28$-$0.66: panels 64S \& 64L} This region is also known as the \object{S 157} H{\sc ii} region (Sharpless
\cite{sharpless59}. The UC H{\sc ii} region identified by KCW 94 (G111.282$-$0.663) is associated with
an extended H{\sc ii} region to the NE $\sim$ 1\arcmin\ in radius and a further compact or 
ultracompact region to the SW (Kurtz et al.~\cite{kwho99}). We identify two sub-mm clumps in our
SCUBA image (G111.282$-$0.665SMM \& G111.281$-$0.670SMM), of which the first is positionally
associated with the UC H{\sc ii} region identified by KCW94 and the second is associated with the
compact H{\sc ii} region identified by Kurtz et al.~(\cite{kwho99}).  

\paragraph{G111.61+0.37: panels 65S \& 65L} This region is also known as the H{\sc ii} region/molecular cloud
complex \object{S 159}. KCW94 identified a single UC H{\sc ii} region in this field, positionally coincident with the
bright sub-mm peak seen in our SCUBA images. Kurtz et al.~(\cite{kwho99}) report the detection of
extended 21 cm emission to the east of the UC H{\sc ii} region and a further compact component is
reported to the west of the UC H{\sc ii} region by Lebr{\' o}n, Rodr{\'{\i}}guez \& Lizano
(\cite{lrl01}). This latter component is positionally coincident with the NW spur of emission
seen in the SCUBA 850 $\mu$m image.

\paragraph{G133.95+1.06: panels 66S \& 66L} This region is the well-known star-forming region
\object{W3(OH)} found in the W3 star-forming complex (Wynn-Williams et al.~\cite{www74}). The UC H{\sc
ii} region observed by KCW94 is found to split up into multiple sources at higher resolution (Baudry et
al.~\cite{bdb88}) and W3 (OH) is also known to harbour a compact cluster of infrared sources (Bica, Dutra
\& Barbuy \cite{bdb2003}) and three compact cores, collectively known as W3(H$_{2}$0), detected by their millimetre-wave continuum
and H$_{2}$O maser emission (Wyrowski et al.~\cite{wswm99}, \cite{whsw97}). The sub-mm peak detected in our SCUBA map
(G133.949+1.064SMM) is coincident with the position of W3(H$_{2}$O). For a recent overview of star formation in the W3 region the interested
reader is referred to Tieftrunk et al.~(\cite{tmwr98}).

\paragraph{G188.79+1.03/G188.77+0.92: panel 89L} Both of these UC H{\sc ii} regions lie within the \object{S 257} H{\sc ii}
region complex, which was recently imaged in the far-infrared by Ghosh et al.~(\cite{ghosh2000}).
The far-infrared maps suggest that G188.79+1.03 may be extended in an east-west direction, which is
consistent with the low-level (3$\sigma$)  emission  seen to the west of the main clump in our
SCUBA 850 $\mu$m image (see Fig.~\ref{fig:scu850}). G188.77+0.92 was not detected in the sub-mm
during this survey and is not shown in Fig.~\ref{fig:scuboth} or \ref{fig:scu850}. Both
G188.79+1.03 and G188.77+0.92 are known to be associated with  molecular outflows (Shepherd \&
Churchwell \cite{sc96}).

\paragraph{G213.88$-$11.84: panel 91L} This region is also known as \object{GGD 14} (G{\' o}mez, Rodr{\'{\i}}guez,
\& Garay \cite{grg2000}) and, in addition to the cometary UC H{\sc ii} region detected by KCW94, hosts a
cluster of ultracompact radio sources that are potentially low-mass pre-main-sequence stars 
(G{\' o}mez, Rodr{\'{\i}}guez, \& Garay \cite{grg2002}). G213.88$-$11.84 is associated with a
molecular outflow (Little, Heaton \& Dent \cite{lhd90}). We identify a single sub-mm clump in our
SCUBA 850 $\mu$m image, although the clump is extended toward the NE (in the direction of the
radio source \object{[GRG2002] VLA 7}) which may indicate the presence of unresolved substructure in our
image.

\section{Analysis and Discussion} \label{sect:analysis}

During the survey we identified three distinct types of object: ultracompact cm-wave sources that are not
associated with any sub-mm emission (``sub-mm quiet objects''), sub-mm clumps that are associated with
ultracompact cm-wave sources (``radio-loud clumps''); and sub-mm clumps that are not associated with any known
ultracompact cm-wave sources (``radio-quiet clumps''). In this section we investigate the likely nature of each
kind of object. We also examine the broad properties of the sample as a whole, focusing upon their morphology and
clustering properties. We postpone a detailed analysis of the physical properties of each sub-mm clump to a later
paper in the series (Thompson et al.~\cite{paper2}), where we present an in-depth study of the temperature, mass
and density of each sub-mm clump.

\subsection{``Sub-mm quiet'' ultracompact radio sources: are they UC H{\sc ii} regions?}
\label{sect:uchii_nosubmm}

From an inspection of Figs.~\ref{fig:scuboth} and \ref{fig:scu850}, it is clear that the vast majority
of ultracompact radio sources  investigated in this study are positionally associated with bright
sub-mm emission and are thus likely to be young UC H{\sc ii} regions embedded within molecular cloud clumps.
Nevertheless we have identified a substantial population of ultracompact radio sources in the fields
surveyed that are not associated with any detected sub-mm emission. For brevity we hence refer to these
objects as ``sub-mm quiet objects''. What is the likely nature of these sub-mm quiet objects? The
standard model of an UC H{\sc ii} region is a  massive YSO embedded within a dusty molecular cocoon that is
optically thick to the visible-UV radiation of the YSO. The compact cm-wave radiation results from 
free-free emission from the ionised gas of the  UC H{\sc ii} region but the bulk of the radiation is emitted
in the sub-mm, far- and mid-infrared regions through dust reprocessing of the shorter-wavelength
radiation by the optically thick cocoon (e.g. Crowther \& Conti \cite{cc2003}; Churchwell
\cite{c2002}). Following this picture sub-mm quiet objects are unlikely to be associated with massive dust
cocoons and thus may not be true UC H{\sc ii} regions.

We investigate this possibility by constraining the upper limit to the mass of any molecular clump
associated with the sub-mm quiet objects.  The VLA surveys possess sufficient sensitivity to detect
UC H{\sc ii} regions excited by B0 stars out to a distance of 20 kpc (Giveon et al.~\cite{gbh2005})
but the sensitivity of our SCUBA survey is such that we would not detect clumps of less than a few
hundred solar masses at this distance. There is thus the potential that the sub-mm quiet objects may
represent a population of distant UC H{\sc ii} regions embedded in relatively low mass clumps.

In the following discussion we refer only to the 850 $\mu$m data, as the signal-to-noise ratio of these
images is typically a factor or 10 higher that that at 450 $\mu$m. A total of 80 radio sources identified
from the literature search described in Sect.~\ref{subsect:sample} were found to be unassociated with 850
$\mu$m emission to a level of 3$\sigma$, or a typical flux limit of 0.4--0.6 Jy/beam. These sub-mm quiet
objects are listed in Table \ref{tbl:uchii_nosubmm} along with  their corresponding SCUBA field, the 850
$\mu$m flux upper limit, the distance to the radio component (if known) and an upper limit to the mass of
any molecular clump that could be associated with the sub-mm quiet objects. Masses were calculated using
the method outlined in Hildebrand (\cite{hildebrand83}) assuming the upper flux limits and distances
given in Table \ref{tbl:uchii_nosubmm},  a dust temperature of 30 K, dust emissivity $\beta=2$ and a mass
coefficient of $C_{850 \mu \rm m} = 50$ g\,cm$^{-2}$ (from Kerton et al.~\cite{kmjb2001}). For cases
where  a kinematic distance ambiguity exists we have calculated the mass upper limits at both the near
and far distances. We have also assumed that each sub-mm quiet object in a single SCUBA field lies at the
same distance.

\begin{table*}[p]
\caption{Ultracompact radio components in the survey fields with undetected sub-mm emission to a level of
3$\sigma$. The radio sources are indicated by their SIMBAD identifier; [WC89] for the WC89a survey,
[KCW94] for the KCW94 survey, [GPSR5] for the Becker et al.~(\cite{b94}) 5GHz Galactic Plane survey;
[GPSR] for the Becker et al.~(\cite{b90}) \& Zoonematkermani et al.~(\cite{z90}) 1.4 GHz Galactic Plane
survey and [GBH05] for the work of Giveon et al.(\cite{gbh2005}). Upper limits to the clump mass have been calculated from the 850 $\mu$m flux upper
limit using the assumptions outlined in Sect.~\ref{sect:uchii_nosubmm}. 
An ellipsis is given for those radio components of unknown distance while for objects with distance
ambiguities both near and far distances and clump masses are given.}
\label{tbl:uchii_nosubmm}
\begin{tabular}{lllccc}\hline\hline
UC H{\sc ii} field & \multicolumn{2}{c}{Known ultracompact radio components} & 850 $\mu$m flux limit & Distance &
Clump mass limit\\
 & & & (Jy/beam) & (kpc) & (M$_{\odot}$) \\\hline
G5.48$-$0.24  & [GBH05] 5.48464-0.23059 & &  0.4 &  14.3 &  219\\
G6.55$-$0.10  & [GBH05] 6.55647-0.09042 & &  0.6 &  16.7 &  472\\
G8.67$-$0.36  & [GBH05] 8.66530-0.34198 & [GBH05] 8.66324-0.34337 &  0.9  & 4.6 &  53\\
\vspace*{0.15cm}        &       [GBH05] 8.66202-0.34038 & [GBH05] 8.66730-0.34437 & & & \\
G10.30$-$0.15 & [WC89] 010.30-0.15A  &   [GBH05] 10.30726-0.14585  & 0.6 & 6.0 &  65 \\
              & [GBH05] 10.30485-0.14319 & [GBH05] 10.32159-0.15542 & & & \\
\vspace*{0.15cm}              & [GBH05] 10.30605-0.14551 & [GBH05] 10.30485-0.14319 & & & \\
G10.62$-$0.38 & [GBH05] 10.59905-0.38336 & [GBH05] 10.59905-0.38336 &   0.9 &  4.8 &  57\\
G110.21+2.63  &[KCW94] 110.209+2.630  & &  0.2 &  0.7 &  0.3\\
G13.19+0.04   & [GBH05] 13.19014+0.04077 & [GBH05] 13.18643+0.04999 &  0.4 &  4.4/12.1 & 21/161\\
 \vspace*{0.15cm}             & [GBH05] 13.19014+0.04077 & & & & \\
G13.87+0.28   & [GBH05] 13.87696+0.28238 & [GBH05] 13.87595+0.28479 & 0.6 &  4.5 &  35 \\
G18.15-0.28   & [GBH05] 18.15158-0.28016 & [GBH05] 18.14744-0.29764 & 0.7 &  4.2 &  34 \\
G19.07$-$0.27 & [WC89] 019.07-0.27       &  [GBH05] 19.06392-0.27384 & 0.5 &  5.4 &  40 \\
        &       [GBH05] 19.06256-0.27062 &  [GBH05] 19.06732-0.28637 & & & \\
\vspace*{0.15cm}        &       [GBH05] 19.07084-0.27694 & & & & \\
G19.49+0.14   & [GBH05] 19.48265+0.16098 & [GBH05] 19.48660+0.15574 & 0.3 &  2.0/14.0 & 3/145 \\
        &       [GBH05] 19.48283+0.15668 & [GBH05] 19.49724+0.13390 & & & \\
\vspace*{0.15cm}        &       [GBH05] 19.48468+0.15613 & & & & \\
G20.08$-$0.14 & [GBH05] 20.07849-0.14609 & &  0.2 &  12.3 &  71 \\
G23.46$-$0.20 & [WC89] 023.46-0.20A      & [GBH05] 23.43595-0.20366 & 0.4 &  9.0 &  82 \\
        &       [GBH05] 23.43821-0.20869 & [GBH05] 23.43876-0.21157 & & & \\
        &       [GBH05] 23.43876-0.21157 & [GBH05] 23.44160-0.21151 & & & \\
\vspace*{0.15cm}        &       [GBH05] 23.44160-0.21151 & & & & \\
G24.68$-$0.16 &   [GBH05] 24.67878-0.15397 & &   0.5 &  6.3/9.1 &  56/117 \\
G26.10$-$0.07 &  [GBH05]26.09201-0.05675 & & 0.4 & \ldots & \ldots \\ 
G25.72+0.05 &   [GBH05] 25.69526+0.03343 & [GBH05] 25.69562+0.03499 & 0.2 &  14.0 &  120 \\
G27.28+0.15 &   [GBH05] 27.28030+0.16537 & [GBH05] 27.27497+0.15800 & 0.5 &  15.2 &  338 \\
        &       [GBH05] 27.27070+0.14850 & [GBH05] 27.27809+0.14873 & & & \\
\vspace*{0.15cm}        &       [GBH05] 27.28271+0.16406 & [GBH05] 27.27569+0.14696 & & & \\
G27.49+0.19 &   [GBH05] 27.49356+0.19212 & [GBH05] 27.49617+0.19389 & 0.4 &  2.5/12.6 & 7/168 \\
\vspace*{0.15cm}        &       [GBH05] 27.49556+0.19105 & & & & \\
G28.60$-$0.36 & [GBH05] 28.59890-0.36738 & [GBH05] 28.60723-0.36382 & 0.3 &  5.2/9.7 &  22/75 \\
        &       [GBH05] 28.59173-0.36057 & [GBH05] 28.60375-0.36693 & & & \\
\vspace*{0.15cm}        &       [GBH05] 28.59560-0.36862 & & & & \\
G28.80+0.17 &   [GBH05] 28.79740+0.17314 & [GBH05] 28.80450+0.16777 & 0.5 &  6.4/8.5  & 62/109\\
G28.83$-$0.23 & [GBH05] 28.82159-0.22657 & [GBH05] 28.81780-0.22737 & 0.4 &  5.1/9.8  & 30/111 \\ 
\vspace*{0.15cm}        &       [GBH05] 28.81948-0.22485 & & & & \\
G30.78$-$0.02 & [GBH05] 30.78458-0.01577 & &  0.8 &  5.5 &  70 \\
G33.13$-$0.09 & [GBH05] 33.14311-0.08555  & &  0.2 &  7.1 &  34\\
G37.36$-$0.23 & [GPSR5] 37.356-0.232  & [GBH05]37.35747-0.23191 & 0.4 & \ldots & \ldots \\  
G37.77$-$0.20 & [GBH05] 37.76595-0.19983 & [GBH05] 37.76913-0.20613 &  0.4  & 8.8  & 83\\
G39.25$-$0.06 & [GBH05] 39.25972-0.05001 & [GBH05] 39.24685-0.06521 &  0.4  & 11.4  & 160\\
\vspace*{0.15cm}        &       [GBH05] 39.25972-0.05001 & & & & \\
G41.52$-$0.04 & [GPSR5] 41.525+0.039 & GPSR5 41.507+0.030 & 0.2 & \ldots & \ldots \\
G41.71+0.11   & [WC89] 041.71+0.11  & & 0.2 & \ldots & \ldots \\
G42.42$-$0.27 & [WC89] 042.42-0.27B & &   0.3 &  5.2 &  21\\
G42.90+0.57   & [WC89] 042.90+0.57A & [WC89] 042.90+0.57B & 0.3 & 5.2 & 21 \\ 
G44.26+0.10   &   [GPSR] 44.261+0.104 &   & $>$0.1 & \ldots & \ldots\\
G188.77+1.07  & [KCW94] 188.770+01.074  & & 0.2 & \ldots & \ldots \\ \hline
\end{tabular}
\end{table*}

It can be seen from Table \ref{tbl:uchii_nosubmm} that the 850 $\mu$m flux upper limits preclude high
mass clumps toward all but the furthest sub-mm quiet objects. Objects closer than a  distance of
10 kpc cannot be embedded in clumps with masses in excess of a few tens of solar masses. The most
extreme source in Table \ref{tbl:uchii_nosubmm} is \object{[KCW94] 110.209+2.630} which at an assumed
distance of 0.7 kpc has an upper limit to the mass of an associated molecular clump of only 0.3
M$_{\odot}$. Even at distances greater than 10 kpc the masses of any associated molecular clumps are
only a few hundred solar masses (472 M$_{\odot}$ in the furthest object in the sample; \object{[GBH05]
6.55647$-$0.09042}). This mass limit is close to the corresponding median mass of the detected sub-mm
clumps in the sample; the median flux and distance in our sample of detected sub-mm clumps is 7 Jy and
5.2 kpc, corresponding to a median mass of 530 M$_{\odot}$. It is possible that those sub-mm quiet
objects with distances in excess of 10 kpc may be part of a relatively low clump-mass population of
embedded UC H{\sc ii} regions.

If the nearby sub-mm quiet objects with distances less than 10 kpc are embedded within  molecular cloud
clumps then these clumps must possess masses of less than a few tens of solar masses. There is
considerable uncertainty regarding the  minimum mass cloud clump required to form a high mass star,
some studies suggest star formation efficiencies of 5--10\% (Clark et al.~\cite{cbzb2005}) or 10--30\%
(Lada \& Lada \cite{lada2003}; and references therein). One thus  might naively expect that a 10
M$_{\odot}$ star should only form in a clump of at least 30--200 M$_{\odot}$. Observational values for
the mass of high-mass star-forming clumps are at the higher end of the scale, ranging from 720
M$_{\odot}$ (Mueller et al.~\cite{msej02}) to 10$^{4}$ M$_{\odot}$ (Hatchell et al.~\cite{hfmtm00}).
The masses of the sub-mm quiet objects are thus below the lower mass end of the observational scale and
close to the minimum clump mass suggested by a simple consideration of star formation efficiency. We
thus conclude that it is unlikely that sub-mm quiet objects nearer than 10 kpc are true UC H{\sc ii} regions,
unless they are embedded within extremely low-mass clumps.

It is possible that the sub-mm quiet objects represent planetary nebulae, which are known to emit cm-wave
radiation at the mJy level and possess relatively low-mass dust envelopes that could fall below our SCUBA
detection limits. An alternate explanation is suggested by the fact that in a significant number of our
SCUBA survey fields (10/32 fields) the sub-mm quiet sources are strongly clustered, particularly objects
from the Giveon et al.~(\cite{gbh2005}) survey. Moreover, the majority of these ``clusters'' are often
found at the peripheries of radio-loud or radio-quiet sub-mm clumps, for example G19.07$-$0.27 in panel
19L,  G27.49+0.19 in panel 34L and G28.60$-$0.36 in panel 76L of Fig.~\ref{fig:scuboth}. The Giveon
et al.~survey, like all of the VLA surveys described in this paper, is a snapshot interferometer survey
with limited $uv$ coverage that significantly limits the largest angular scale visible in the survey
data. The bright components of ``resolved-out'' large angular scale structures often appear as clusters
of apparent point sources and it is  possible  that the clustered Giveon et al.~(\cite{gbh2005}) objects
trace  bright knots within extended structures rather than clusters of UC H{\sc ii} regions.

\begin{figure}
\centering
\includegraphics*[scale=0.45,trim=110 30 150 50]{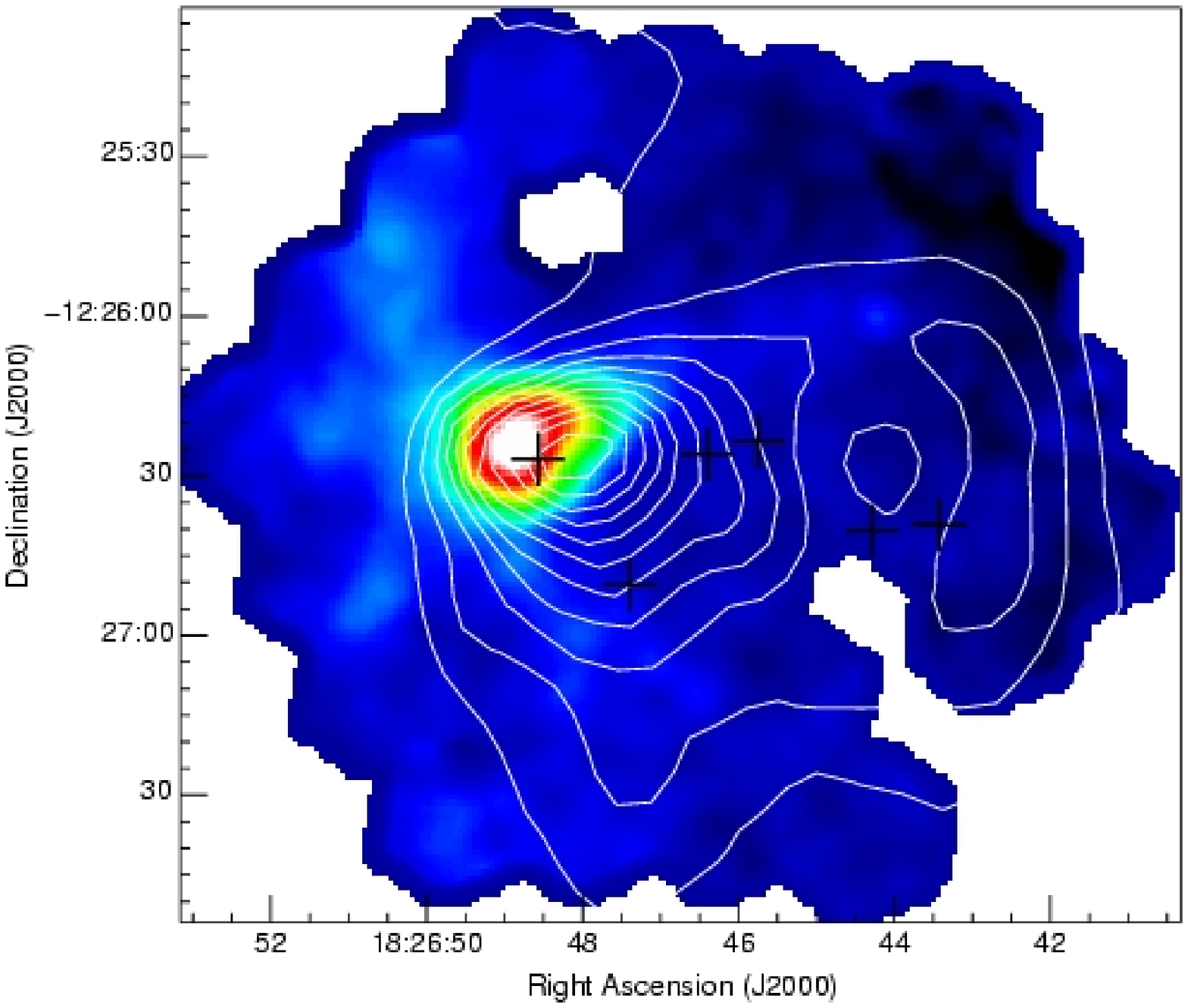}\\
\includegraphics*[scale=0.45,trim=110 30 150 50]{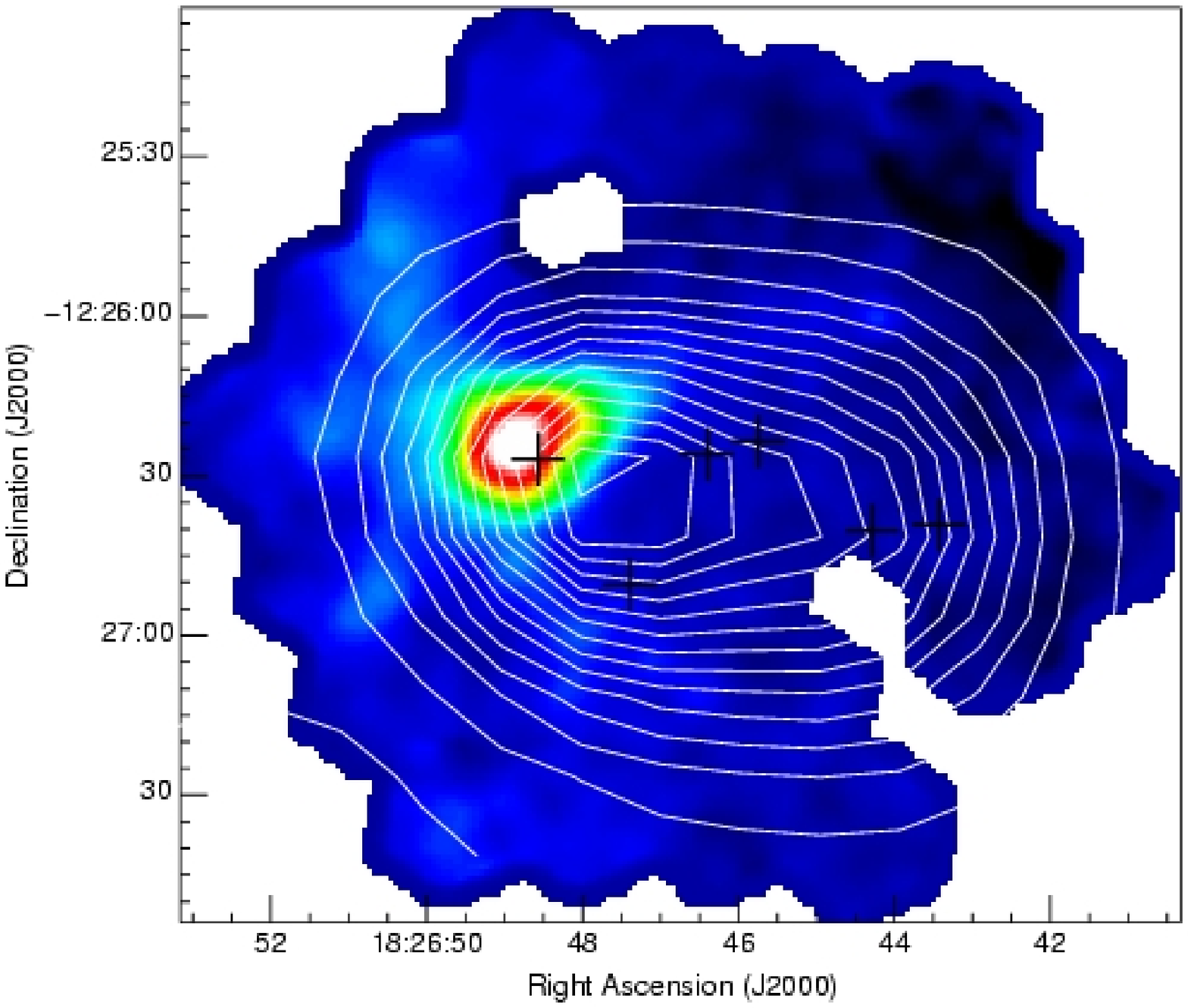}
\caption{850 $\mu$m SCUBA images of the UC H{\sc ii} field G19.07$-$0.27 overlaid with countours of  
MSX 8 $\mu$m emission (\emph{top}) and NVSS 20 cm emission (\emph{bottom}). Crosses indicate the
position of ultracompact radio components identified from the WC89a, Becker et al.~(\cite{b94}) and
Giveon et al.~(\cite{gbh2005}) surveys.}
\label{fig:g1907}
\end{figure}

This hypothesis is bolstered by the fact that a number of the clustered Giveon et al.~objects are
correlated with extended MSX 8 $\mu$m emission signifying PAH emission from the fringes of an extended
H{\sc ii} region. As an example we present MSX 8 $\mu$m and NRAO VLA Sky Survey (NVSS) 20 cm images of the
SCUBA field G19.07$-$0.27 in Fig.~\ref{fig:g1907}. NVSS data were used by Kurtz et al.~(\cite{kwho99})
to identify extended continuum emission toward a small sample of UC H{\sc ii} regions, as the longer
wavelength and more compact VLA configuration of NVSS considerably improves its sensitivity to
large-scale structure. Fig.~\ref{fig:g1907} reveals that the sub-mm clump G19.077-0.289SMM lies at the
eastern edge of an extended region of 8 $\mu$m and 20 cm emission, which probably marks the position of
an extended H{\sc ii} region. The cluster of Giveon et al.~(\cite{gbh2005}) objects lies within this extended
H{\sc ii} region and are more than likely bright components of the extended free-free emission of the H{\sc ii}
region rather than individual UC H{\sc ii} regions.

The overall appearance of the G19.07$-$0.27 region in the sub-mm, mid-IR and radio is consistent with the
hierarchical structure hypothesis of Kim \& Koo (\cite{kk01}, \cite{kk03}) for H{\sc ii} regions with
both ultracompact and extended components. As seen in Fig.~\ref{fig:g1907} the ``sub-mm loud''
ultracompact radio source associated with G19.077-0.289SMM lies to the south-west of the sub-mm peak, on
the same face as the extended mid-IR and radio emission traced by MSX and NVSS. It is thus possible that
the sub-mm loud ultracompact component corresponds to the position of an embedded O-star(s)  whose H{\sc
ii} region is generally constrained by the high ambient clump density within the clump (and hence has an
ultracompact component) but has broken free from the clump into the lower density region to the west to
form a champagne flow H{\sc ii} region. It is unclear what fraction of the ultracompact cm-wave sources
contained in the WC89a, KCW94 or Giveon et al.~(\cite{gbh2005}) catalogues are extended objects similar
to G19.07$-$0.27. Nevertheless the Kim \& Koo (\cite{kk01}) hierarchical structure hypothesis offers a
plausible explanation for these clustered sub-mm quiet objects. Further investigation of the UC H{\sc ii}
region catalogues is necessary to determine the fraction of extended interlopers in the present UC H{\sc
ii} region catalogues  and whether the extended H{\sc ii} regions are consistent with the Kim \& Koo
model.

\subsection{Radio-loud and radio-quiet sub-mm clumps}
\label{sect:rlrq}

The sub-mm clumps that were detected in the survey split into two types; those that are positionally associated
with ultracompact radio sources and those that are not associated with any detected ultracompact radio sources.
The former are referred to as ``radio-loud'' clumps and the latter as ``radio-quiet'' clumps. We define
positionally associated as the position of the radio source  lying within the lowest 850 $\mu$m contour bounding
the clump  (the lowest contour corresponds to the 3$\sigma$ level in each image, see Sect.~\ref{sect:images}). It
is important to note that all of our target fields have been observed with the VLA as part of the WC89a, KCW94,
Becker et al.~(\cite{b94}) or Giveon et al.~(\cite{gbh2005}) surveys and so the radio-quiet clumps are truly
radio-quiet (to within the flux limits set by the VLA surveys) rather than simply radio-unobserved. We discuss
the likely individual nature of the radio-loud and radio-quiet clumps in more detail in Sects.~\ref{sect:rl} and
\ref{sect:rq} but in this section we will briefly dwell upon the properties of the population of sub-mm clumps as
a whole.

\subsubsection{Sub-mm colours and ``luminosities''}

The sub-millimetre wavelength radiation detected from the clumps originates from thermal emission
from dust grains contained within the molecular gas of the clumps and is an excellent optically
thin tracer of the column density within the clumps (e.g.~Hildebrand \cite{hildebrand83}). The
wavelength dependence of the sub-mm emission depends upon both the temperature of the dust and its
emissivity coefficient $\beta$. The standard procedure in the determination of the column density
(or mass) of a sub-mm clump, its temperature and grain emissivity is to fit a greybody model to the
broad spectrum SED of the clump (e.g.~Dent, Matthews \& Ward-Thompson \cite{dmwt98}). Usually a
combination of IRAS far-infared and millimetre/sub-millimetre measurements provide the SED, but for
many of the clumps detected in our survey (particularly those SCUBA fields with multiple sub-mm
components) the IRAS data is hopelessly confused due to the large beam at 100 $\mu$m
($\sim$2\arcmin\ FWHM).

\begin{figure}
\begin{center}
\includegraphics[angle=-90,scale=0.6]{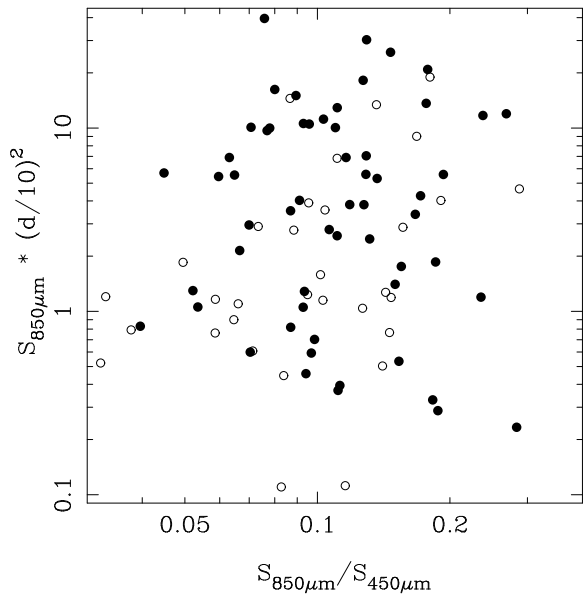}
\caption{Sub-mm ``colour-luminosity'' diagram. Radio-loud clumps are 
indicated by filled circles and radio-quiet clumps are
shown by open circles.}
\label{fig:submm_cl}
\end{center}
\end{figure}

As part of Paper II we present greybody and radiative transfer analyses for those clumps that are not confused over the
large scale of the IRAS beam, but here we present an investigation of the sub-mm properties of the
wider sample as a whole. One of the standard techniques used to investigate the properties of large
stellar samples is the colour-magnitude diagram. This approach is not a standard tool in the study
of star-forming cores, but was recently adapted by Crowther \& Conti (\cite{cc2003}) to investigate
the mid-infrared properties of UC H{\sc ii} regions. We adapt their technique for the sub-millimetre by
using the distance-scaled 850 $\mu$m flux of the clumps as a surrogate for the clump luminosity  and the 850$\mu$m/
450 $\mu$m flux ratio for the sub-mm colour, renaming the approach as a colour-luminosity diagram to
avoid confusion.

Assuming that clumps are optically thin at both 450 and 850 $\mu$m, their sub-mm colour should depend
solely upon the source-averaged dust temperature and grain emissivity. Thus by plotting a sub-mm
colour-luminosity diagram we are able to investigate different regions in the dust
temperature/emissivity parameter space as a function of clump luminosity. The sub-mm
colour-luminosity diagram of the sub-mm clumps detected in the survey is shown in
Fig.~\ref{fig:submm_cl}. Only those clumps with unambiguous distances are included in
Fig.~\ref{fig:submm_cl}. As many of the radio-quiet clumps lack known kinematic distances we assume
that radio-quiet clumps located within the same SCUBA field as a radio-loud clump with a known
distance lie at that known distance (i.e.~the radio-loud and radio-quiet clumps are part of the same
complex). 

Two insights are immediately apparent from the colour-luminosity diagram shown in Fig.~\ref{fig:submm_cl}. The
first is that, in general, the radio-loud clumps tend to have a higher distance-scaled 850 $\mu$m flux than the
radio-quiet clumps.  The median distance-scaled 850 $\mu$m flux for radio-loud clumps is 3.8 Jy, compared to
1.2 Jy for the radio-quiet clumps. A Kolmogorov-Smirnov test also indicates that the two distance-scaled flux
distributions are unlikely to be drawn from the same population, with a probability of only 2.5\% for a single
flux distribution hypothesis. This may imply that the radio-quiet clumps form a separate lower-luminosity
population to the radio-loud UC H{\sc ii} region-containing clumps (note that the typical UC H{\sc ii} detection
limit from WC89a is of a B0.5 star at 5 kpc). The radio-quietness of these clumps may perhaps thus be due to the
fact that the clumps are  forming lower-mass stars rather than high-mass protostars in a pre-UC H{\sc ii} region
phase. It must however be cautioned that, as the radio-quiet sample is almost certainly comprised of a mixed
population of massive star-forming clusters in a pre-UC H{\sc ii} phase and low- to intermediate-mass star-forming
clusters, it is impossible to make conclusions about the nature of the entire sample of radio-quiet clumps from
this data.

The second insight from  the sub-mm colour-luminosity diagram is that both radio-quiet and radio-loud
clumps possess a similar range of sub-mm colours. This implies that all the clumps in the diagram
possess a similar range of temperatures and dust emissivities. The sub-mm properties of the 
radio-quiet clumps closely resemble those of the radio-loud clumps, suggesting that the clumps form a
fairly homogenous population. It is intriguing to speculate that the main difference between the two
is the presence or absence of an UC H{\sc ii} region as this naturally gives rise to the implication that
the clumps are in different evolutionary states. However, the tendency for the radio-quiet clumps to
exhibit generally lower distance-scaled 850 $\mu$m fluxes implies that the radio-quiet clump
population may also contain clumps that are forming lower-mass stars rather than high mass protostars
in a pre-UC H{\sc ii} region phase. We will dwell further upon the nature of the radio-loud and radio-quiet clumps
in Sects.~\ref{sect:rl} and \ref{sect:rq}.

\subsubsection{Morphology}

\begin{figure}
\begin{center}
\includegraphics[angle=-90,scale=0.6]{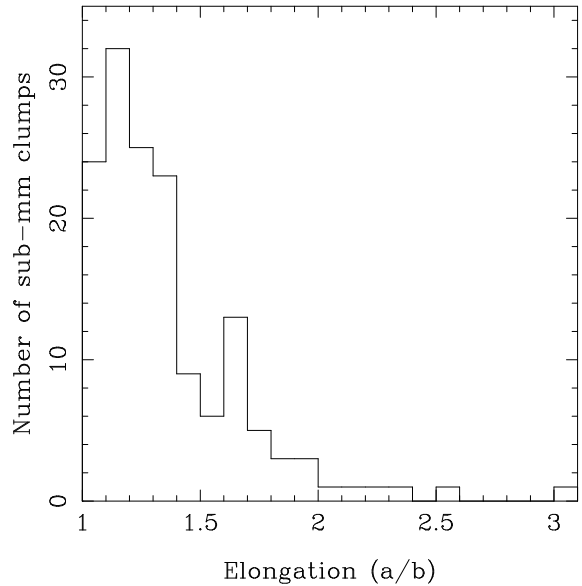}
\caption{Histogram of 850 $\mu$m clump elongation}
\label{fig:elongation}
\end{center}
\end{figure}

\begin{figure}
\begin{center}
\includegraphics[angle=-90,scale=0.6]{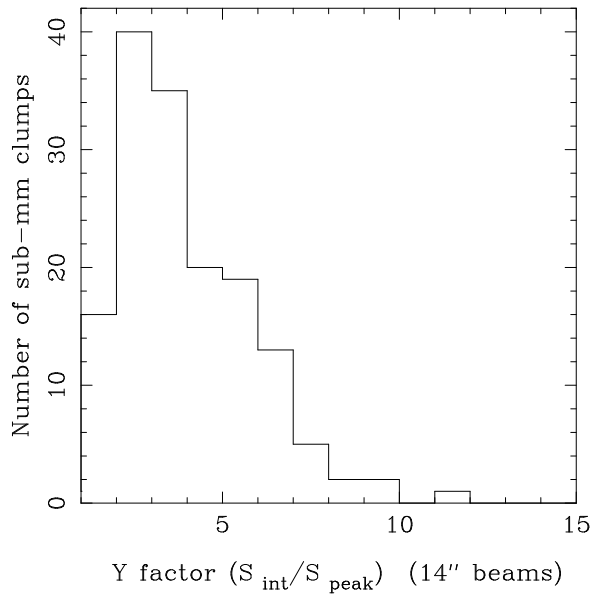}
\caption{Histogram of 850 $\mu$m clump Y factor (integrated flux divided by peak flux). For an unresolved
source the Y-factor is obviously equal to one.}
\label{fig:yfactor}
\end{center}
\end{figure}

We examined the morphology of the sub-mm clumps detected in the survey by measuring their half-power
diameters along both major and minor axes. The major and minor axes of the clumps were initially
estimated by Gaussian fits. It was found during the fitting procedure that a significant fraction of
the clumps could not be accurately fitted by Gaussians and so for consistency all major and minor
axes were directly estimated from the 50\% contours of each clump. Due to the better signal-to-noise
of the 850 $\mu$m images we restrict the following discussion solely to the morphology of the clumps
at this wavelength. The half-power diameters of the clumps are more sensitive to the distribution of
the central regions of the clumps rather than the low-lying flux in their outer regions. To
investigate the general extent of the sub-mm emission and compare this to the recent survey of
candidate high-mass protostellar objects (HMPOs) of Williams et al.~(\cite{wfs2004})  we also measured the
850 $\mu$m flux Y-factor (the ratio of integrated to peak flux) for each clump in the survey..  
We plot histograms of the clump elongation ($a/b$), and
Y-factor ($S_{\rm integrated}/S_{\rm peak}$) in  Figs.~\ref{fig:elongation}
and \ref{fig:yfactor} respectively.

The majority of the sub-mm clumps display small elongations, with a median elongation of 1.4,
suggesting a predominantly spherical or low axis-ratio population. This of course assumes
that the clumps observed at 850 $\mu$m represent single objects. This assumption seems to generally
hold for the clumps in this survey: elongated clumps with high signal-to-noise 450 $\mu$m data remain
as elongated single structures in the higher resolution 450 $\mu$m images (e.g. G33.13$-$0.09 in
panel 43S)  and there is no correlation of elongation with distance,as would be expected if the more
distant clumps are unresolved linear chains or ridges.

We do not
see any differences between the morphology of radio-loud or radio-quiet clumps, histograms of the
elongation of each population resemble each other closely and  each sample possesses a similar median
elongation (1.3 and 1.5 for radio-loud and radio-quiet respectively). The radio-quiet clumps are
marginally larger than the radio-loud clumps, with median major and minor axes of 30\arcsec\ and
19\arcsec\ for radio-quiet clumps as compared to 24\arcsec\ and 19\arcsec\ for the radio-loud clumps
respectively.  

The Y-factors of the sample as a whole (plotted in Fig.~\ref{fig:yfactor}) follow a distribution that
is peaked at a value of $\sim$3. As in the histograms of elongation there is little difference
between the histograms and median values of the Y-factor of radio-loud and radio-quiet clumps. The
median value of the Y-factor for all sub-mm clumps detected in the survey is close to the value for
candidate HMPOs observed by Williams et al.~(\cite{wfs2004}). This may indicate that, as the clumps
span a range in distance from 1--14 kpc, they could possess a scale-free envelope structure as
suggested for candidate HMPOs by Williams et al.~(\cite{wfs2004}). As mentioned previously we do not
see any evidence that the objects in our study fragment at higher resolution, unlike the objects of
Williams et al.~(\cite{wfs2004}). However absence of evidence is not evidence of absence and caution
must be applied to the hypothesis that the sub-mm clumps possess a scale-free envelope structure.
Nevertheless we may at least draw the qualitative conclusion that, like the candidate HMPOs of
Williams et al.~(\cite{wfs2004}), a significant fraction of the mass of the clumps lies outside the
central beam. If the radio-loud clumps in our survey represent embedded UC H{\sc ii} regions which are
more evolved than the candidate HMPOs of Williams et al.~(\cite{wfs2004}) then we may also conclude
that the fraction of the mass in the outer region of the clump compared to that in the central beam
does not evolve appreciably between the HMPO and UC H{\sc ii} stages.

\subsubsection{Clustering properties}

A substantial number of UC H{\sc ii} regions from the survey display emission from more than one sub-mm clump in
the SCUBA field of view (for example G11.94$-$0.62 in panel 13L, G29.96$-$0.02 in panel 37L and G79.30+0.28 in
panel 59L). Almost half of the SCUBA fields  that were imaged during the survey (44/105) are associated with two
or more sub-mm clumps. Many of these ``neighbouring'' clumps have not been observed in suitable kinematic
distance tracers and so it is not possible to determine whether these clumps are located in the same molecular
cloud complex, or are a chance alignment of objects at different distances. Nevertheless the data that
are so far
available for individual cases (e.g.~G34.26+0.15, Hunter et al.~\cite{hnb98}; G10.30$-$0.15 and G29.96$-$0.02,
Thompson et al.~in preparation and G30.78$-$0.02, Motte et al.~\cite{msl03}) suggests that a number of the
multiple clumps are clusters of objects found within the same molecular cloud complex. Several of the
neighbouring sub-mm clumps are associated with star formation tracers: either UC H{\sc ii} regions
(e.g.~G28.29$-$0.36 in panel 36L) or H$_{2}$O masers (e.g.~G11.94$-$0.62 in panel 13L; Hofner \& Churchwell \cite{hc96}),
although a considerable number are radio-quiet.  For an assumed distance of 5 kpc each 2\arcmin\ diameter SCUBA
fields represents a $\sim$3 pc diameter region, similar to the parsec-scale condensations within Giant Molecular
Clouds (Blitz \cite{blitz93}). There is thus the important implication that these ``neighbouring'' sub-mm clumps
may represent other sites of star formation within the same molecular cloud complex containing the
recently-formed YSO(s) exciting the UC H{\sc ii} region. 

We quantify the likelihood of finding ''neighbouring'' clumps within a SCUBA field using the
companion clump fraction (CCF), a statistic more often used to investigate the multiplicity of
stars, but which has been used to investigate the multiplicity of candidate
HMPOs (Williams et al.~\cite{wfs2004}). The CCF gives a measure of the average number of companions per sub-mm
clump and is given by the formula:

\begin{equation}
CCF = \frac{B + 2T + 3Q}{S + B + T + Q},
\end{equation}

where S, B, T and Q are the number of single, double (binary), triple and quadruple clumps in our survey,
respectively. We
find a CCF for the sub-mm clumps detected in our survey of 0.90$\pm$0.07, where the error was determined
from simple Poisson counting statistics. This value is moderately higher
than that reported by Williams et al.~(\cite{wfs2004}) and, as the two surveys are at similar depths and of
objects at similar distances, more than likely reflects the initial selection  criteria  of the sample upon
which the Williams et al.~study is based (Sridharan et al.~\cite{srid2002}). The Sridharan et
al.~(\cite{srid2002}) sample were selected to be isolated, in that the sources were non-detections in large-beam
radio surveys and thus are not associated with UC H{\sc ii} regions nearer than a few arcminutes. Our UC H{\sc ii} region
survey sample does not possess  such selection criteria, which is reflected in the higher CCF that
we determine for the sub-mm clumps in our sample. As we did not select our sample on the grounds of
isolation it can be argued that the higher CCF is more representative of the multiplicity of sub-mm clumps
in massive star-forming regions than the Williams et al.~result. We have shown that it is highly likely for
the sub-mm clumps in massive star-forming regions to possess at least one near neighbour (within a projected
distance of a few pc). This suggests that massive YSOs or UC H{\sc ii} regions are ideal places to search for
nearby, possibly less evolved companions and that the environments of UC H{\sc ii} regions are clustered active
sites of massive star formation.

\begin{figure}
\begin{center}
\includegraphics[angle=-90,scale=0.6]{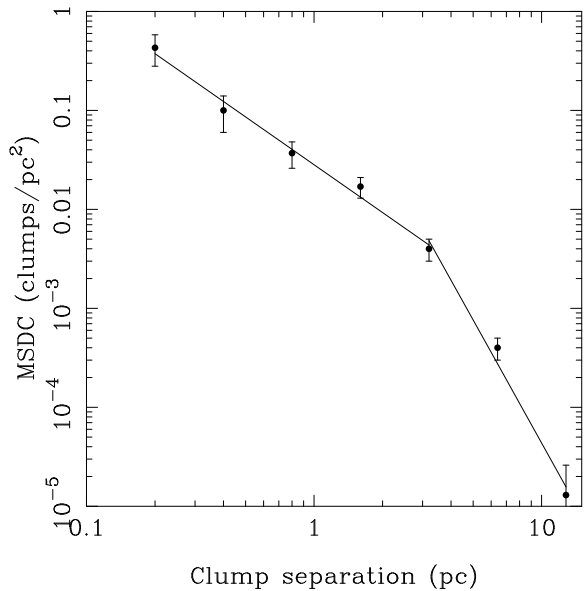}
\caption{The mean surface density of companions for all the 850 $\mu$m clumps detected in the survey. Error
bars are determined from simple Poisson counting statistics. Solid lines indicate linear
least-squared fits to the first 5 and last 3 data points respectively.}
\label{fig:msdc}
\end{center}
\end{figure}

We also determined the surface density of the sub-mm clumps using the mean surface density of companions
(MSDC) technique outlined in Williams et al.~(\cite{wfs2004}), so that we could compare the surface density
of clumps found toward more evolved UC H{\sc ii} regions  to that measured for the relatively more isolated and
potentially less evolved candidate HMPO clumps. We determined the MSDC by measuring the linear distance
between pairs of clumps in the SCUBA images and scaling the number of linear distance pairs in annuli of
fixed radius by the area of the distance annulus to obtain a value for the surface density. Clumps located
in the same SCUBA field were assumed to lie at the same heliocentric distance. We present a plot of surface
density versus linear separation in Fig.~\ref{fig:msdc}.

Fig.~\ref{fig:msdc} shows that the MSDC of the sub-mm clumps follows two power laws, with a break at $\sim$3
pc. At small linear separations the index of the power law is -1.6, close to the value of -1.7 found for
candidate HMPO clumps by Williams et al.~(\cite{wfs2004}). But for linear separations over 3 pc the MSDC
declines steeply with a power law index of -4.1. The underlying cause behind this break is unclear; it is
possibly due to incomplete sampling at large linear separations, but as the numbers of distance pairs in
each distance bin are relatively constant apart from the very largest value we do not feel that this is a
likely explananation. The decline in surface density over a radius of 3 pc may reflect a natural size scale
for the extent of massive star-forming clusters in molecular clouds. If this is the case, one would
also expect to see a similar effect in the candidate HMPO sub-mm clumps, particularly as the selection
criteria of the HMPO sample select against radio-loud companions within several arcminutes (or several pc at
the typical heliocentric distance of the HMPOs). This effect is not seen within the Williams et
al.~(\cite{wfs2004}) sample, whose surface density is consistent with a single unbroken power law of index
-1.7. 

Our sample contains objects at a wide range of heliocentric distances (and thus a large range of linear
distance pairs) and is limited at both small linear distances by the angular resolution of the JCMT and at
large linear distances by the limited field of view of SCUBA. Our flux-limited, rather than mass-limited,
sample may also cause a bias towards detecting bright, relatively nearby clumps at large linear separations
or may underestimate the number of faint companion clumps. It is thus possible that the break in the power
law is caused by a selection bias in our sample to particular ranges of linear distances. However, our
sample of clumps observed towards UC H{\sc ii} regions is very similar to that observed by Williams et
al.~(\cite{wfs2004}) in terms of heliocentric distance and both surveys have a corresponding 850 $\mu$m 
flux limit. We thus think it unlikely that the different MSDC distributions are caused by underlying
selection biases in each sample. Further wide-field observations of a large sample of massive star-forming regions are required
to more accurately the determine surface density distribution of sub-mm clumps and investigate whether the
apparent steep decline in surface density at large separations is indicative of a natural size scale for
regions of massive star formation.


\subsection{The likely nature of radio-loud sub-mm clumps}
\label{sect:rl}

Here we examine those sub-mm clumps associated with ultracompact radio sources (the radio-loud clumps)
in order to determine whether their observed properties are consistent with massive clumps containing 
embedded UC H{\sc ii} regions. As we have already demonstrated, those ultracompact radio sources that are
not associated with sub-mm emission (sub-mm quiet objects) are unlikely to be young embedded UC H{\sc ii}
regions unless they lie at considerable heliocentric distance ($\ge$ 10 kpc). So, do the radio-loud
sub-mm clumps fit the standard picture of young UC H{\sc ii} regions that are still embedded within their
natal molecular cloud cores?

\begin{figure*}
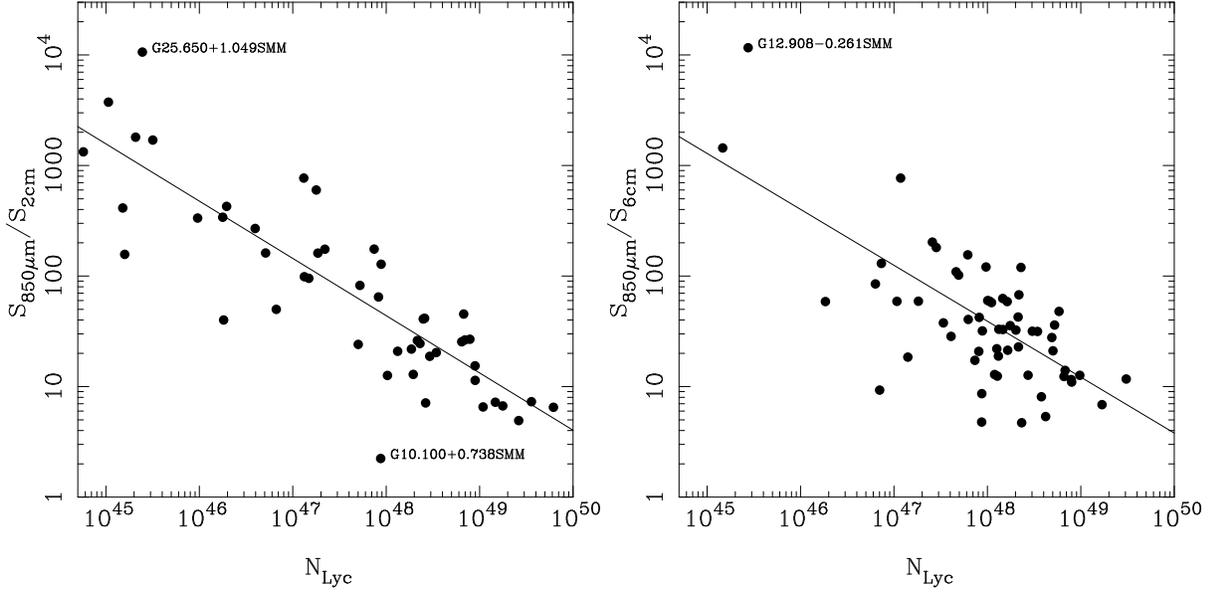

\begin{centering}
\includegraphics[angle=-90,scale=0.6]{4853f8a.ps}
\includegraphics[angle=-90,scale=0.6]{4853f8b.ps}
\caption{Comparison between the radio to sub-mm ratio (\emph{left:} 850 $\mu$m vs 2 cm; 
\emph{right:} 850 $\mu$m vs 6 cm) and the Lyman photon flux inferred from the radio free-free
emission of the UC H{\sc ii} regions. Solid lines represent linear least-squared fits to the data. Particular 
outliers to the fits are indicated and are discussed in the text.}
\label{fig:submm_radio}
\end{centering}
\end{figure*}

The standard model of UC H{\sc ii} regions as discussed by e.g.~Crowther \& Conti (\cite{cc2003}) or
Churchwell (\cite{c2002}) is of a small and young H{\sc ii} region photoionised by one or more high-mass
stars and surrounded by a massive optically thick cocoon of dust and gas.  Clearly the bright sub-mm
clumps observed in our survey fit the picture of massive envelopes of dust and gas with sufficient
dust to be optically thick to short-wavelength emission, but is the sub-mm emission consistent with
being reprocessed radiation from the embedded star(s) responsible for exciting UC H{\sc ii} regions? If this
were the case then a clear relation ought to exist between the sub-mm luminosity of the clumps (which
arises from the reprocessed stellar luminosity)  and the cm-wave radio emission of the associated UC
H{\sc ii} regions (which depends upon the UV luminosity of the stars). This relation has been shown to exist
for a small number of UC H{\sc ii} regions from the WC89a and KCW94 catalogues by Crowther \& Conti
(\cite{cc2003}) using VLA cm-wave and IRAS 100 $\mu$m flux data. Crowther \& Conti's approach was
unfortunately limited by confusion within the large 100 $\mu$m beam of IRAS. Using our SCUBA images we
can almost entirely remove these confusion issues and considerably enlarge the sample of UC H{\sc ii}
regions over that of Crowther \& Conti (\cite{cc2003}). 

In Fig.~\ref{fig:submm_radio} we present plots of the 850 $\mu$m/cm-wave  flux ratio  against the
number of Lyman continuum photons ($N_{\rm Lyc}$) for each radio-loud clump detected in our survey
whose heliocentric distance is both known and unaffected by a near/far kinematic ambiguity. Lyman
fluxes were calculated from the integrated radio fluxes of ultracompact radio sources associated with
the sub-mm clumps as published in WC89a, KCW94, Becker et al.~\cite{b94} or Giveon et
al.~\cite{gbh2005} using the standard equation relating free-free continuum brightness to the Lyman
continuum flux (e.g.~Carpenter et al.~\cite{css90}) and assuming that each radio source represents an
UC H{\sc ii} region. Where  multiple ultracompact radio sources were found to be associated with a single
sub-mm clump, the flux of each was summed to provide a total radio flux before evaluating the Lyman
continuum flux. Diagrams are plotted for both 2 and
6 cm data, as many of the objects observed in the WC89a survey do not possess 6cm data.

Caution must however be applied to the comparison of the derived Lyman flux from each VLA survey. The VLA surveys of WC89a,
KCW94, Becker et al.~(\cite{b94}) and Giveon et al.~(\cite{gbh2005}) were taken at different wavelengths and so all possess different
selection effects, flux limits, synthesised beam sizes and sensitivity to extended structures. The 6 cm  fluxes are mostly drawn from
the sample of WC89a which was based on the single-dish point sources from Wink, Altenhoff \& Metzger (\cite{wam82}) with a 6 cm
sensitivity of $\sim$0.5 Jy and is therefore intrinsically biased against weak 6 cm emitters.  The IRAS-selected sample of KCW94 
which appears in the 2cm plot (Fig. \ref{fig:submm_radio}a) is however only limited by the sensitivity of their VLA observations. 
This explains the lack of low $N_{\rm Lyc}$ sources at 6cm in Fig.~\ref{fig:submm_radio}

\begin{figure}
\begin{centering}
\includegraphics[angle=-90,scale=0.6]{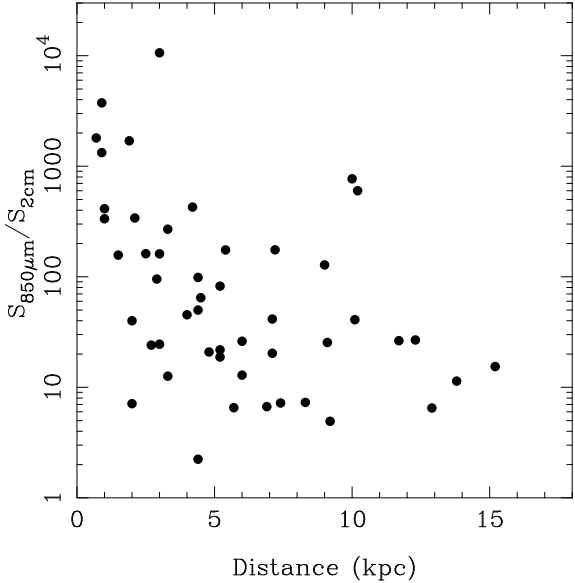}
\caption{The 2 cm to 850 $\mu$m flux ratio of UC H{\sc ii} regions as a function of distance. Note the trend toward higher
values of $S_{850\mu{\rm m}}/S_{2{\rm cm}}$ at small distances which may indicate that the 2 cm fluxes of nearby UC H{\sc ii} regions are
underestimated due to their over-resolution by the VLA.}
\label{fig:sub-mm_radio_dist}
\end{centering}
\end{figure}

It is also possible that as the shorter wavelength data are less sensitive to extended structure (for the same $uv$ coverage and array
configuration) the Lyman flux calculated from fluxes measured at this wavelength may be lower than that calculated from
longer-wavelength data. We illustrate this effect in Fig.~\ref{fig:sub-mm_radio_dist}, where the 850 $\mu$m/2 cm flux ratio is plotted
against distance. As can be seen in Fig,~\ref{fig:sub-mm_radio_dist} there is a trend towards higher flux ratio values at smaller
distances. Partly this is due to selection against faint radio sources at large distances, but this effect may also be caused by
selection against nearby (i.e.~within $\sim$ 2 kpc) UC H{\sc ii} regions with low flux ratios (\& high $N_{\rm Lyc}$ or that the 2 cm
fluxes of nearby UC H{\sc ii} regions are underestimated due to the extended flux sensitivity issue described earlier. 

We caution against overinterpretation of Fig.~\ref{fig:submm_radio} but note that  Fig.~\ref{fig:sub-mm_radio_dist} reveals that the
bias toward high values of  S$_{850\mu{\rm m}}$/S$_{2\rm cm}$ only becomes marked for UC H{\sc ii} regions with S$_{850\mu{\rm
m}}$/S$_{2\rm cm} \ga$ 300 or $N_{\rm Lyc} \la 5 \times 10^{46}$ s$^{-1}$. It can thus be seen that the correlation between the
sub-mm/radio flux ratio seen in Fig.~\ref{fig:submm_radio} holds for all except low values of N$_{\rm Lyc}$ or high values of the
sub-mm/cm-wave flux ratio.    

 Bearing in mind the caveats discussed in the previous paragraph, both plots in Fig.~\ref{fig:submm_radio} show a clear linear
relation between the 850 $\mu$m/radio flux ratio and the Lyman continuum flux of the embedded UC H{\sc ii} regions. This linear
relationship extends over more than four orders of magnitude in the number of Lyman photons ($N_{\rm Lyc}$) and almost three orders of
magnitude in the 850 $\mu$m/radio flux ratio. There are fewer objects with low Lyman flux  in the 6 cm plot and the scatter is larger
at 6 cm as compared to the 2 cm plot, but in general both  plots agree closely. Fig.~\ref{fig:submm_radio} also shows that we do not
see objects with either low $N_{\rm Lyc}$ \& low 850 $\mu$m flux or high $N_{\rm Lyc}$ \& high 850 $\mu$m flux. This is unlikely to be
caused by any of the selection effects discussed earlier, as both the VLA surveys and our own SCUBA survey have sufficient sensitivity
to detect these types of objects (if they exist) in these regions of the plot, either at low $N_{\rm Lyc}$ or at low 850 $\mu$m flux. 
A power-law fit to the data in both diagrams (following the form
$y=A\,10^{bx}$) yields values for $A$ of $10^{26.51\pm2.01}$ and $10^{25.88\pm3.32}$ at 2 and 6 cm respectively; and  $b$ of
$-$0.52$\pm$0.04 and $-$0.51$\pm$0.07, with straight-line fit coefficients of 0.9 and 0.7 for 2 and 6 cm respectively. 

Fig.~\ref{fig:submm_radio} clearly shows that the radio-loud sub-mm clumps in our survey are in excellent
agreement with the standard model of embedded UC H{\sc ii} regions and that the clumps are predominantly heated by
their embedded massive stars. Hereafter in this paper we will refer to these ultracompact radio components as UC
H{\sc ii} regions. The scatter in both diagrams is likely to be caused  by the selection effects mentioned
earlier, the absorption
of Lyman continuum photons within the UC H{\sc ii} regions by dust (which would move points leftwards and upwards)
or by the presence of stars of lower mass that contribute to the total IR flux but do not contribute to the
Lyman flux (which would move points vertically upwards). We indicate two specific clumps in particular 
whose outlying positions in Fig.~\ref{fig:submm_radio} may be due to these effects
(G25.650+1.049SMM \& G12.908$-$0.261SMM). We also indicate the object G10.100+0.738SMM in
Fig.~\ref{fig:submm_radio}, whose outlying position at the bottom of the diagram is unlikely to be caused by
either selection effects, Lyman photon absorption or the presence of lower-mass stars. This
object is almost certainly a misclassified UC H{\sc ii} region. Walsh et al.~(\cite{walsh03}) classify this object
as a planetary nebula, noting that the position of the sub-mm clump coincides with the position of the
well-known  planetary nebula \object{NGC 6537}. The position of G10.100+0.738SMM in
Fig.~\ref{fig:submm_radio} confirms this hypothesis.

\subsection{Positional association of UC H{\sc ii} regions and methanol masers with sub-mm clumps}

The positional association of ultracompact radio components and sub-mm clumps has already been discussed in 
Sect.~\ref{sect:rlrq} where we identified both radio-loud and radio-quiet clumps; and in Sect.~\ref{sect:rl}
where we showed that ``sub-mm loud'' radio components are strongly consistent with being UC H{\sc ii} regions. The
angular resolution of our SCUBA images not only
identifies whether the UC H{\sc ii} regions in our survey are associated with sub-mm
clumps, but also serves to study the (projected) position of the cm-wave radio emission with respect to the peak of
their associated sub-mm clump. The location of UC H{\sc ii} regions within their embedding clumps potentially allows
us to investigate the characteristics of the birthplaces of the massive OB stars that excite the UC H{\sc ii} regions.
For example, are the massive stars born within dense central cores within the clumps or are their birthsites
offset from the density peak of the clump as suggested by the hierarchical model of Kim \& Koo (\cite{kk01})? 

\begin{figure}
\begin{center}
\includegraphics[angle=-90,scale=0.6]{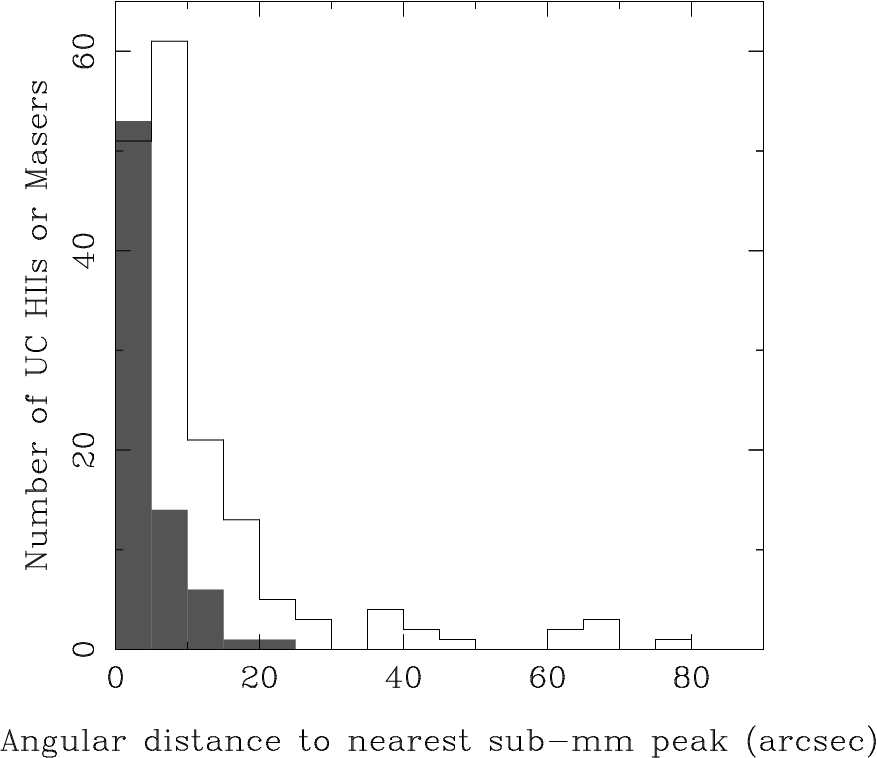}
\caption{Histogram of the angular distance of the UC H{\sc ii} regions to the nearest sub-mm peak (unfilled
histogram) plotted over a histogram of the angular distance of the methanol masers from Walsh et
al.~\cite{walsh03} to their nearest sub-mm
peak (shaded histogram). Note the difference in the two populations in terms of the central peak and the
width of the distribution. }
\label{fig:uchii_disthist}
\end{center}
\end{figure}

In Figure \ref{fig:uchii_disthist} we present a histogram of the angular distance between UC H{\sc ii} regions and the peak position of the closest sub-mm clump. For comparison we also plot
a similar histogram of the angular distance between methanol masers and sub-mm clumps from Walsh et
al.~(\cite{walsh03}), which is shown as a shaded histogram. As in previous sections we have restricted our
analysis to the higher signal-to-noise 850 $\mu$m data.  As can be seen in Figure \ref{fig:uchii_disthist} the
two angular distance distributions are markedly different; methanol masers follow a very tightly peaked
distribution whereas the UC H{\sc ii} regions show a dip at small distances with a much broader overall
distribution.

The tight correlation of masers with sub-mm peaks is evidenced by Walsh et al's detection statistics; all but
one of the 84 masers surveyed was associated with sub-mm emission and 83\% were found within 5\arcsec\ of a
sub-mm peak (Walsh et al.~\cite{walsh03}). Figure \ref{fig:uchii_disthist} shows that the FWHM of the methanol
maser angular distance distribution is $\sim$5\arcsec. Methanol masers are thus predominantly found toward the
peak positions of sub-mm clumps, which suggests that the masers trace deeply embedded star formation at or very
near (at a median distance of 5 kpc, 5\arcsec\ corresponds to a projected linear distance of 0.1 pc) to the
centre of the dense dusty clumps traced by the sub-mm emission. The dip in the ultracompact radio component
distribution shows that the majority of UC H{\sc ii} regions are found approximately twice as far from the sub-mm peak
positions as methanol masers, i.e. UC H{\sc ii} regions are more likely to be offset from the centre of their dense
dusty embedding clumps.The width of the ultracompact radio distribution is also twice as large as that of the
methanol maser distribution.

The observed dip in the UC H{\sc ii} region distribution is small compared to the FWHM of the JCMT
beam, but is twice as large as the typical pointing accuracy that was achieved during the observations
($\le$5\arcsec). We examined our data carefully to exclude systematic errors in the pointing by comparing
histograms for sources that were obtained on different nights. These histograms all displayed a similar
distribution, indicating that systematic pointing effects are not the cause of the observed dip. We also note
that the Walsh et al.~(\cite{walsh03}) data do not display a similar distribution, even though they  were
obtained under similar conditions and with the same instrument. Finally, exactly the same distribution was
observed from independent data: the angular distance between the UC H{\sc ii} positions determined by  Walsh et
al.~(\cite{wbh98}) as part of their ATCA maser survey and the sub-mm peaks observed by Walsh et
al.~(\cite{walsh03}). We thus conclude that the difference between the methanol maser and ultracompact radio
component angular distance distributions is likely to be a real effect. A similar tight correlation between the
positions of methanol masers and their associated sub-mm clumps is reported by Beuther et
al.~(\cite{beuther2002}), who found an average linear separation of
just 0.03 pc between the masers and their associated millimetre continuum peaks

Why are methanol masers and ultracompact H{\sc ii} regions found at predominantly different distances from the peak of
their associated sub-mm clumps? One clue  may lie in the different evolutionary states that are traced by
methanol masers and UC H{\sc ii} regions.  Simple models of maser formation (Walsh et al.~\cite{wbh98}; Codella \&
Moscadelli \cite{cm2000}) suggest that methanol masers are initially formed  in the dense molecular environments
of luminous pre-ZAMS high-mass stars, radiatively pumped by the strong mid-IR emission from the deeply embedded
stars, then are destroyed as an UC H{\sc ii} region forms and expands around the young high-mass star. If this is the
case then methanol masers should trace younger objects than UC H{\sc ii} regions. 

Discounting projection effects, which should be negligible in our large sample, we identify three possibilities
to explain the relative absence of UC H{\sc ii} regions in the centres of sub-mm clumps:

\begin{enumerate}
\item{UC H{\sc ii} regions in the densest regions of sub-mm clumps are quenched for some fraction ($\sim$1/2) of their lifetime, perhaps by
accretion.}                          
\item{UC H{\sc ii} regions are preferentially formed away from the dense central regions of the sub-mm clumps.}                                           
\item{UC H{\sc ii} regions or their direct precursors clear enough gas and dust to destroy the submm                         
peaks in which they formed, shifting the observed SCUBA peak away from the UC H{\sc ii} region.}                                                                             
\end{enumerate}
  
The former two hypotheses explain the absence of UC H{\sc ii} regions from the dense central cores of sub-mm
clumps in terms of reducing the number of observable UC H{\sc ii} regions at high densities; either quenching
shortens the lifetime of the UC H{\sc ii} phase, or  UC H{\sc ii} regions are mostly formed some distance from the
local density peak, as suggested by the Kim \& Koo (\cite{kk01}) hierarchical model. However, neither of these
hypotheses can explain the difference in distribution between methanol masers and ultracompact radio components.
If methanol masers are indeed the precursors to UC H{\sc ii} regions then they should share the same angular
distance distributions. Reducing the number of observed UC H{\sc ii} regions at the sub-mm peak positions does not
resolve the issue of why the overall angular distance distribution of UC H{\sc ii} regions is twice as broad as
that of methanol masers, nor why methanol masers are tightly correlated with the sub-mm peak positions.

The clearing hypothesis may provide a better answer. As UC H{\sc ii} regions expand they begin to clear away their
surrounding gas and dust, ultimately resulting in the eventual destruction of the molecular cloud in which the
high mass star was born. As the UC H{\sc ii} expands in its early stages the first region to be cleared should be
the dense dusty core within which the high mass star was born. These cores are typically $\le$0.1 pc in diameter
(Kurtz et al.~\cite{kcchw00}) and would hence be unresolved by SCUBA. As this core is destroyed the peak position
of the larger clump observed by SCUBA  (which does not resolve the individual cores comprising the clump)
would shift to the next most dense core (i.e.~next highest column density) within the clump, effectively changing
the UC H{\sc ii}/sub-mm peak distance. This effect may be complicated by the fact that the UC H{\sc ii}
region will still be surrounded by a shell of warm dust, but as long as the column density of the next most dense
core is sufficiently higher than that of the shell surrounding the UC H{\sc ii}, then the observed sub-mm peak
position will be offset from that of the UC H{\sc ii}. Hence, the methanol maser phase  of high-mass star formation is
tightly correlated with the peak of the SCUBA sub-mm clump, then the apparent peak of the clump shifts as the UC
H{\sc ii} develops and destroys the dense core within which it is embedded. The width of the dip in the UC H{\sc
ii} region distribution yields an estimate of the clearing radius, i.e.~5\arcsec\ or 0.1 pc at 5 kpc.

The difficulty with this hypothesis rests upon whether the relatively small diameter UC H{\sc ii} regions can clear
dust on these scales. The diameters of UC H{\sc ii} regions measured by WC89 \& KCW94 are typically 2--4\arcsec, but
more recent observational results show that UC H{\sc ii} can in fact possess much larger ionised regions on arcminute
scales (e.g. Kurtz et al.~\cite{kwho99}; Kim \& Koo \cite{kk01}; Ellingsen et al.~\cite{esk2005}). The dust is
more difficult to clear than gas, but the dust sublimation boundary of UC H{\sc ii} regions can be as much as 0.3--0.4
times the ionised gas radius initially and throughout the expansion phase (Inoue \cite{inoue2002}; Arthur et
al.~\cite{arthur2004}). In addition, the analysis of Franco et al.~(\cite{franco2000}) suggests that in
strongly-peaked density distributions ($\ge r^{-3/2}$) the ionisation front of the UC H{\sc ii} region undergoes a
runaway expansion that expands at $\sim$ 0.5 pc/Myr or more. From modelling of clumps associated with UC H{\sc ii}
regions we know that their density distributions are steeper than $r^{-3/2}$ (Hatchell \& van der Tak
\cite{hvdt03}) and so it is likely that the UC H{\sc ii} regions in these steep density distributions rapidly destroy
and clear their dense birthplaces, shifting the observed SCUBA sub-mm  peak to the next highest column
density peak within the clump. 

Corroborating observational evidence to this hypothesis may be found in high-resolution interferometric studies
of UC H{\sc ii} regions and their molecular environment. The edges of UC H{\sc ii} regions are often truncated by the
presence of nearby dense hot molecular cores, e.g.~the well-known UC H{\sc ii} regions G29.96$-$0.02 and G9.62+0.19
(Cesaroni et al.~\cite{cch94}). Confirmation of the small effect uncovered in our SCUBA imaging using higher
resolution millimetre or sub-millimetre interferometry is a priority to determine whether clearing by the UC H{\sc ii}
regions is  indeed responsible.



\subsection{Radio-quiet sub-mm clumps: are they precursors to UC H{\sc ii} regions?}
\label{sect:rq}

We have already shown (in Sect.~\ref{sect:rl}) that the radio-loud clumps are strongly consistent with the
standard model of embedded UC H{\sc ii} regions, but what is the likely nature of the radio-quiet clumps? A cursory
inspection of Figs.~\ref{fig:scuboth} \& \ref{fig:scu850} and Table \ref{tbl:uchiis} does not reveal any
immediate differences between the fluxes or morphology of radio-quiet and radio-loud clumps, with perhaps the
only difference that the radio-quiet clumps possess on average lower peak and integrated sub-mm fluxes that the
radio-louds.  There are six main possibilities for the nature of the radio-quiet clumps:

\renewcommand{\labelenumi}{\emph{\roman{enumi})}}

\begin{enumerate}

\item{the clumps contain embedded UC H{\sc ii} regions that are too compact and/or optically thick to
have been detected in the VLA surveys;}
\item{conversely, the radio-quiet
clumps may contain diffuse H{\sc ii} regions that are extended on scales larger than that to
which the VLA surveys were sensitive;} 
\item{the clumps are not gravitationally
bound and will thus not form stars;} 
\item{the clumps are star-forming, but are
forming low- or intermediate-mass stars that do not produce H{\sc ii} regions rather than OB stars;}
\item{the clumps are
star-forming and have not yet formed massive OB stars;}
\item{ the clumps are
star-forming and contain dusty UC H{\sc ii} regions in which absorption from the dust reduces
the UV photon flux and hence the free-free emission below the detection levels of the VLA
surveys.}
\end{enumerate}
\renewcommand{\labelenumi}{\emph{\alph{enumi})}}

We briefly examine each of these possibilities in turn, to try to ascertain the likely nature
of the radio-quiet clumps.

\paragraph{\emph{i)} Undetected ultracompact H{\sc ii} regions:\\}

With the current data it is not possible to completely rule out scenarios in which the
radio-quiet clumps contain embedded UC H{\sc ii} regions that were not detected in the VLA
surveys of WC89a and KCW94. From the radio flux upper limits of WC89a and KCW94 (typically
0.4--1mJy/beam at 2 cm) we may evaluate a lower limit to the radius of a detectable
optically thick UC H{\sc ii} region using Equations 4 \& 5 from Molinari et
al.~(\cite{mol2000}). Assuming an upper limit to the 2 cm flux of 1 mJy  and a synthesised
beam FWHM of 0\farcs5, we derive a lower limit to the detectable radius of an UC H{\sc ii} of
$r_{\rm u} \ge 6.7 \times 10^{-8}\,\, d$ pc, where $d$ is the distance to the UC H{\sc ii}
region in pc.

Unfortunately, the distances to many of the radio-quiet clumps are highly uncertain, as
there is often no kinematic tracer (radio-recombination line or maser) associated with the
clump. If we assume that the radio-quiet clumps are at the same typical distance as those
clumps containing UC H{\sc ii} regions then from Table \ref{tbl:samplelist}, we expect the upper
bound on the distance to be $\sim$ 14 kpc, at which distance the VLA surveys are only
sensitive to UC H{\sc ii} regions $\ge 9 \times 10^{-4}$ pc. For a more fiducial distance of 5
kpc this upper limit falls to $3 \times 10^{-4}$ pc. In comparison, the initial
Str\"omgren radius for an O6 star surrounded by a pure hydrogen nebula of density $n = 2
\times 10^{7}$ cm$^{-3}$ is $\sim$ 10$^{-3}$ pc (de Pree et al.~\cite{dprg95}).


Assuming all the radio-quiet clumps are populated by UC H{\sc ii} regions at or above their dust-free
Str\" omgen radius, the 2cm upper limits rule out essentially all O~stars and B0 stars in all but
the furthest 20\% of sources.  Later-type B stars are not ruled out: B0.5 stars in all but the 20\%
closest sources, and later-type B stars at all survey distances, would fall below the 2cm flux
limit.  Of course, the time to produce a Str\" omgren sphere is very short and UC H{\sc ii} regions
subsequently expand, making them more observable, and although the expansion is inhibited over that
expected by simple pressure arguments, WC89a see few sources at their initial
Str\" omgren radii.  On the other hand, both dust absorption and higher densities reduce the
detectability by reducing the Str\" omgren radius.  In the case of 90\% UV absorption by dust, we
would also fail to detect O9.5~stars in the more distant half of the sample.   The closer the radio-quiet clump, the higher the probability of
containing no UC H{\sc ii}  or a very young and therefore compact UC H{\sc ii}.  We thus conclude that it is
possible for the radio-quiet clumps to contain late O or early B-type stars. A wider investigation of the 
kinematic distances of the radio-quiet clumps is required to reveal their potential to host massive stars or
undetected UC H{\sc ii} regions.

\paragraph{\emph{ii)} extended H{\sc ii} regions:\\}

Due to the fact that interferometers such as the VLA screen out extended radio emission, the converse that some
H{\sc ii} regions are too extended to be detected is also true. The largest angular scales that the WC89a and
KCW94 surveys were sensitive to is $\sim$ 10\arcsec\ at 2 cm. Because both of these surveys were carried out in
snapshot mode, the sparse $uv$ coverage of the resulting data means that the largest angular scale should be
considered to be an upper limit. H{\sc ii} regions larger than 10\arcsec\ with a smooth surface brightness
distribution would not have been detected, although bright components of clumpy H{\sc ii} regions would possibly
have been detected as a ``cluster'' of UC H{\sc ii} regions.  This effect may be seen in some of the KCW94 UC
H{\sc ii} regions, which merge to form a larger H{\sc ii} region in more compact VLA configuration images (Kurtz
et al.~\cite{kwho99}), or those clustered objects in the Giveon et al.~(\cite{gbh2005}) catalogue which were
discussed in more detail in Sect.~\ref{sect:uchii_nosubmm}. However, it is unlikely that all the radio-quiet
clumps are associated with extended H{\sc ii} regions and in any case these objects inform us as to the prior
massive star formation that has taken place in the region, not deeply embedded star-formation that may be 
currently taking place. These radio-quiet clumps fall into categories \emph{v)} or \emph{vi)} below.

\paragraph{\emph{iii)} Unbound clumps, and \emph{iv)} Low mass star formation:\\}

The Jeans mass for a 50K, $n=10^4 \hbox{ cm}^{-3}$ molecular cloud is about 80~\msun,
which produces a 850 $\mu$m flux of $\sim1\hbox{ Jy}$ at 5~kpc.  On this basis, a few
weak, nearby, radio-quiet clumps could be unbound (about 15\% of the sample), but the
remaining 85\% fall above this limit.

It is more difficult to rule out the presence of low-mass star forming clusters in our sample, particularly for
the less luminous objects. Two examples of low-mass star forming clusters are \object{NGC 1333} and
\object{L1448}, which were
recently observed at 850 $\mu$m as part of a survey of the Perseus molecular cloud (Hatchell et
al.~\cite{perseus}). Both NGC 1333 and L1448 contain several hundred M$_{\odot}$ of dust and gas within a few
pc$^{2}$, and their resulting 850 $\mu$m fluxes are of the order 200--300 Jy. This implies that clusters of this
type could have been detected out to a distance of $\sim$5.5 kpc in our survey (assuming a typical detection
limit of $<$ 1 Jy at 850 $\mu$m). We thus conclude that the weaker radio-quiet clumps detected in our survey may
be forming clusters of low-mass stars. It is difficult to point to specific examples as many of the radio-quiet
clumps do not possess measured kinematic distances. Follow-up observations to determine the distance (and hence
luminosity)  of these objects are a priority.

\paragraph{\emph{iv)} pre-UC H{\sc ii}\ massive protostars and \emph{v)} young massive stars with
dust-quenched UC H{\sc ii}:\\}

Of all of these possibilities the last two are by far the most intriguing, as they imply
that the radio-quiet clumps may contain a population of still-accreting massive stars with
newly-formed UC H{\sc ii}, and the long sought-after massive protostellar precursors to UC H{\sc ii} 
regions.

With the current body of observational data it is not possible to completely determine the nature
of the radio-quiet clumps. A full investigation of the SCUBA-detected clumps near UC H{\sc ii}, using
molecular line observations to probe the temperatures, chemical composition, structure and dynamics
of the cores, will be reported at a later date.


\section{Summary and Conclusions}
\label{sect:concl}

We present the results of a sub-mm continuum imaging survey of UC H{\sc ii} regions, performed with the SCUBA
bolometer array on the JCMT. A total of 105 IRAS sources from the UC H{\sc ii} region catalogues of Wood \& Churchwell
(\cite{wc89a}) and Kurtz, Churchwell \& Wood (\cite{kcw94}) were mapped at 450 \& 850 $\mu$m using SCUBA's
jiggle-mapping mode. We detected 155 sub-mm clumps within the SCUBA images and identify three kinds of object
within our survey: ``sub-mm-quiet objects'' which are ultracompact cm-wave radio sources that are not associated
with sub-mm emission; ``radio-loud'' sub-mm clumps which are associated with ultracompact cm-wave sources; and
``radio-quiet'' sub-mm clumps which are not associated with detectable cm-wave emission. A number of IRAS point
sources (14 in total) were found to exhibit no emission at 850 $\mu$m greater than 0.1--0.3 Jy/beam. We draw the
following conclusions from our survey:

\begin{enumerate}

\item{The sub-mm-quiet ultracompact radio components from the WC89a, KCW94, Becker et al.~(\cite{b90}) and
Giveon et al.~(\cite{gbh2005}) surveys are unlikely to be true embedded UC H{\sc ii} regions unless they are located at
distances greater than 10 kpc. We preclude embedding clump masses in excess of a few solar masses
for objects at distance less than 10 kpc. We suggest that the sub-mm quiet objects are likely either planetary
nebulae or bright knots embedded within larger H{\sc ii} regions.}

\item{The sub-mm clumps as a whole possess similar sub-mm colours and distance-scaled fluxes, which
suggests that they are drawn from similar populations and have similar masses, luminosities and
temperatures. There is a tendency for radio-quiet sub-mm clumps to exhibit lower distance-scaled fluxes
which may indicate that these objects are, on average, of lower luminosity than radio-loud clumps.}

\item{The sub-mm clumps display small elongations with a median value of 1.4, which suggests a predominantly
spherical or low axis ratio population. The ratio of peak to source-integrated flux follows a similar
distribution to the High-Mass Protostellar Object (HMPO) candidates observed by Williams et al.~(\cite{wfs2004}),
which may indicate that the fraction of the mass in the outer regions of the clumps does not evolve significantly
between the HMPO and UC H{\sc ii} region phases. We also find that there is little difference between the
morphology of radio-quiet and radio-loud clumps.}

\item{The clumps are moderately clustered. Out of 105 2\arcmin\ diameter images we detected 155 clumps. The
Companion Clump Fraction (CCF) of the clumps is 0.90$\pm$0.07, which is higher than the value of 0.65 found for
HMPOs by Williams et al.~(\cite{wfs2004}). We suggest that the difference is real, and the selection bias against
nearby UC H{\sc ii}s in the Williams et al. study results in a lower clump companion fraction.}

\item{The mean surface density of clump companions (MSDC) follows a broken power-law distribution, with a break
at a clump separation of 3 pc. This may reflect a natural size scale for the extent of massive star-forming
clusters in molecular clouds, although our sample of objects at a wide range of heliocentric distances and
flux-limited data could introduce selection biases against distant or faint companions. Further wide-field
observations of a more uniform distance selected sample are required to address this issue more rigorously.}

\item{Radio-loud clumps are shown to be strongly consistent with the standard model of embedded UC H{\sc ii} regions
(e.g.~Crowther \& Conti \cite{cc2003}; Churchwell \cite{c2002}). The Lyman continuum fluxes are in excellent
agreement with the 850 $\mu$m to cm-wave flux ratio, showing that the YSOs exciting the UC H{\sc ii} regions are
responsible for heating the sub-mm clumps. We also identify one object within the survey whose 850 $\mu$m flux
is inconsistent with the standard model (G10.100+0.738SMM) and suggest that this sub-mm emission from this
object originates from the planetary nebula \object{NGC 6537}.}

\item{We show that UC H{\sc ii} regions are predominantly found at an angular distance of $\geq$10\arcsec\ from the
sub-mm peak of their associated sub-mm clump. The distribution of these angular distances is clearly different
to that of methanol masers and their associated sub-mm clumps derived from the data of Walsh et
al.~(\cite{walsh03}). We suggest that the most likely hypothesis to explain the differences in the two
distributions is that UC H{\sc ii} regions efficiently clear their immediate surroundings of dense material, 
shifting the sub-mm peak observed by SCUBA to the next highest column density peak within the clump. We estimate
a value for the clearing radius of $\sim$0.1 pc, for a median distance in our sample of 5 kpc.}

\item{We speculate upon the nature of the radio-quiet clumps and conclude that they are unlikely to contain
embedded O-stars. We cannot rule out the presence of embedded B-stars, particularly in the furthest objects.
Only the closest 15\% of radio-quiet clumps possess sub-mm fluxes that imply they are either gravitationally
unbound or likely to be forming only low mass stars, and hence we conclude that the majority of the sample are
massive and luminous enough to form intermediate or high-mass stars. We do not have sufficient data to conclude
that the radio-quiet clumps are in a pre-UC H{\sc ii} region phase, but we argue that their characteristics are
suggestive of such a stage. We will present a full investigation of these clumps at a later date.}

\end{enumerate}

\begin{acknowledgements}

The authors would like to thank all of the anonymous visiting observers and JCMT support staff who undertook the
flexibly scheduled observations. We would also like to thank the referee for the many useful suggestions and
careful reading of the paper. This research made use of data products from the Midcourse Space Experiment obtained
from the NASA/ IPAC Infrared Science Archive, which is operated by the Jet Propulsion Laboratory, California
Institute of Technology, under contract with the National Aeronautics and Space Administration.. MSX data
processing was funded by the Ballistic Missile Defense Organization with additional support from NASA Office of
Space Science.  This research would not have been possible without the SIMBAD astronomical database service
operated at CCDS, Strasbourg, France  and the NASA Astrophysics Data System Bibliographic Services.  JH is
supported at Exeter by a PPARC AF.

\end{acknowledgements}

\clearpage
\onecolumn
\renewcommand{\thefootnote}{\alph{footnote}} 
\setcounter{table}{1} 
\begin{longtable}{llllcccc}
\caption{Identifications, coordinates, peak and integrated fluxes of each submillimetre source
detected in the survey. Multiple sources located in a single jiggle-map  are
indicated by labels following the UC H{\sc ii} region field name. These labels are the same as those
used in Figs.~\ref{fig:scuboth} \& \ref{fig:scu850}. 450 $\mu$m peak flux measurements indicated
by a dagger ($^{\dag}$) are measured from smoothed images and the flux density unit in this case
is Janskys per 14\arcsec\ beam. The resolution of the unsmoothed images is 8\arcsec\ at 450 $\mu$m
and 14\arcsec\ at 850 $\mu$m. Upper limits for non-detections at 450 $\mu$m are indicated.  Where
it was not possible to measure a 450 $\mu$m flux, either due to the smaller FOV of the short
wavelength array or calibration problems this is indicated by an ellipsis (\ldots). Quoted errors
include calibration uncertainties of 30\% at 450 $\mu$m and 10\% at 850 $\mu$m. All source
positions were measured from the 850 $\mu$m images unless otherwise stated.} \label{tbl:uchiis} \\

\hline
UCH{\sc ii} field & Source & RA (J2000) & Dec (J2000) & \multicolumn{2}{c}{Peak flux (Jy/beam)}& \multicolumn{2}{c}{Integrated flux density (Jy)}\\
 & & (hr min sec) & (\degr\,\,\arcmin\,\,\arcsec) & 450 $\mu$m & 850 $\mu$m & 450 $\mu$m & 850 $\mu$m\\\hline
\endfirsthead

\caption{continued.}\\
\hline\hline
UCH{\sc ii} field & Source & RA (J2000) & Dec (J2000) & \multicolumn{2}{c}{Peak flux (Jy/beam)}& \multicolumn{2}{c}{Integrated flux density (Jy)}\\
 & & (hr min sec) & (\degr\,\,\arcmin\,\,\arcsec) & 450 $\mu$m & 850 $\mu$m & 450 $\mu$m & 850 $\mu$m\\\hline
\endhead

\hline
\endfoot

G1.13$-$0.11   & G1.125$-$0.108SMM & 17:48:41.7 & $-$28:01:48 & $26.3\pm$ 9.0$^{\dag}$ & $2.8\pm$ 0.4& $174.3\pm 55.2$  & $13.4\pm 2.8$ \\
G4.41+0.13     & G4.417+0.127SMM\setcounter{footnote}{0}\footnotemark & 17:55:18.0 & $-$25:04:41 & $<8.1$$^{\dag}$  & $1.2\pm$ 0.1& $<8.1$ & $2.0\pm 0.3$ \\
G5.48$-$0.24 (a)   & G5.477$-$0.246SMM & 17:59:03.5 & $-$24:20:55 & $<9.7$ & $0.7\pm$ 0.1& $<9.7$ & $4.4\pm 0.5$ \\
G5.48$-$0.24 (b)  & G5.479$-$0.254SMM & 17:59:05.5 & $-$24:21:03 & $<9.7$ & $0.6\pm$ 0.1& $<9.7$ & $3.0\pm 0.4$ \\
G5.89$-$0.39   & G5.887$-$0.394SMM & 18:00:31.0 & $-$24:04:01 & $55.7\pm$ 17.1& $19.4\pm$ 2.0& $250.6\pm 80.2$ & $46.5\pm 5.3$ \\
G5.97$-$1.17 (a)  & G5.972$-$1.159SMM & 18:03:37.1 & $-$24:22:13 & \ldots & $1.3\pm$ 0.2& \ldots & $4.9\pm 0.8$ \\
G5.97$-$1.17 (b)  & G5.974$-$1.175SMM & 18:03:41.1 & $-$24:22:37 & $20.8\pm$ 6.7$^{\dag}$ & $3.6\pm$ 0.4& $69.7\pm 21.7$ & $16.4\pm 2.5$ \\
G6.55$-$0.10   & G6.551$-$0.099SMM & 18:00:50.3 & $-$23:20:36 & $<7.8$ & $1.8\pm$ 0.3& $<7.8$ & $11.1\pm 2.1$ \\
G8.14+0.23 (a)    & G8.143+0.219SMM\setcounter{footnote}{1}\footnotemark & 18:03:02.9 & $-$21:48:08 & $25.8\pm$ 7.9$^{\dag}$ & $4.2\pm$ 0.4& $91.8\pm 27.8$ & $26.4\pm 2.9$ \\
G8.14+0.23 (b)    & G8.140+0.224SMM\setcounter{footnote}{1}\footnotemark & 18:03:01.5 & $-$21:48:10 & $23.6\pm$ 7.2$^{\dag}$ & \ldots & $152.5\pm 46.0$ & \ldots \\
G8.14+0.23 (c)    & G8.135+0.245SMM & 18:02:56.0 & $-$21:47:47 & $13.1\pm$ 4.2$^{\dag}$  & $1.7\pm$ 0.2& $73.7\pm 22.2$ & $7.0\pm 0.8$ \\
G8.14+0.23 (d)    & G8.136+0.212SMM & 18:03:03.5 & $-$21:48:42 & \ldots & $1.1\pm$ 0.1& \ldots & $2.4\pm 0.3$ \\
G8.67$-$0.36 (a)  & G8.683$-$0.368SMM\setcounter{footnote}{4}\footnotemark & 18:06:23.4 & $-$21:37:09 & \ldots & $4.1\pm$ 0.5& \ldots & $11.9\pm 2.3$ \\
G8.67$-$0.36 (b)  & G8.671$-$0.357SMM & 18:06:19.4 & $-$21:37:27 & $54.5\pm$ 16.6& $12.2\pm$ 1.3& $291.3\pm 92.3$ & $32.3\pm 5.0$ \\
G9.62+0.20     & G9.621+0.194SMM & 18:06:15.1 & $-$20:31:34 & $38.4\pm$ 11.8& $9.3\pm$ 1.0& $168.1\pm 53.3$ & $21.7\pm 3.8$ \\
G9.88$-$0.75 (a)  & G9.880$-$0.751SMM & 18:10:19.2 & $-$20:45:31 & $14.4\pm$ 4.7$^{\dag}$ & $1.9\pm$ 0.2& $104.3\pm 35.3$ & $13.8\pm 1.9$ \\
G9.88$-$0.75 (b)  & G9.872$-$0.750SMM & 18:10:18.0 & $-$20:45:57 & $11.8\pm$ 4.0$^{\dag}$ & $1.5\pm$ 0.2& $91.2\pm 28.7$ & $8.1\pm 1.1$ \\
G10.10+0.74    & G10.100+0.738SMM   & 18:05:13.4 & $-$19:50:34 & $<5.1$ & $0.8\pm$ 0.1& $<5.1$ & $1.0\pm 0.1$ \\
G10.15$-$0.34  & G10.152$-$0.345SMM & 18:09:21.8 & $-$20:19:30 & $17.0\pm 6.1$$^{\dag}$ & $2.2\pm 0.4$& $83.1\pm 28.0$  & $10.6\pm 2.4$ \\
G10.30$-$0.15 (a) & G10.299$-$0.149SMM & 18:08:56.2 & $-$20:06:02 & $28.5\pm 8.7$& $6.3\pm 0.7$& $254.0\pm 78.3$ & $27.9\pm 3.5$ \\
G10.30$-$0.15 (b) & G10.291$-$0.138SMM & 18:08:52.7 & $-$20:06:09 & $12.7\pm$ 4.1& $1.9\pm$ 0.3& $43.3\pm 13.3$ & $4.4\pm 0.7$ \\
G10.62$-$0.38  & G10.624$-$0.385SMM & 18:10:29.0 & $-$19:55:51 & $120.3\pm$ 36.2& $27.5\pm$ 2.8& $730.5\pm 256.9$ & $65.3\pm 9.1$ \\
G10.84$-$2.59  & G10.842$-$2.594SMM & 18:19:12.7 & $-$20:47:31 & $24.2\pm$ 7.5& $5.6\pm$ 0.6& $96.6\pm 30.7$ & $14.8\pm 3.3$ \\
G11.11$-$0.34 (a) & G11.110$-$0.401SMM & 18:11:32.4 & $-$19:30:44 & $16.2\pm$ 5.1& $2.7\pm$ 0.3& $92.1\pm 28.2$ & $15.8\pm 3.3$ \\
G11.11$-$0.34 (b) & G11.117$-$0.415SMM & 18:11:36.2 & $-$19:30:48 & \ldots & $0.9\pm$ 0.2& \ldots & $4.9\pm 0.7$ \\
G11.94$-$0.62 (a) & G11.936$-$0.618SMM & 18:14:01.4 & $-$18:53:29 & $29.9\pm$ 9.1& $5.8\pm$ 0.6& $337.0\pm 106.5$ & $20.1\pm 2.8$ \\
G11.94$-$0.62 (b) & G11.918$-$0.615SMM & 18:13:58.7 & $-$18:54:23 & \ldots & $3.9\pm$ 0.4& \ldots & $10.9\pm 1.6$ \\
G12.43$-$0.05 (a) & G12.429$-$0.050SMM & 18:12:55.0 & $-$18:11:14 & $<3.9$$^{\dag}$  & $0.4\pm$ 0.1& $<3.9$  & $2.4\pm 0.3$ \\
G12.43$-$0.05 (b) & G12.439$-$0.060SMM & 18:12:58.5 & $-$18:10:58 & $<3.9$$^{\dag}$  & $0.4\pm$ 0.1& $<3.9$  & $2.5\pm 0.3$ \\
G12.43$-$0.05 (c) & G12.437$-$0.055SMM & 18:12:57.0 & $-$18:10:58 & $<3.9$$^{\dag}$  & $0.4\pm$ 0.1& $<3.9$  & $3.9\pm 0.4$ \\
G12.90$-$0.25 (a) & G12.908$-$0.261SMM & 18:14:39.6 & $-$17:52:04 & $53.5\pm$ 16.1$^{\dag}$ & $7.0\pm$ 0.8& $254.3\pm 80.3$  & $22.1\pm 3.5$ \\
G12.90$-$0.25 (b) & G12.900$-$0.243SMM & 18:14:34.6 & $-$17:51:56 & $13.3\pm$ 4.3$^{\dag}$ & $1.5\pm$ 0.3& $53.3\pm 17.2$ & $3.8\pm 0.7$ \\
G13.19+0.04   (a) & G13.177+0.059SMM & 18:14:01.2 & $-$17:28:43 & \ldots & $3.7\pm$ 0.5& \ldots & $11.0\pm 2.3$ \\
G13.19+0.04   (b) & G13.186+0.028SMM & 18:14:09.2 & $-$17:29:06 & \ldots & $0.7\pm$ 0.3& \ldots & $1.8\pm 0.3$ \\
G13.87+0.28    & G13.873+0.279SMM & 18:14:36.3 & $-$16:45:44 & $16.3\pm$ 5.2& $3.8\pm$ 0.5& $100.1\pm 33.5$ & $16.7\pm 4.9$ \\
G15.04$-$0.68 (a) & G15.026-0.674SMM\setcounter{footnote}{1}\footnotemark & 18:20:23.1 & $-$16:11:58 & $33.1\pm$ 10.4& $12.0\pm$ 1.5& $416.9\pm 127.0$ & $65.2\pm 9.0$ \\
G15.04$-$0.68 (b) & G15.031-0.676SMM\setcounter{footnote}{1}\footnotemark & 18:20:24.1 & $-$16:11:43 & $33.9\pm$ 10.6& \ldots & $139.4\pm 43.9$ & \ldots \\
G15.04$-$0.68 (c) & G15.031-0.669SMM\setcounter{footnote}{1}\footnotemark & 18:20:22.7 & $-$16:11:32 & $45.6\pm$ 14.0& \ldots & $343.5\pm 105.3$ & \ldots \\
G18.15$-$0.28  & G18.149$-$0.286SMM & 18:25:01.8 & $-$13:15:35 & $17.3\pm$ 5.4$^{\dag}$ & $2.0\pm$ 0.3& $118.7\pm 40.7$ & $4.7\pm 1.1$ \\
G18.30$-$0.39 (a) & G18.302$-$0.286SMM\setcounter{footnote}{1}\footnotemark & 18:25:41.8 & $-$13:10:19 & $17.3\pm$ 5.4$^{\dag}$ & $2.4\pm$ 0.4& $111.2\pm 34.0$ & $16.7\pm 0.3$ \\
G18.30$-$0.39 (b) & G18.298$-$0.392SMM\setcounter{footnote}{1}\footnotemark & 18:25:42.2 & $-$13:10:37 & $20.5\pm$ 6.4$^{\dag}$ & \ldots & $143.9\pm 1.1$ & \ldots \\   
G19.07$-$0.27  & G19.077$-$0.289SMM & 18:26:49.0 & $-$12:26:25 & $21.7\pm$ 6.6& $4.1\pm$ 0.4& $110.6\pm 35.9$ & $13.1\pm 2.8$ \\
G19.49+0.14   (a) & G19.487+0.136SMM & 18:26:03.7 & $-$11:52:45 & $4.6\pm$ 1.8$^{\dag}$ & $0.7\pm$ 0.2& $13.3\pm 5.0$ & $1.8\pm 0.6$ \\
G19.49+0.14   (b) & G19.474+0.158SMM & 18:25:57.6 & $-$11:52:49 & $4.0\pm$ 1.7$^{\dag}$ & $0.9\pm$ 0.2& $7.6\pm 2.4$ & $2.1\pm 0.5$ \\
G19.49+0.14   (c) & G19.486+0.148SMM & 18:26:01.1 & $-$11:52:29 & $9.4\pm$ 3.1$^{\dag}$ & $1.0\pm$ 0.2& $24.2\pm 9.1$ & $2.8\pm 0.7$ \\ 
G19.61$-$0.23  & G19.609$-$0.235SMM & 18:27:38.2 & $-$11:56:38 & $84.6\pm$ 25.4& $19.8\pm$ 2.0& $191.7\pm 59.4$ & $26.2\pm 3.0$ \\
G20.08$-$0.14 (a) & G20.081$-$0.136SMM & 18:28:10.5 & $-$11:28:50 & $46.3\pm$ 13.9& $8.2\pm$ 0.8& $77.5\pm 23.9$ & $13.8\pm 1.6$ \\
G20.08$-$0.14 (b) & G20.073$-$0.143SMM & 18:28:11.0 & $-$11:29:26 & $<3.0$ & $0.4\pm$ 0.1& $<3.0$ & $0.8\pm 0.1$ \\
G20.99+0.09   (a) & G20.990+0.089SMM & 18:29:05.0 & $-$10:34:13 & $<2.7$$^{\dag}$  & $0.6\pm$ 0.1& $<2.7$ & $3.7\pm 0.4$ \\
G20.99+0.09   (b) & G20.981+0.094SMM & 18:29:02.8 & $-$10:34:36 & $7.3\pm$ 2.4$^{\dag}$ & $1.2\pm$ 0.1& $25.4\pm 8.7$ & $7.2\pm 1.0$ \\
G22.76$-$0.49  & G22.756$-$0.485SMM & 18:34:28.0 & $-$09:16:12 & $<2.4$ & $0.4\pm$ 0.1& $<2.4$ & $1.6\pm 0.3$ \\
G23.46$-$0.20  & G23.454$-$0.203SMM & 18:34:45.1 & $-$08:31:15 & $10.8\pm$ 3.4$^{\dag}$ & $1.2\pm$ 0.2& $14.0\pm 4.3$ & $1.3\pm 0.2$ \\
G23.71+0.17    & G23.708+0.169SMM & 18:33:53.5 & $-$08:07:25 & $14.7\pm$ 4.5& $2.9\pm$ 0.3& $56.2\pm 17.3$ & $15.1\pm 2.9$ \\
G23.87$-$0.12 (a) & G23.866$-$0.114SMM & 18:35:12.1 & $-$08:06:49 & $10.6\pm$ 3.6$^{\dag}$ & $0.9\pm$ 0.1& $42\pm 12.7$ & $6.0\pm 1.0$ \\
G23.87$-$0.12 (b) & G23.872$-$0.122SMM\setcounter{footnote}{2}\footnotemark & 18:35:14.6 & $-$08:06:43 & $12.1\pm$ 4.0$^{\dag}$ & $1.1\pm$ 0.2& $78.7\pm 23.9$ & $5.2\pm 0.7$ \\
G23.87$-$0.12 (c) & G23.866$-$0.124SMM\setcounter{footnote}{2}\footnotemark & 18:35:14.3 & $-$08:07:07 & \ldots & $1.4\pm$ 0.2& \ldots & $4.7\pm 0.7$ \\
G23.87$-$0.12 (d) & G23.861$-$0.125SMM & 18:35:14.0 & $-$08:07:24 & $11.3\pm$ 3.7$^{\dag}$ & $1.3\pm$ 0.2& $28.9\pm 9.2$ & $2.8\pm 0.4$ \\
G23.96+0.15   (a) & G23.944+0.159SMM & 18:34:21.9 & $-$07:55:08 & $12.1\pm$ 4.0$^{\dag}$ & $1.2\pm$ 0.2& $38.7\pm 11.7$ & $2.5\pm 0.4$ \\
G23.96+0.15   (b) & G23.955+0.148SMM & 18:34:25.5 & $-$07:54:52 & $22.1\pm$ 6.8$^{\dag}$ &  $2.0\pm$ 0.2& $52.5\pm 16.2$ & $6.9\pm 0.8$ \\
G23.96+0.15   (c) & G23.952+0.152SMM & 18:34:24.3 & $-$07:54:55 & $23.5\pm$ 7.2$^{\dag}$ & $2.3\pm$ 0.3& $87.1\pm 26.5$ & $7.7\pm 1.0$ \\
G23.96+0.15   (d) & G23.949+0.163SMM\setcounter{footnote}{0}\footnotemark & 18:34:21.7 & $-$07:54:47 & $11.3\pm$ 3.7$^{\dag}$ & $1.2\pm$ 0.2& $58.3\pm 17.5$ & $2.2\pm 0.3$ \\
G24.47+0.49   (a) & G24.469+0.495SMM\setcounter{footnote}{0}\footnotemark & 18:34:08.4 & $-$07:17:52 & $14.9\pm$ 4.7$^{\dag}$ & $1.2\pm$ 0.2& $115.1\pm 36.1$ & $5.7\pm 0.7$ \\
G24.47+0.49   (b) & G24.474+0.487SMM & 18:34:10.7 & $-$07:17:52 & $14.5\pm$ 4.6$^{\dag}$ & $1.7\pm$ 0.2& $65.0\pm 20.2$ & $12.4\pm 1.6$ \\
G24.47+0.49   (c) & G24.476+0.492SMM\setcounter{footnote}{0}\footnotemark & 18:34:09.7 & $-$07:17:37 & $18.2\pm$ 5.7$^{\dag}$ & $1.4\pm$ 0.2& $112.0\pm 36.2$ & $3.7\pm 0.4$ \\
G24.68$-$0.16 (a) & G24.674$-$0.152SMM & 18:36:50.3 & $-$07:24:50 & $46.2\pm$ 14.0$^{\dag}$ & $4.0\pm$ 0.4& $275.8\pm 89.6$ & $17.4\pm 2.4$ \\
G24.68$-$0.16 (b) & G24.668$-$0.161SMM & 18:36:51.6 & $-$07:25:24 & $14.0\pm$ 4.6$^{\dag}$ & $1.1\pm$ 0.2& $86.5\pm 27.7$ & $9.0\pm 1.1$ \\
G24.68$-$0.16 (c) & G24.683$-$0.150SMM & 18:36:50.9 & $-$07:24:18 & $13.1\pm$ 4.3$^{\dag}$ & $0.9\pm$ 0.2& $20.4\pm 6.3$ & $3.0\pm 0.4$ \\
G25.38$-$0.18  & G25.382$-$0.184SMM & 18:38:15.7 & $-$06:47:58 & \ldots & $2.7\pm$ 0.3& \ldots & $12.4\pm 1.7$ \\
G25.65+1.05    & G25.650+1.049SMM & 18:34:21.1 & $-$05:59:43 & $45.8\pm$ 13.8& $8.6\pm$ 0.9& $258.8\pm 84.5$ & $28.7\pm 3.9$ \\
G25.72+0.05  (a)  & G25.710+0.043SMM & 18:38:03.4 & $-$06:24:16 & $21.9\pm$ 6.8$^{\dag}$ & $2.2\pm$ 0.3& $85.4\pm 29.0$ & $7.4\pm 1.8$ \\
G25.72+0.05  (b)  & G25.714+0.046SMM & 18:38:03.2 & $-$06:23:56 & $12.7\pm$ 4.1$^{\dag}$ & $1.0\pm$ 0.2& $64.7\pm 21.4$ & $2.9\pm 0.4$ \\
G26.54+0.42  (a)  & G26.529+0.422SMM & 18:38:12.8 & $-$05:30:10 & $<4.8$$^{\dag}$  & $0.5\pm$ 0.2& $<4.8$ & $1.6\pm 0.2$ \\
G26.54+0.42  (b)  & G26.545+0.413SMM\setcounter{footnote}{2}\footnotemark & 18:38:16.5 & $-$05:29:34 & $8.5\pm$ 3.0$^{\dag}$ & $1.0\pm$ 0.2& $41.9\pm 12.7$ & $6.5\pm 1.5$ \\
G26.54+0.42  (c)  & G26.548+0.422SMM\setcounter{footnote}{2}\footnotemark & 18:38:15.0 & $-$05:29:08 & \ldots & $0.7\pm$ 0.2& \ldots & $4.0\pm 0.7$ \\
G27.28+0.15  (a)  & G27.267+0.146SMM & 18:40:33.4 & $-$04:58:22 & $7.8\pm$ 2.8$^{\dag}$ & $0.6\pm$ 0.2& $42.6\pm 13.4$ & $5.8\pm 0.8$ \\
G27.28+0.15  (b)  & G27.280+0.144SMM & 18:40:35.2 & $-$04:57:46 & $13.1\pm$ 4.2$^{\dag}$ & $1.7\pm$ 0.2& $33.4\pm 11.1$ & $5.9\pm 0.8$ \\
G27.28+0.15  (c)  & G27.285+0.149SMM & 18:40:34.9 & $-$04:57:20 & $13.8\pm$ 4.4$^{\dag}$ & $1.7\pm$ 0.2& $45.5\pm 15.2$ & $8.2\pm 1.2$ \\
G27.28+0.15  (d)  & G27.296+0.151SMM & 18:40:35.5 & $-$04:56:41 & $8.2\pm$ 2.9$^{\dag}$ & $1.0\pm$ 0.2& $23.2\pm 7.4$ & $3.9\pm 0.7$ \\
G27.49+0.19  (a)  & G27.494+0.187SMM & 18:40:49.7 & $-$04:45:10 & $11.4\pm$ 3.8$^{\dag}$ & $1.2\pm$ 0.2& $24.5\pm 8.4$ & $4.6\pm 1.1$ \\
G27.49+0.19  (b)  & G27.504+0.193SMM & 18:40:49.4 & $-$04:44:28 & $<3.6$$^{\dag}$ & $0.6\pm$ 0.2& $<3.6$ & $2.5\pm 0.5$ \\
G28.20$-$0.05  & G28.200$-$0.051SMM & 18:42:58.4 & $-$04:14:01 & $61.6\pm$ 18.5& $9.4\pm$ 0.9& $244.9\pm 77.8$ & $19.6\pm 2.6$ \\
G28.29$-$0.36 (a) & G28.305$-$0.388SMM\setcounter{footnote}{0}\footnotemark & 18:44:22.2 & $-$04:17:39 & $13.6\pm$ 4.3$^{\dag}$& $1.1\pm$ 0.1& $44.6\pm 14.1$ & $4.2\pm 0.6$ \\
G28.29$-$0.36 (b) & G28.301$-$0.384SMM\setcounter{footnote}{0}\footnotemark & 18:44:20.7 & $-$04:17:44 & $14.4\pm$ 4.5$^{\dag}$& $1.4\pm$ 0.2& $48.9\pm 15.9$ & $4.1\pm 0.6$ \\
G28.29$-$0.36 (c) & G28.288$-$0.365SMM\setcounter{footnote}{2}\footnotemark & 18:44:15.3 & $-$04:17:55 & $19.5\pm$ 6.0$^{\dag}$& $2.2\pm$ 0.2& $126.3\pm 41.7$ & $11.8\pm 1.7$ \\
G28.29$-$0.36 (d) & G28.283$-$0.363SMM\setcounter{footnote}{2}\footnotemark & 18:44:14.3 & $-$04:18:07 & \ldots & $1.8\pm$ 0.2& \ldots & $5.4\pm 0.7$ \\
G28.60$-$0.36 (a) & G28.596$-$0.362SMM & 18:44:48.4 & $-$04:01:24 & $<5.1$ & $0.6\pm$ 0.1& $<5.1$ & $5.2\pm 0.9$ \\
G28.60$-$0.36 (b) & G28.603$-$0.374SMM\setcounter{footnote}{0}\footnotemark & 18:44:51.8 & $-$04:01:21 & $<5.1$ & $0.7\pm$ 0.1& $<5.1$ & $3.5\pm 0.4$ \\
G28.60$-$0.36 (c) & G28.602$-$0.380SMM & 18:44:53.0 & $-$04:01:36 & $<5.1$ & $0.5\pm$ 0.1& $<5.1$ & $2.2\pm 0.3$ \\
G28.80+0.17   (a) & G28.806+0.180SMM & 18:43:15.7 & $-$03:35:22 & \ldots & $1.1\pm$ 0.1& \ldots & $7.6\pm 1.1$ \\
G28.80+0.17   (b) & G28.812+0.168SMM & 18:43:18.8 & $-$03:35:19 & \ldots & $1.6\pm$ 0.2& \ldots & $6.3\pm 0.8$ \\
G28.83$-$0.23  & G28.834$-$0.209SMM & 18:44:42.0 & $-$03:44:30 & $<5.1$ & $1.0\pm$ 0.2& $<5.1$ & $2.0\pm 0.5$ \\
G29.96$-$0.02 (a) & G29.956$-$0.017SMM & 18:46:04.0 & $-$02:39:22 & $60.0\pm$ 18.0& $11.5\pm$ 1.2& $200.4\pm 62.1$ & $19.2\pm 2.3$ \\
G29.96$-$0.02 (b) & G29.938$-$0.012SMM & 18:46:01.0 & $-$02:40:12 & $3.6\pm$ 1.3& $0.6\pm$ 0.2& $9.6\pm 3.1$ & $1.4\pm 0.3$ \\
G29.96$-$0.02 (c) & G29.972$-$0.011SMM & 18:46:04.4 & $-$02:38:20 & $4.8\pm$ 1.6& $0.7\pm$ 0.2& $15.0\pm 4.9$ & $1.9\pm 0.5$ \\
G29.96$-$0.02 (d) & G29.964$-$0.011SMM & 18:46:03.7 & $-$02:38:48 & $8.3\pm$ 2.6& $1.1\pm$ 0.2& $20.4\pm 7.0$ & $2.1\pm 0.6$ \\
G30.54+0.02    & G30.535+0.020SMM & 18:46:59.5 & $-$02:07:26 & $18.9\pm$ 5.7$^{\dag}$& $2.0\pm$ 0.3& $75.0\pm 24.9$ & $5.3\pm 1.2$ \\
G30.78$-$0.02 (a) & G30.785$-$0.022SMM & 18:47:35.9 & $-$01:55:13 & $31.5\pm$ 9.9$^{\dag}$& $2.9\pm$ 0.4& $130.8\pm 45.1$ & $9.6\pm 1.7$ \\
G30.78$-$0.02 (b) & G30.786$-$0.028SMM\setcounter{footnote}{0}\footnotemark & 18:47:37.4 & $-$01:55:22 & $19.4\pm$ 6.5$^{\dag}$ & $2.1\pm$ 0.3& $106.6\pm 35.7$ & $7.1\pm 1.4$ \\
G31.28+0.06    & G31.280+0.061SMM & 18:48:12.5 & $-$01:26:31 & $53.9\pm$ 16.4& $6.6\pm$ 0.7& $224.5\pm 73.1$ & $24.9\pm 6.6$ \\
G31.40$-$0.26 (a) & G31.396$-$0.258SMM & 18:49:33.2 & $-$01:29:05 & $52.6\pm$ 15.9& $6.5\pm$ 0.7& $160.4\pm 51.1$ & $10.4\pm 1.4$ \\
G31.40$-$0.26 (b) & G31.387$-$0.269SMM & 18:49:34.7 & $-$01:29:52 & $<5.1$ & $0.7\pm$ 0.1& $<5.1$ & $1.9\pm 0.5$ \\
G32.80+0.19    & G32.797+0.190SMM & 18:50:30.8 & +00:02:01 & $77.2\pm$ 23.3& $12.0\pm$ 1.2& $314.0\pm 99.2$ & $23.8\pm 3.1$ \\
G33.13$-$0.09  & G33.132$-$0.094SMM & 18:52:08.2 & +00:08:09 & $37.0\pm$ 11.3& $4.8\pm$ 0.5& $117.9\pm 37.6$ & $13.7\pm 2.1$ \\
G33.92+0.11    & G33.914+0.109SMM & 18:52:50.4 & +00:55:27 & $42.6\pm$ 12.9& $6.1\pm$ 0.6& $226.0\pm 68.7$ & $21.0\pm 2.5$ \\
G34.26+0.15  (a)  & G34.257+0.152SMM & 18:53:18.8 & +01:14:56 & $226.9\pm$ 68.1& $55.7\pm$ 5.6& $1464.2\pm 454.3$ & $189.4\pm 19.1$ \\
G34.26+0.15  (b)  & G34.244+0.133SMM & 18:53:21.5 & +01:13:42 & \ldots & $2.7\pm$ 0.3& \ldots & $1.5\pm 0.2$ \\
G34.26+0.15  (c)  & G34.257+0.164SMM & 18:53:16.2 & +01:15:14 & $33.6\pm$ 10.3& $7.4\pm$ 0.7& $255.3\pm 78.8$ & $24.4\pm 2.5$ \\
G35.02+0.35    & G35.024+0.349SMM & 18:54:00.6 & +02:01:15 & $49.9\pm$ 15.1& $6.3\pm$ 0.6& $128.3\pm 41.0$ & $10.0\pm 1.2$ \\
G35.05$-$0.52 (a) & G35.039$-$0.505SMM & 18:57:04.9 & +01:38:43 & \ldots & $0.5\pm$ 0.1& \ldots & $1.2\pm 0.1$ \\
G35.05$-$0.52 (b) &G35.053$-$0.518SMM & 18:57:09.3 & +01:39:04 & \ldots & $1.1\pm$ 0.2& \ldots & $3.0\pm 0.5$ \\
G35.05$-$0.52 (c) & G35.049$-$0.513SMM & 18:57:07.6 & +01:39:00 & \ldots & $0.6\pm$ 0.1& \ldots & $2.8\pm 0.4$ \\
G35.57+0.07    & G35.576+0.067SMM & 18:56:01.4 & +02:23:00 & \ldots & $1.6\pm$ 0.2& \ldots & $10.2\pm 1.5$ \\
G35.58$-$0.03  & G35.578$-$0.031SMM & 18:56:22.7 & +02:20:26 & \ldots & $4.9\pm$ 0.5& \ldots & $9.9\pm 1.6$ \\
G37.55$-$0.11  & G37.545$-$0.113SMM & 19:00:16.3 & +04:03:11 & \ldots & $2.2\pm$ 0.3& \ldots & $7.4\pm 1.8$ \\
G37.77$-$0.20  & G37.764$-$0.216SMM & 19:01:02.3 & +04:12:03 & \ldots & $2.0\pm$ 0.2& \ldots & $13.7\pm 1.8$ \\
G37.87$-$0.40  & G37.873$-$0.400SMM & 19:01:53.7 & +04:12:47 & \ldots & $4.7\pm$ 0.5& \ldots & $15.1\pm 2.6$ \\
G39.25$-$0.06  & G39.254$-$0.060SMM & 19:03:13.2 & +05:35:47 & \ldots & $1.4\pm$ 0.2& \ldots & $6.1\pm 0.9$ \\
G41.74+0.10    & G41.743+0.097SMM & 19:07:15.7 & +07:52:47 & \ldots & $0.9\pm$ 0.2& \ldots & $2.2\pm 0.6$ \\
G42.42$-$0.27 (a) & G42.421$-$0.261SMM & 19:09:48.6 & +08:19:00 & $6.2\pm$ 2.4$^{\dag}$ & $0.6\pm$ 0.1& $73.3\pm 23.7$ & $4.3\pm 1.3$ \\
G42.42$-$0.27 (b) & G42.434$-$0.260SMM & 19:09:49.8 & +08:19:44 & $19.6\pm$ 6.1$^{\dag}$& $1.8\pm$ 0.2& $26.4\pm 8.8$ & $2.6\pm 0.5$ \\
G43.18$-$0.52 (a) & G43.179$-$0.519SMM & 19:12:09.0 & +08:52:11 & $22.3\pm$ 6.9& $3.9\pm$ 0.4& $124.1\pm 38.9$ & $13.2\pm 2.3$ \\
G43.18$-$0.52 (b) & G43.158$-$0.518SMM & 19:12:06.4 & +08:51:05 & $<5.1$ & $0.6\pm$ 0.1& $<5.1$ & $1.2\pm 0.2$ \\
G43.24$-$0.05  & G43.237$-$0.047SMM & 19:10:33.9 & +09:08:21 & $27.1\pm$ 8.3& $4.7\pm$ 0.5& $104.7\pm 32.6$ & $13.3\pm 1.7$ \\
G43.80$-$0.13  & G43.795$-$0.127SMM & 19:11:53.9 & +09:35:50 & $62.5\pm$ 18.8& $7.1\pm$ 0.7& $133.7\pm 42.2$ & $13.8\pm 1.7$ \\
G45.07+0.13    & G45.071+0.131SMM & 19:13:22.3 & +10:50:51 & $20.3\pm$ 6.3& $6.7\pm$ 0.7& $80.2\pm 27.7$ & $15.5\pm 1.8$ \\
G45.12+0.13    & G45.122+0.131SMM & 19:13:28.1 & +10:53:34 & $12.0\pm$ 3.9$^{\dag}$ & $5.0\pm$ 0.5& $103.5\pm 37.3$ & $24.6\pm 2.7$ \\
G45.47+0.05  (a)  & G45.467+0.045SMM & 19:14:25.8 & +11:09:28 & \ldots & $4.5\pm$ 0.5& \ldots & $16.0\pm 1.8$ \\
G45.47+0.05  (b)  & G45.455+0.060SMM & 19:14:21.3 & +11:09:15 & \ldots & $2.8\pm$ 0.3& \ldots & $8.9\pm 1.1$ \\
G69.54$-$0.98  & G69.541$-$0.975SMM & 20:10:09.1 & +31:31:37 & $62.8\pm$ 18.9& $8.9\pm$ 0.9& $470.2\pm 150.2$ & $32.9\pm 4.7$ \\
G70.29+1.60    & G70.293+1.600SMM & 20:01:45.6 & +33:32:44 & $53.6\pm$ 16.1& $13.6\pm$ 1.4& $256.4\pm 79.3$ & $37.6\pm 4.3$ \\
G76.18+0.13  (a)  & G76.184+0.082SMM\setcounter{footnote}{0}\footnotemark & 20:23:58.3 & +37:37:29 & $8.1\pm$ 2.9$^{\dag}$ & $1.1\pm$ 0.2& $83.9\pm 30.7$ & $2.7\pm 0.9$ \\
G76.18+0.13  (b)  & G76.188+0.097SMM\setcounter{footnote}{2}\footnotemark & 20:23:55.2 & +37:38:10 & $11.8\pm$ 3.8$^{\dag}$ & $1.1\pm$ 0.2& $128.5\pm 44.7$ & $6.7\pm 1.9$ \\
G76.18+0.13  (c)  & G76.192+0.090SMM\setcounter{footnote}{0}\footnotemark & 20:23:57.7 & +37:38:09 & \ldots & $0.8\pm$ 0.2& \ldots & $2.7\pm 0.6$ \\
G76.38$-$0.62 (a) & G76.385$-$0.623SMM & 20:27:27.5 & +37:22:49 & $24.5\pm$ 7.6$^{\dag}$ & $3.4\pm$ 0.4& $192.3\pm 62.9$ & $17.0\pm 1.9$ \\
G76.38$-$0.62 (b) & G76.382$-$0.620SMM & 20:27:26.1 & +37:22:46 & $26.8\pm$ 8.3$^{\dag}$ & $3.6\pm$ 0.4& $48.2\pm 15.0$ & $6.1\pm 0.7$ \\
G76.38$-$0.62 (c) & G76.386$-$0.637SMM\setcounter{footnote}{3}\footnotemark & 20:27:31.1 & +37:22:25 & $12.4\pm$ 4.2$^{\dag}$ &  \ldots & $53.1\pm 18.7$ & \ldots \\
G77.97$-$0.01 (a) & G77.953+0.005SMM & 20:29:31.9 & +39:01:10 & $8.8\pm$ 3.0$^{\dag}$ & $0.8\pm$ 0.1& $18.5\pm 6.9$ & $2.6\pm 0.5$ \\
G77.97$-$0.01 (b) & G77.962$-$0.008SMM & 20:29:36.8 & +39:01:09 & $5.5\pm$ 2.2$^{\dag}$ & $0.7\pm$ 0.1& $9.3\pm 3.4$ & $1.7\pm 0.4$ \\
G78.44+2.66  (a)  & G78.438+2.660SMM & 20:19:39.1 & +40:56:39 & $25.2\pm$ 7.7$^{\dag}$ & $1.7\pm$ 0.2& $181.3\pm 59.8$ & $9.7\pm 1.2$ \\
G78.44+2.66  (b)  & G78.446+2.659SMM & 20:19:40.7 & +40:57:02 & $21.2\pm$ 6.5$^{\dag}$ & $1.4\pm$ 0.2& $119.5\pm 37.1$ & $7.0\pm 0.9$ \\
G79.30+0.28  (a)  & G79.296+0.281SMM & 20:32:29.2 & +40:16:02 & $12.2\pm$ 4.0$^{\dag}$ & $1.1\pm$ 0.1& $89.3\pm 31.0$ & $6.2\pm 0.9$ \\
G79.30+0.28  (b)  & G79.316+0.279SMM\setcounter{footnote}{4}\footnotemark & 20:32:33.8 & +40:16:55 & $10.7\pm$ 3.5$^{\dag}$ & $1.5\pm$ 0.2& $35.6\pm 11.2$ & $3.4\pm 0.5$ \\
G79.30+0.28  (c)  & G79.306+0.278SMM & 20:32:31.8 & +40:16:23 & $7.9\pm$ 2.8$^{\dag}$ & $0.7\pm$ 0.1& $51.0\pm 15.6$ & $3.8\pm 0.5$ \\
G79.32+1.31  (a)  & G79.308+1.307SMM & 20:28:09.7 & +40:52:50 & $16.6\pm$ 5.2$^{\dag}$ & $1.4\pm$ 0.2& $59.2\pm 22.0$ & $4.9\pm 0.8$ \\
G79.32+1.31  (b)  & G79.319+1.312SMM & 20:28:10.3 & +40:53:32 & \ldots & $0.8\pm$ 0.1& \ldots & $1.1\pm 0.2$ \\
G79.32+1.31  (c)  & G79.324+1.291SMM & 20:28:16.5 & +40:53:01 & $7.4\pm$ 2.7$^{\dag}$ & $0.7\pm$ 0.1& $19.9\pm 6.9$ & $2.3\pm 0.3$ \\
G80.87+0.42    & G80.865+0.419SMM & 20:36:52.5 & +41:36:22 & $16.6\pm$ 5.2$^{\dag}$ & $3.4\pm$ 0.4& $193.2\pm 63.5$ & $13.6\pm 1.9$ \\
G106.80+5.31 (a)  & G106.794+5.313SMM & 22:19:16.7 & +63:18:42 & $30.0\pm$ 9.5& $6.8\pm$ 0.7& $433.1\pm 156.7$ & $48.7\pm 7.7$ \\
G106.80+5.31 (b)  & G106.800+5.314SMM & 22:19:18.9 & +63:18:59 & $18.7\pm$ 6.3& $4.7\pm$ 0.6& $173.7\pm 54.7$ & $9.3\pm 1.1$ \\
G109.87+2.11   & G109.871+2.113SMM & 22:56:18.3 & +62:01:44 & $76.6\pm$ 23.1& $14.5\pm$ 1.5& $679.5\pm 217.6$ & $75.7\pm 9.9$ \\
G110.21+2.63   & G110.211+2.609SMM & 22:57:07.3 & +62:37:22 & $<5.4$ & $0.6\pm$ 0.3& $<5.4$ & $7.1\pm 2.5$ \\
G111.28$-$0.66 (a) & G111.282$-$0.665SMM\setcounter{footnote}{2}\footnotemark & 23:16:04.1 & +60:01:55 & $22.0\pm$ 6.7$^{\dag}$ & $2.4\pm$ 0.3& $150.6\pm 45.9$  & $13.1\pm 2.2$ \\
G111.28$-$0.66 (b) & G111.281$-$0.670SMM\setcounter{footnote}{2}\footnotemark & 23:16:04.4 & +60:01:38 & \ldots & $1.5\pm$ 0.2& \ldots& $5.8\pm 0.6$ \\
G111.61+0.37   & G111.612+0.374SMM & 23:15:31.3 & +61:07:12 & $15.9\pm$ 4.9& $2.9\pm$ 0.3& $163.6\pm 52.1$ & $14.9\pm 3.2$ \\
G133.95+1.06   & G133.949+1.064SMM & 02:27:04.7 & +61:52:23 & $78.5\pm$ 23.6& $20.4\pm$ 2.1& $482.0\pm 148.6$ & $62.1\pm 8.7$ \\
G138.30+1.56   & G138.296+1.557SMM & 03:01:32.3 & +60:29:16 & $35.2\pm$ 10.7$^{\dag}$ & $2.8\pm$ 0.4& $183.1\pm 57.0$$^{\dag}$  & $19.6\pm 6.9$ \\
G188.79+1.03   & G188.795+1.031SMM & 06:09:07.2 & +21:50:38 & \ldots & $2.1\pm$ 0.4& \ldots & $8.4\pm 3.7$ \\
G188.95+0.92   & G188.950+0.884SMM & 06:08:53.5 & +21:38:13 & \ldots & $5.7\pm$ 0.6& \ldots & $24.4\pm 6.8$  \\
G213.88$-$11.84 & G213.880$-$11.836SMM & 06:10:50.8 & $-$06:11:48 & \ldots & $6.0\pm$ 0.6& \ldots & $31.8\pm 3.9$ \\
\footnotetext[1]{A noisy bolometer clipped from the final rebinned image is located close to this source and thus the measured flux is a lower limit to the true
flux.}
\footnotetext[2]{These sources are separable in the unsmoothed 450 $\mu$m image but blended in the 850 $\mu$m image. The quoted 850 $\mu$m flux is that of
the blended single source. The coordinates quoted are measured from the 450 $\mu$m image.}
\footnotetext[3]{These sources are separable in the 850 $\mu$m images but blended into one source in the smoothed 450 $\mu$m images. The quoted 450 $\mu$m
flux is the flux of the blended single source.}
\footnotetext[4]{A noisy bolometer clipped from the final rebinned 850 $\mu$m image lies over this source and it was not possible to masure any sensible
peak or integrated 850 $\mu$m flux.}
\footnotetext[5]{This sources lies on the edge of the field of view and thus one or more of the quoted fluxes is a lower limit to the true flux.}
\end{longtable}

\end{document}